\documentclass[aps,prd,twocolumn,superscriptaddress,nofootinbib]{revtex4-1}
\pdfoutput=1
\usepackage{aas_macros}
\usepackage[normalem]{ulem}
\usepackage[utf8]{inputenc}
\usepackage[english]{babel}
\usepackage{amsmath}
\usepackage{amsfonts}
\usepackage{amssymb}
\usepackage{listings}
\usepackage{lipsum}
\usepackage{multirow}
\usepackage{datetime}
\usepackage{graphicx}
\usepackage{mathtools}
\usepackage{mathrsfs}
\usepackage{dcolumn}
\usepackage{comment}
\usepackage{multirow}

\usepackage{color}   %May be necessary if you want to color links
\usepackage{xcolor}

\usepackage{hyperref}
\hypersetup{
    colorlinks=true, 
    pdfborder = {0 0 0.5 [3 3]},
    anchorcolor=black,
    citecolor=blue,
    linktoc=all,    
    linktocpage=true,
    linkcolor=red,
	urlcolor=blue
}

\newcommand{\vect}[1]{\boldsymbol{\mathbf{#1}}}

\begin{document}

\title{Dark photon superradiance: Electrodynamics and multimessenger signals}

\author{Nils Siemonsen}
\email[]{nsiemonsen@perimeterinstitute.ca}
\affiliation{Perimeter Institute for Theoretical Physics, Waterloo, Ontario N2L 2Y5, Canada}
\affiliation{Arthur B. McDonald Canadian Astroparticle Physics Research Institute, 64 Bader Lane, Queen's University, Kingston, Ontario K7L 3N6, Canada}
\affiliation{Department of Physics \& Astronomy, University of Waterloo, Waterloo, Ontario N2L 3G1, Canada}
\author{Cristina Mondino}\email[]{cmondino@perimeterinstitute.ca}\affiliation{Perimeter Institute for Theoretical Physics, Waterloo, Ontario N2L 2Y5, Canada}
\author{Daniel Ega\~na-Ugrinovic}\email{degana@perimeterinstitute.ca}\affiliation{Perimeter Institute for Theoretical Physics, Waterloo, Ontario N2L 2Y5, Canada}
\author{Junwu Huang}\email{jhuang@perimeterinstitute.ca}\affiliation{Perimeter Institute for Theoretical Physics, Waterloo, Ontario N2L 2Y5, Canada}
\author{Masha Baryakhtar}\email[]{mbaryakh@uw.edu}
\affiliation{Physics Department, University of Washington, Seattle, WA 98195-1560, USA}
\author{William E.\ East}\email{weast@perimeterinstitute.ca}\affiliation{Perimeter Institute for Theoretical Physics, Waterloo, Ontario N2L 2Y5, Canada}

\date{\today}

\begin{abstract}
We study the electrodynamics of a kinetically mixed dark photon cloud that forms through superradiance  around a spinning black hole, and design strategies to search for the resulting multimessenger signals. A dark photon
superradiance cloud sources a rotating dark electromagnetic field which,
through kinetic mixing, induces a rotating visible electromagnetic field.
Standard model charged particles entering this field 
initiate a transient phase of particle production that populates a plasma
inside the cloud and leads to a system which shares
qualitative features with a pulsar magnetosphere. We study the
electrodynamics of the dark photon cloud with resistive
magnetohydrodynamics methods applicable to highly magnetized plasma,
adapting techniques from simulations of pulsar magnetospheres.  We identify
turbulent magnetic field reconnection as the main source of dissipation and
electromagnetic emission, and compute the peak luminosity from clouds around solar-mass black holes to be as large as $10^{43}\,{\rm erg/s}$ for observationally-allowed dark photon parameter space. The emission is expected to have a significant X-ray component and to potentially be 
periodic, with period set by the dark photon mass. The luminosity is comparable to the brightest X-ray sources in the Universe,  allowing for searches at
distances of up to hundreds of Mpc with existing telescopes. We discuss observational
strategies, including targeted electromagnetic follow-ups of solar-mass black hole mergers and targeted continuous gravitational wave searches
of anomalous pulsars.

\end{abstract}

\maketitle
\tableofcontents

\section{Introduction}

Ultralight fields arise in abundance in Beyond the Standard Model (SM) theories of particle physics. The most well-known and well-motivated such particle is the QCD axion~\cite{Weinberg1978,Wilczek1978}, proposed to solve the  discrepancy between the observed and predicted magnitude of the neutron electric dipole moment arising from CP violation in the strong sector of the SM~\cite{Peccei:1977hh}. Beyond the QCD axion, light bosonic fields have been found to be ubiquitous in string theory \cite{Arvanitaki:2009fg,Svrcek:2006yi,Abel:2008ai,Goodsell:2009xc}, and provide excellent candidates for the dark matter particle or a dark matter mediator \cite{Preskill:1982cy,Abbott:1982af, Dine:1982ah,Graham:2015rva, Nelson:2011sf,Arias:2012az,Essig:2013lka,Adams:2022pbo,Antypas:2022asj}, making this class of particles one of the most exciting candidates for new physics.

Black hole (BH) superradiance~\cite{Zeldovich1971,Misner1972,Starobinskii1973,Detweiler:1980uk,Bekenstein:1998nt, Brito:2015oca} is a unique mechanism that enables searches for weakly interacting ultralight bosons~\cite{Arvanitaki:2009fg,Arvanitaki:2010sy} that relies only on the boson's gravitational interaction.  If a new light boson with Compton wavelength of order the BH horizon size exists in the theory---whether or not there is an initial abundance of the particle in the environment---the BH will spin down and source macroscopic, coherent, gravitationally-bound states of ultralight bosons~\cite{Arvanitaki:2009fg,Arvanitaki:2010sy,Brito:2015oca}. These bosonic ``clouds"  carry up to several percent of the BH's initial mass, and have an energy density comparable to that of neutron star matter for stellar mass BHs~\cite{Arvanitaki:2010sy,East:2017ovw,East:2018glu}. 
The resulting large energy density of the cloud has time-dependent components, rotating around the BH axis at a frequency fixed by the particle mass, resulting in coherent, monochromatic gravitational  wave (GW) radiation that depletes the cloud over parametrically longer times~\cite{Arvanitaki:2010sy,Yoshino:2013ofa,Arvanitaki:2014wva,1969ApJ...157..869G}.

The signatures of GW emission and BH spindown have been proposed to constrain and search for ultra-light bosons  \cite{Arvanitaki:2010sy,Rosa:2011my,Pani:2012bp,Pani:2012vp,Yoshino:2013ofa,Brito:2013wya,Arvanitaki:2014wva,Brito:2014wla,Brito:2015oca,Arvanitaki:2016qwi,East:2017mrj,Brito:2017wnc,Brito:2017zvb,Baryakhtar:2017ngi,East:2018glu,Baumann:2019eav,Siemonsen:2019ebd,Brito:2020lup,Zhu:2020tht}. Bosons in the   $10^{-13}-10^{-11}$~eV range can lead to up to thousands of GW signals originating from our Galaxy alone \cite{Arvanitaki:2014wva,Arvanitaki:2016qwi,Brito:2017wnc,Brito:2017zvb}. Blind continuous wave searches for monochromatic GW from scalar boson clouds \cite{KAGRA:2021tse,KAGRA:2022osp,Palomba:2019vxe,DAntonio:2018sff,Dergachev:2019wqa,Zhu:2020tht}, as well as stochastic searches for an excess of GW power from spin-0 \cite{Tsukada:2018mbp} and spin-1  \cite{Tsukada:2020lgt} boson clouds around yet undiscovered BHs have been  carried out with LIGO-Virgo-KAGRA (LVK)~\cite{TheLIGOScientific:2014jea,TheVirgo:2014hva,kagra} data. These searches have produced some constraints; however, a robust underlying BH natal spin distribution is needed to conclusively exclude  particle parameter space. Another search strategy is to follow up BHs with a measured mass and spin, which are newly born from binary BH mergers~\cite{Arvanitaki:2016qwi}; then the expected signals can be precisely computed, and a conclusive search performed. Directed searches for continuous GWs from a potential scalar boson cloud around Cygnus X-1 have also been carried out \cite{Sun:2019mqb}. Currently, the sensitivity of GW searches is not sufficient to see follow-up signals from spin-0 bosons around binary BH remnants \cite{Isi:2018pzk}, but they are promising for spin-1 bosons in upcoming observation runs, and especially in next-generation observatories~\cite{Chan:2022dkt,Siemonsen:2022yyf}. 

BH spin measurements have set constraints on ultralight bosons using measurements of BH properties from X-ray binaries \cite{Arvanitaki:2014wva,Cardoso:2018tly,Baryakhtar:2020gao,Mehta:2020kwu}, as well as measurements of binary BH constituents using LVK observations \cite{Cardoso:2018tly,Ng:2019jsx,Ng:2020ruv}. The latter have produced constraints on spin-0 bosons of a factor of $\sim 2$ in mass~\cite{Ng:2019jsx,Ng:2020ruv}, and we  expect slightly stronger constraints for spin-1 bosons, although such an analysis has not been carried out for the full data set. The X-ray binary measurements depend on BH accretion disk modeling, which may introduce additional systematics~\cite{Li:2004aq,DiskPaper}. In addition, while the gravitational interaction of the BH superradiance cloud and the accretion disk does not significantly affect the constraints \cite{Arvanitaki:2014wva}, non-gravitational interactions of the cloud can perturb the disk dynamics,  invalidating the constraints on the dark photon mass from BH spin measurements for the parameters considered in this paper. 

While the gravitational aspects of superradiance have been studied extensively, making contact with particle physics models of ultralight spin-0 or spin-1 particles can dramatically change this picture. For spin-0 axions, the relevant interaction at the next-order after the mass term is  a quartic coupling (see e.g. \cite{Arvanitaki:2010sy,Yoshino:2012kn,Yoshino:2015nsa, Gruzinov:2016hcq,Fukuda:2019ewf}). This results in energy exchange between levels in the cloud~\cite{Gruzinov:2016hcq,Baryakhtar:2020gao,Omiya:2022gwu}, slowing down the spin extraction and resulting in lower-frequency gravitational waves from transitions and axion wave emission \cite{Baryakhtar:2020gao}.  More complicated dark sectors can result in production of new dark states \cite{Baryakhtar:2017ngi,Fukuda:2019ewf,Mathur:2020aqv,East:2022ppo,East:2022rsi,Cannizzaro:2022xyw}. Beyond interactions within the dark sector itself, interactions with SM particles can lead to additional energy loss channels~\cite{Sen:2018cjt}, although for axion-like particles these are subdominant to the dynamics of self-interactions~\cite{Baryakhtar:2020gao}.

\begin{figure*}[t!]
\centering
\includegraphics[width=0.46\textwidth]{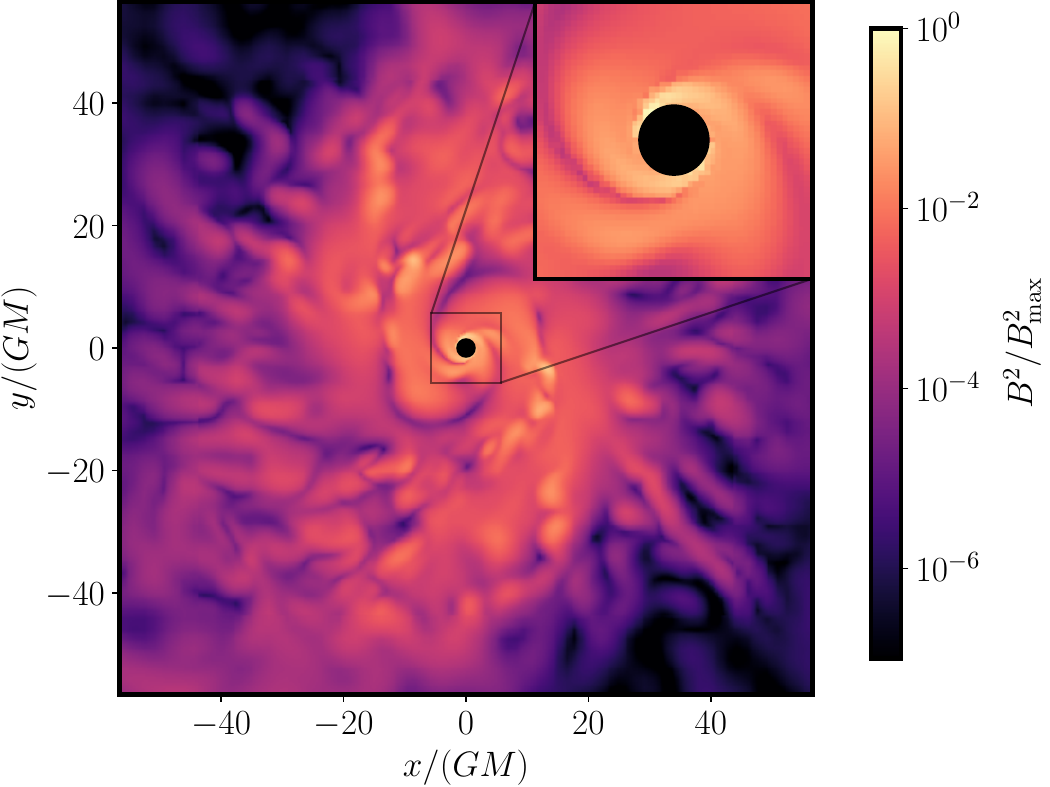}
\hspace{.5cm}
\includegraphics[width=0.5\textwidth]{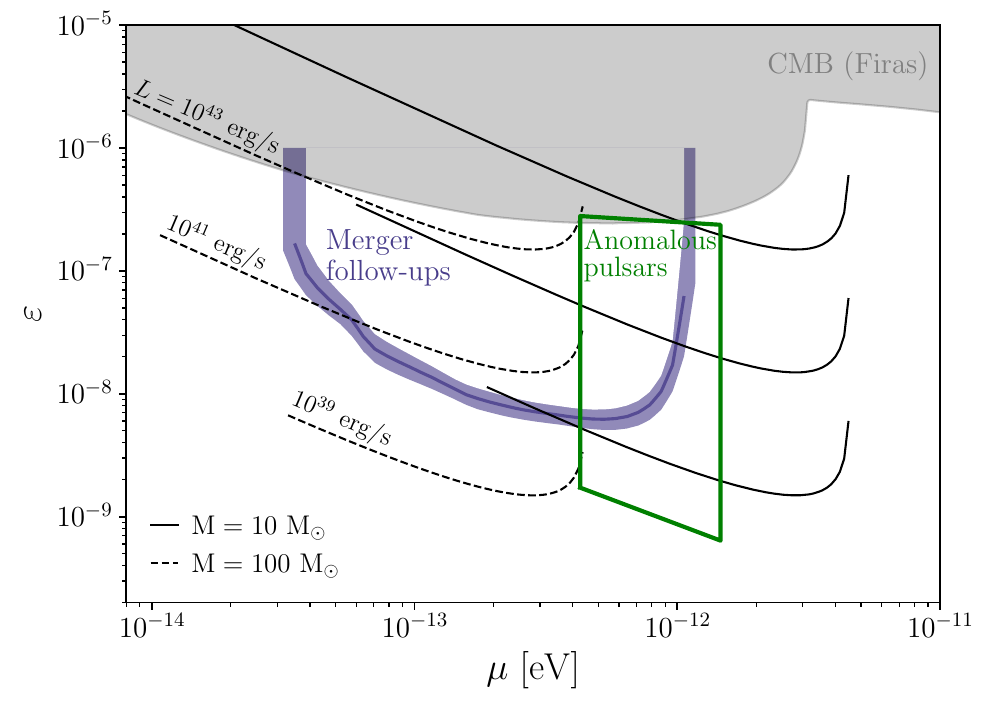}
\label{fig:eps_mu}
\caption{\textit{(left)} We show the visible magnetic field strength $B^2$ (normalized by its maximum) in the equatorial plane of the central BH of mass $M$, dimensionless spin $a_*=0.86$, and a dark photon mass $\mu=0.3/(GM)$. The dark photon of the superradiance cloud forces the pair plasma into a circular motion resulting in magnetic field line twisting, which is released through magnetic field line reconnection, resulting in a turbulent plasma state (shown here) and efficient energy dissipation into the plasma, driving the luminous electromagnetic emissions from the system. \textit{(right)} Kinetically mixed dark photon parameter space of interest in this work. The solid (dashed) black lines are contours of constant electromagnetic luminosity emitted from the superradiance cloud around a BH of mass 10 (100) $\mathrm{M}_{\odot}$ and initial spin $a_*=0.9$. The region above the blue contour is relevant for electromagnetic follow-ups of compact binary mergers, discussed in Sec.~\ref{sec:EWfromMerger} (the shaded band on top of the contour is due to uncertainties on the merger rate).
The area within the green contour is of interest for continuous gravitational waves searches targeted on anomalous pulsars, as described in Sec.~\ref{sec:GWfromPulsar}. The gray shaded region is excluded by existing measurements of the CMB spectrum by COBE/FIRAS \cite{Fixsen:1996nj,Caputo:2020bdy}.}
\label{fig:summary}
\end{figure*}

In this work, we focus on studying the effects of the lowest-order interactions one can write down for spin-1 dark photons: kinetic mixing with the SM photon~\cite{Okun:1982xi,Holdom:1985ag}.
In the presence of such a mixing, the huge energy density of the cloud picks up a visible electromagnetic field component that interacts directly with electrons, leading to cascade production of charged particles and to the formation of a plasma. To study the plasma dynamics, we analyze an isolated, relativistic superradiance dark photon cloud, and compute the evolution of the visible electric and magnetic fields using a resistive-magnetohydrodynamic description, valid in the limit
of a strongly magnetized, tenuous plasma, that we adapt from simulations of pulsar magnetospheres. See Fig.~\ref{fig:summary} for an example visualization of the resulting magnetic field strength around a rotating BH.

Our simulations show that the resulting system is a luminous multimessenger source: a BH system which emits an enormous electromagnetic flux, up to several orders of magnitude brighter than pulsars and magnetars. This radiation is generated by turbulent field and plasma dynamics in the superradiance cloud, and is expected to have a large high-energy component.  We find partial evidence for an intrinsic periodicity set by the mass of the dark photon particle, giving rise to a novel object that we call a ``new pulsar."  In Fig.~\ref{fig:summary}, we show the parameter space of dark photon particles and the expected  peak luminosity for illustrative BHs  as a function of dark photon mass and kinetic mixing parameter.  

 Our results motivate  a variety of novel astrophysical searches to discover these systems. These include electromagnetic follow-ups of BH mergers that result in rotating BHs, most promising in the X-ray and radio bands. Another target is  gravitational wave follow-ups of pulsars with coincident frequencies or positive frequency drifts, which could be superradiance cloud signals lasting thousands of years or more; see Fig.~\ref{fig:summary}. In much of the parameter space, we find that the evolution of the cloud is still dominated by its gravitational dynamics, making the overall evolution free of electromagnetic modeling uncertainties. At small dark photon masses and large kinetic mixings, the electromagnetic emission has a larger power than the GWs, where we also have an exceptionally bright sources.

Some aspects of dark photon superradiance with a non-zero kinetic mixing have been explored in \cite{Caputo:2021efm}, and superradiance of the SM photon itself has been treated in ~\cite{Pani:2012vp,Blas:2020kaa,Cannizzaro:2020uap,Cannizzaro:2021zbp}. Ours is the first work to consistently take into account the dynamics of the SM plasma that is automatically generated by the  kinetically-mixed cloud. The interactions with the plasma completely alter the behavior of the visible electromagnetic fields in the vicinity of the BH and the resulting signatures.  

This paper is organized as follows; in Sec.~\ref{sec:sr}, we review gravitational spin-1 superradiance. In Sec.~\ref{sec:cloudsummary}, we provide an executive summary of the  dynamics of kinetically mixed superradiance, which we explore in detail in the subsequent sections. In Sec.~\ref{sec:SFQED}, we describe the processes by which an isolated dark photon cloud generates its own plasma density. In Sec.~\ref{sec:fields}, we study the dynamics of the coupled system of electromagnetic fields and charged currents. In Sec.~\ref{sec:luminosity}, we describe the key electromagnetic emission mechanisms, including electromagnetic radiation and power dissipation in the plasma due to turbulent dynamics. In Sec.~\ref{sec:observation}, we summarize the observational signatures and propose several detection strategies for this new class of astrophysical object, concluding and outlining future directions in Sec.~\ref{sec:conclusions}.

This work spans the areas of particle physics, strong field electrodynamics, gravity, and astrophysical systems, thus introducing much notation, some non-standard; we collect definitions in App.~\ref{app:notation}.  We describe details of the numerical simulations of the superradiance cloud and electromagnetic fields and currents  in App.~\ref{app:procaconstruction} and \ref{app:numericalsetup}, respectively. We present the resistive current prescription in App.~\ref{app:resistivitytest}, the small conductivity regime in App.~\ref{app:lowconductivityregime}, and aspects of the dark photon basis in App.~\ref{appendix:darkphoton}. We use the mostly-plus metric signature $(-,+,+,+)$ and natural units, with $\hbar=c=1$ and  non-reduced Planck mass $M_{\mathrm{pl}} = 1/\sqrt{G}$.

\section{Black hole superradiance for vector fields}
\label{sec:sr}

We begin by reviewing the BH superradiance of a massive vector (spin-1) boson that interacts predominantly through gravity. The kinetic and mass terms for this dark photon $A'^{\mu}$ are given by
\begin{equation}
    \mathcal{L'}=-\frac{1}{4}F'_{\mu\nu} F'^{\mu\nu}-\frac{1}{2}\mu^2 A'^{\mu}A'_{\mu} .
    \label{eq:procalagrangian}
\end{equation}
We assume that the dark photon mass $\mu$ arises from the Stueckelberg mechanism \cite{2009esuf} so in what follows, we do not discuss any dynamics that could originate from a Higgs sector \cite{East:2022ppo,East:2022rsi}. 

The superradiant instability is a purely gravitational process that can lead to the production of an exponentially large number of massive bosons around spinning BHs by extracting the BH's energy and angular momentum. The bosons occupy hydrogenic clouds characterized by a gravitational fine-structure constant $\alpha \equiv r_g \mu = G M \mu$, a principal quantum number $n$, and total, orbital, and magnetic angular momentum numbers $j$, $l$, and $m$. The total and orbital angular momentum can differ due to the boson's intrinsic spin: $j\in\{l-1,l,l+1\}$. Amongst the different cloud levels, the fastest-growing one for vector bosons is the $(j,n,l,m)=(1,1,0,1)$ mode. Given its dynamical dominance, we focus for brevity exclusively on the study of this level. This will be sufficient for exploring the features that we wish to highlight in this work. The  boson's energy in this level at leading-order in the gravitational coupling is given by
\begin{equation}
    \omega \simeq \mu \bigg(1-\frac{\alpha^2}{2}\bigg) .
\label{eq:energy}
\end{equation}

After the birth of the source BH, the number of dark photons in the cloud grows exponentially at a leading-$\alpha$ rate that, for our dynamically dominant mode, is set by
\begin{equation}
\Gamma_{\mathrm{SR}}\equiv \tau_{\mathrm{SR}}^{-1} \simeq 4 \alpha^{7}(\Omega_{\mathrm{BH}}-\omega) \simeq 4 a_* \alpha^6 \mu ,
\label{eq:rate}
\end{equation}
where $a_*$ is the BH's dimensionless spin and $\Omega_{\mathrm{BH}}$ its angular velocity
\begin{equation}
\Omega_{\mathrm{BH}}=\frac{1}{2}\left(\frac{a_*}{1+\sqrt{1-a_*^2}}\right)r_g^{-1}  .
\label{eq:omegabh}
\end{equation}
In the last equality of \eqref{eq:rate}, we approximated   $\Omega_{\mathrm{BH}}\gg \omega$ and took large-spin BHs $(1-a_*^2)\ll 1$. 
If, on the other hand, the BH's spin is small so that its angular velocity falls below the boson's energy
\begin{equation}
\Omega_{\mathrm{BH}} \leq \omega, 
\label{eq:srcondition}
\end{equation}
then $(j,n,l,m)=(1,1,0,1)$ superradiance does not occur. Equation~\eqref{eq:srcondition} implies a maximum possible value for the fine-structure constant (saturated for maximally spinning BHs)
\begin{equation}
\alpha \lesssim 1/2  .
\end{equation}
The above condition, together with the strong suppression of the superradiant rate at small $\alpha$ [see \eqref{eq:rate}] indicate that superradiance is most effective for gravitational couplings of order $\alpha \sim 10^{-1}$ or, equivalently, boson masses $\mu \sim 0.1/r_g$, which for stellar BHs corresponds to $\mu \sim 10^{-12}\, \mathrm{eV}$. 

\begin{figure*}[t]
    \includegraphics[width=0.42\textwidth]{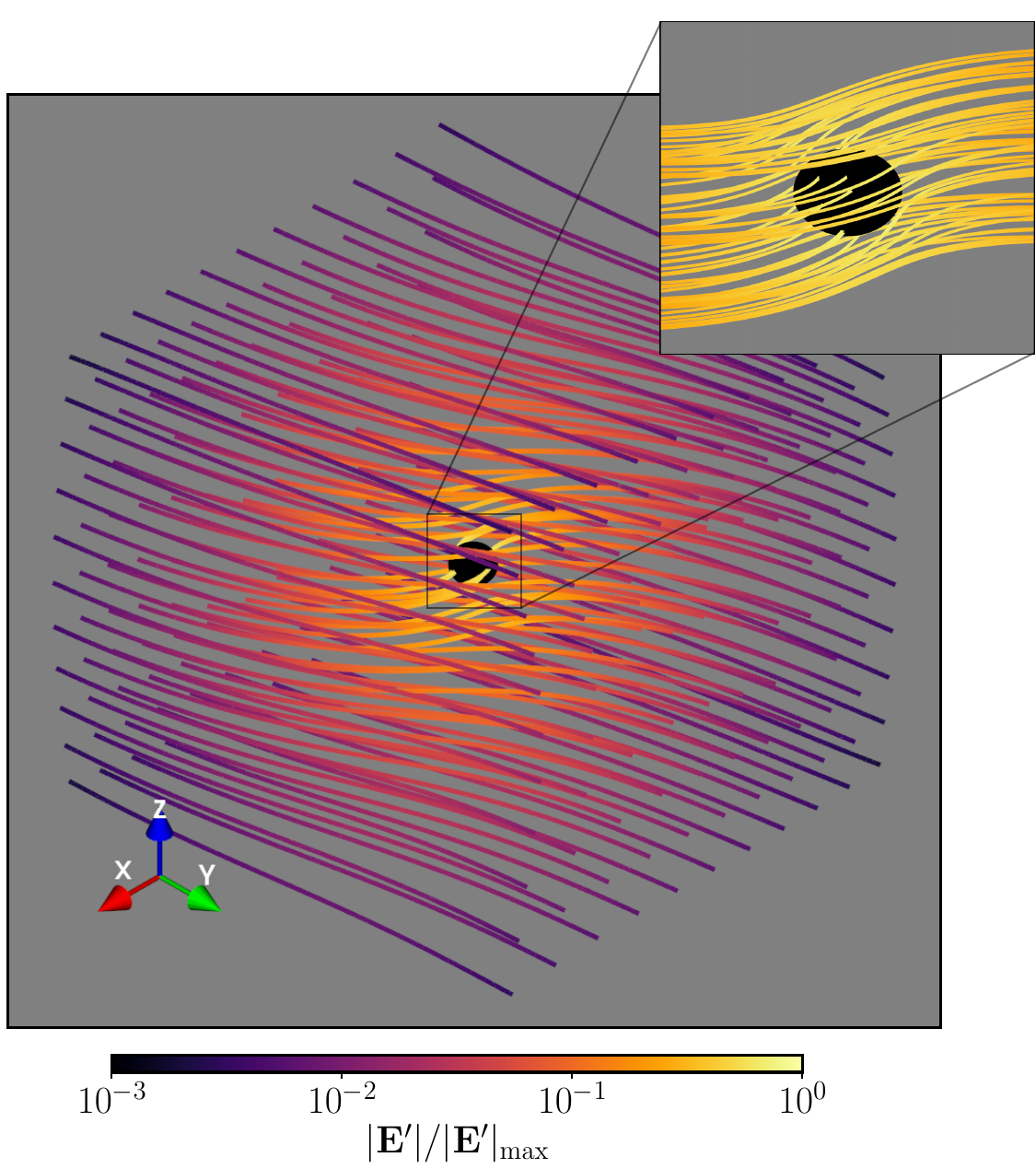}
    \hspace{1cm}
    \includegraphics[width=0.42\textwidth]{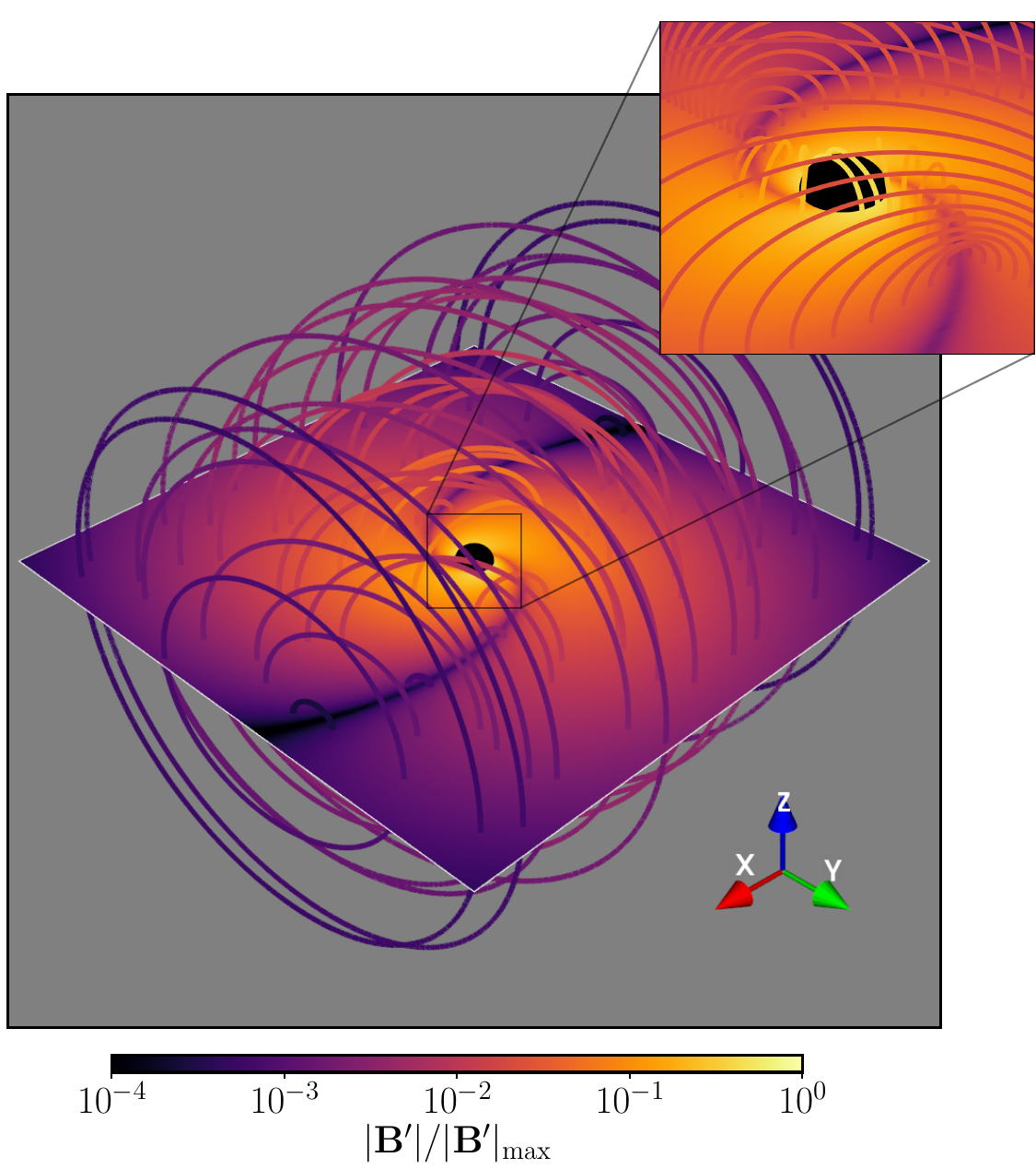}
    \caption{We plot representative sets of field lines of the electric
    \textit{(left)} and magnetic fields \textit{(right)} of the superradiant
    cloud around the central BH in Kerr-Schild coordinates (see App.~\ref{app:procaconstruction} for details). The $m=1$ cloud is
    characterized by $\alpha=0.3$, while the BH has a corresponding spin of
    $a_*=0.86$ (further details can be found in Table~\ref{tab:clouds}). The BH
    spin-axis points in the $z$-direction. Color indicates the field strength
    along each field line, normalized by the respective maximum field strength.
    On the right, we also plot the magnetic field strength inside the
    equatorial plane of the BH.}
    \label{fig:vaccumFieldlines}
\end{figure*}

The growth of the cloud stops when sufficient spin has been extracted so that the condition Eq. \eqref{eq:srcondition} is saturated. The number of dark photons in the cloud can reach $10^{77}$ or more for a $10$ solar mass BH, with the cloud mass
\begin{equation}
  M_{c} \simeq 10^{-2} \bigg(\frac{\Delta a_*}{0.1}\bigg)\bigg(\frac{\alpha}{0.1}\bigg) M  ,
  \label{eq:mcloudapprox}
\end{equation}
 for $\alpha\ll 1$, where $\Delta a_*$ is the difference between the initial BH spin and the final spin which saturates the superradiance condition. The cloud mass  reaches up to $10\%$ of the mass of the BH for large $\alpha$ and high initial BH spin~\cite{East:2017ovw}. The vector field profile around the BH, on the other hand, 
is given at leading order in the fine-structure constant by
\begin{align}
\nonumber A'_0 & = \frac{\sqrt{M_c}}{\sqrt{\pi}\mu^{2} r_c^{5/2}} e^{-r/r_c} \sin\theta\sin(\omega t -\phi) , \\
\vect{A'} & = -\frac{\sqrt{ M_c}}{\sqrt{\pi}\mu r_c^{3/2}} e^{-r/r_c} \lbrace \cos\omega t, \sin\omega t, 0 \rbrace  ,
\label{eq:srprofiles}
\end{align}
where $r_c = {r_g}/{\alpha^2}$ is the cloud's characteristic Bohr radius and we have taken the BH spin direction to lie along the $z$-axis. From Eq. \eqref{eq:srprofiles}, we see that the dark electric and magnetic fields $\vect{E'}\equiv -\nabla A'_0 -  \partial_t \vect{A'}$ and $\vect{B'}\equiv \nabla \times \vect{A'}$ are in a proportion $|\vect{B'}|/|\vect{E'}| \sim \alpha$, so the cloud is electrically dominated. At leading order in $\alpha$, the electric field corresponding to the potential Eq. \eqref{eq:srprofiles} is unidirectional and equatorially oriented, and rotates on this plane at a frequency $\omega\simeq \mu$, while the magnetic field lines form concentric tori around the BH with a common axis perpendicular to the electric field direction and passing through the BH. Both fields decay exponentially away from the BH. We show these features in Fig. \ref{fig:vaccumFieldlines}, where we present exact (in the test field limit) solutions for both the electric and magnetic fields, obtained by numerically solving the vector's equations of motion in the BH's Kerr metric. We refer the reader to Sec.~\ref{sec:fields} and App.~\ref{app:procaconstruction}  for details on the simulations. We note that, close to the BH, the exact field solutions differ from the ones obtained from the approximations \eqref{eq:srprofiles} due to corrections that arise at higher-order in the gravitational coupling. 

Following its formation, the cloud decays via GW emission, which is the main observable signature of superradiance clouds composed of massive bosons that interact with the SM solely by gravitation. The GW emission power is given by 
\begin{equation}
    P_{\mathrm{GW}} \simeq 17 \frac{\alpha^{10}}{G}\bigg(\frac{M_{c}(t)}{M}\bigg)^2  \label{eq:powgw}
\end{equation}
in the $\alpha \ll 1$ limit \cite{Baryakhtar:2017ngi,Siemonsen:2019ebd}. The quadratic dependence of the emission power on the cloud mass leads to a power-law decay of the cloud set by
\begin{equation}
M_c(t)=\frac{M_{c}(t_0)}{1+(t-t_0)/\tau_{\mathrm{GW}}} ,  \label{eq:mct}
\end{equation}
where $\tau_{\mathrm{GW}}$ is the gravitational-wave decay timescale, which is given by 
\begin{equation}
    \tau_{\mathrm{GW}} \simeq \frac{G M}{17 \alpha^{11} \Delta a_*} \sim 30 \, \mathrm{days} \bigg(\frac{0.1}{\Delta a_*}\bigg) \bigg(\frac{0.1}{\alpha}\bigg)^{11} \bigg(\frac{M}{10 M_{\odot}}\bigg)  .
    \label{eq:tgw}
\end{equation}

\section{Kinetically-mixed superradiance clouds: an overview}\label{sec:cloudsummary}

So far, we have discussed a theory where vector bosons lack non-gravitational interactions.
Going beyond this minimal setup, dark photons may interact with the SM at the renormalizable level via kinetic mixing with the SM U(1) gauge boson. In an effective theory below the electroweak scale, this interaction mixes the dark and SM photons via a Lagrangian term $\mathcal L \supset \varepsilon F'_{\mu\nu}F^{\mu\nu}/2$, where $\varepsilon$ is a parameter that quantifies the mixing~\cite{Okun:1982xi,Holdom:1985ag}.
This term can be equivalently written as a mass-mixing term by performing the field redefinition,
$A'_\mu\rightarrow A'_\mu+\varepsilon A_\mu \equiv {A}'_\mu$, which results in the Lagrangian
\begin{align}
\begin{aligned}
    \mathcal{L}=& \ -\frac{1}{4}{F}_{\mu\nu}{F}^{\mu\nu}-\frac{1}{4}{F}'_{\mu\nu}{F}'^{\mu\nu}\\
    & \ -\frac{\mu^2}{2}{A}'_\mu {A}'^\mu-\varepsilon\mu^2 {A}'_\mu {A}^\mu+I_\mu {A}^\mu ,
    \label{eq:interlagrangian}
\end{aligned}
\end{align}
where $I^{\mu}$ is the four-dimensional spacetime current.
This choice of fields is referred to as the interaction basis; other choices of basis are discussed in App.~\ref{appendix:darkphoton}. Due to the mass mixing, the dark photon field acts as a source current for the visible fields and vice-versa, as can be seen either from the equations of motion
\begin{eqnarray}
    \nabla_\alpha {F}^{\alpha\beta}&=&-I^\beta+\varepsilon\mu^2{A}'^\beta ,
    \label{eq:FieldeqInteraction1}
    \\
    \nabla_\alpha {F}'^{\alpha\beta}&=&\mu^2{A}'^\beta+\varepsilon\mu^2 {A}^\beta ,
    \label{eq:FieldeqInteraction2}
\end{eqnarray}
at leading order in the kinetic mixing parameter, or from the energy-momentum conservation relations, which manifestly show exchange of energy between the vector fields
\begin{equation}
\begin{aligned}
    \nabla_\alpha {T}^{\alpha\beta}= & \ -{F}^{\beta\gamma}(I_\gamma-\varepsilon \mu^2 {A}'_\gamma),\\
    \nabla_\alpha {T}'^{\alpha\beta}= & \ \varepsilon\mu^2 {F}'^{\beta\gamma}{A}_\gamma \, .
    \label{eq:energymomentumconservation}
\end{aligned}
\end{equation}

\begin{figure*}[t]
    \centering
    \includegraphics[width=\textwidth]{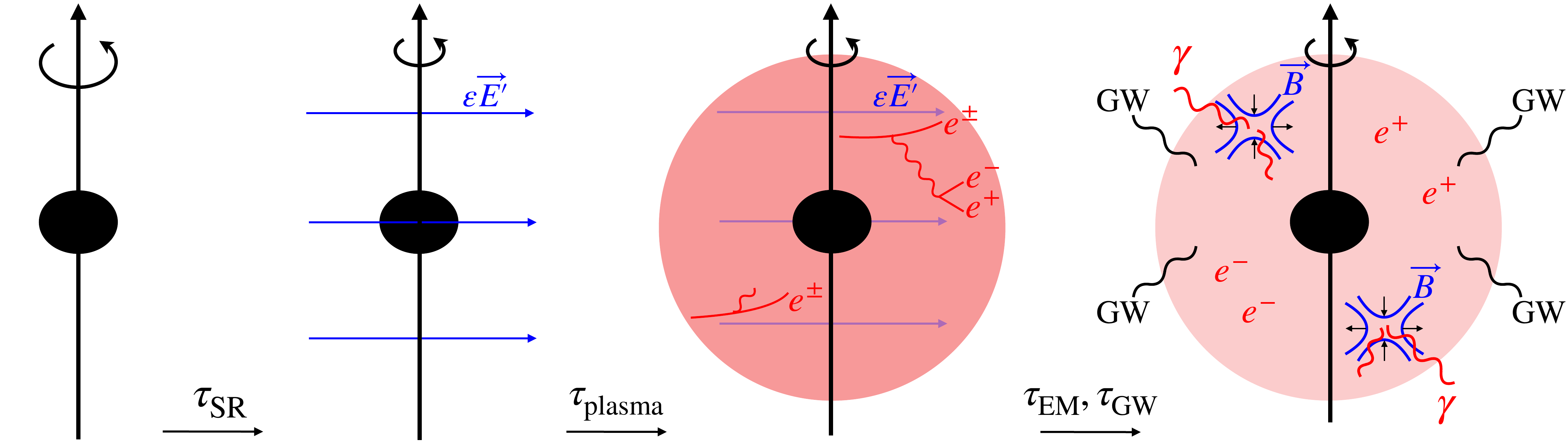}
    \caption{Schematic depiction of the evolution of a kinetically-mixed dark photon superradiance cloud. Starting from a spinning BH ({\it{left}}), a vector superradiance cloud forms on a timescale $\tau_{\mathrm{SR}}$ ({\it{center-left}}).The visible electric field sourced by the cloud accelerates environmental charged particles, leading to cascade production of electrons and positrons on a timescale $\tau_{\mathrm{plasma}}$; note that 
 $\tau_{\mathrm{SR}}\gg\tau_{\mathrm{plasma}}$ and the cascade production occurs and completes before the superradiance instability completes ({\it{center-right}}). The cloud finally decays by GW emission on a timescale $\tau_\mathrm{GW}$, and by transferring energy to the plasma, which loses energy through electromagnetic emission on a timescale $\tau_\mathrm{EM}$({\it{right}}). See text for further details.}

    \label{fig:schematics}
\end{figure*}

In the context of superradiance, the kinetic mixing term allows for the superradiance cloud to source electromagnetic fields\footnote{The induced coupling to SM electrons also results in a higher-dimensional self-interaction term for the dark photons of the Euler-Heisbenberg Lagrangian. Approximately extrapolating the results of self-interacting scalars \cite{Baryakhtar:2020gao}, we estimate that the induced quartic coupling would start to affect the growth of the cloud for $\varepsilon \gtrsim \mathcal{O}(1)$, a value far greater than relevant for the dynamics discussed here, and that is already excluded.}. Our objective in this work is to study the corresponding electrodynamics, and we numerically solve Maxwell's equations with a superradiant source term in curved spacetime. This task is technically complex, but most of our results can be understood in simple physical terms, so to guide the reader through the discussion presented in the following sections, we will begin here by providing a simplified overview of our findings. 

The evolution of a kinetically-mixed dark-photon superradiance cloud can be separated into several stages that are schematically depicted in Fig. \ref{fig:schematics}. Starting with a spinning BH (leftmost panel), these stages correspond to the initial growth of the cloud (center-left panel),  creation of a conducting plasma via particle acceleration and pair-creation due to the visible electric field induced by the cloud (center-right panel), and the establishment of an electromagnetic field and plasma configuration, which decays as the plasma radiates electromagnetically and the cloud emits GWs (rightmost panel). The initial growth of the cloud was reviewed in Sec.~\ref{sec:sr} for non-interacting dark photons.
From Eq.~\eqref{eq:FieldeqInteraction2}, we see that 
the inclusion of kinetic mixing affects the dynamics of the superradiance cloud at order $\varepsilon$, which in turn leads to effects at order $\varepsilon^2$ in the visible fields via Eq.~\eqref{eq:FieldeqInteraction1}. 
Here, we will limit ourselves to computing the visible fields at leading (linear) order in $\varepsilon$, so in what follows we ignore the effects of kinetic mixing on the growth of the superradiance cloud.
Thus, the evolution of the cloud is governed purely by the gravitational dynamics as in the previous section (with a growth timescale $\tau_{\mathrm{SR}}$ given by Eq.~\eqref{eq:rate}), and we may move on to the description of the creation of the plasma by the cloud-induced electric field. See also the text below Eq.~\eqref{eq:tsr_over_tpl} for a discussion of the negligible sub-leading corrections to the dark photon mass.

As discussed in Sec.~\ref{sec:sr}, the cloud is dominated by an equatorially-oriented dark electric field, with a direction that rotates in the equatorial plane at frequency $\mu/(2\pi) \sim 10^2\, \mathrm{Hz}\, (\mu/10^{-12}\,\mathrm{eV})$.
Due to kinetic mixing, the superradiance fields act as a source term in the visible field equations of motion, Eq.~\eqref{eq:FieldeqInteraction1}, and induce a visible electric field that is equal to the dark electric field times the mixing parameter. As the cloud grows through superradiance, the visible field grows concurrently. A fully-formed cloud would, in the absence of charged particles, have a visible electric field of magnitude
\begin{align}
 &|\varepsilon \textbf{E}'|\simeq
 \frac{\varepsilon \sqrt{\Delta a_*} \, \alpha^{5/2} \mu  }{\sqrt{G}}  \nonumber \\ 
 &\simeq 2 \cdot 10^{13}\,\,
 \mathrm{V/m}\, \sqrt{\Delta a_*}\bigg(\frac{\varepsilon}{10^{-7}}\bigg)
  \bigg(\frac{\alpha}{0.1}\bigg)^{5/2} \bigg(\frac{\mu}{10^{-12}\,\mathrm{eV}}\bigg)
  \label{eq:background}
\end{align}
at distances of order the Bohr radius from the BH ($r_c \sim 1/\alpha \mu$). These large fields, however, cannot be achieved due to plasma screening.

Before the cloud reaches its full size, and when visible fields are still only a fraction of the value \eqref{eq:background}, environmental charged particles are accelerated to ultra-relativistic velocities. As illustrated in the middle-right panel in Fig. \ref{fig:schematics}, the rotation of the electric field with the cloud curves the charged particle trajectories, which then emit synchrotron photons. 
These photons, in turn, interact non-perturbatively with the background electric field and produce additional electrons and positrons.
The charge acceleration and pair production processes repeat in a cascade, until a conducting plasma is created. 
This mechanism is reminiscent of  cascade production of electron-positron pairs in the strong magnetic field around supermassive BHs described by Blandford-Znajek \cite{Blandford:1977ds}, with the important difference that our system is electrically instead of magnetically dominated. An extended discussion of the plasma creation will be presented in Sec.~\ref{sec:SFQED}. The effects of the plasma on the dark photon interactions with the visible fields are discussed in App.~\ref{appendix:darkphoton}.

Once the plasma is created, the electrodynamics can be studied by encoding the microscopic particle physics in an effective conductivity $\sigma$,  which allows for the computation of spatial plasma currents from the electromagnetic fields using Ohm's law,
\begin{align}
\mathbf{J}= \sigma(\mathbf{E}+\mathbf{v}\times \mathbf{B})  ,
\label{eq:ohmlaw}
\end{align}
where $\mathbf{v}$ is the plasma fluid velocity. This Ohmic prescription, which will be studied in detail in Sec.~\ref{sec:fields}, is commonly used in magnetohydrodynamics \cite{eckart1940thermodynamics}, and has been proposed to treat dissipative currents in pulsars \cite{Komissarov:2005xc,Gruzinov:2007se,Gruzinov:2008um,Li:2011zh,Palenzuela:2012my}. Pulsar magnetospheres are highly conducting,  $\sigma/\omega \gg 1$ (where $\omega$ is the plasma angular frequency), so that in the bulk of the magnetosphere electromagnetic fields are shorted out and combine to cancel the Lorentz forces entering Eq.~\eqref{eq:ohmlaw}, resulting in a so-called ``force-free'' system that is mostly dissipationless \cite{1969ApJ...157..869G,Blandford:2002bp,meier2004ohm}.
Dissipative effects, however, do arise on specific two-dimensional planes called current sheets of thickness $\sim 1/\sigma$, which may be modelled using Eq.~\eqref{eq:ohmlaw}.

Our kinetically-mixed superradiance cloud shares many similarities with pulsars, and due to screening is also expected to have force-free regions. In fact, from  numerical simulations presented in Sec.~\ref{sec:fields}, we will see that if the plasma is highly conducting, $\sigma/\mu \gg 1$, plasma currents effectively redistribute charge to screen the rotating electric field induced by superradiance  $\varepsilon  \textbf{E}'$. This leads to charge being separated into a  dipole-like distribution, with a characteristic density that at the cloud radius is approximately given by,
\begin{align}
 \label{eq:density}
\rho &\simeq    \varepsilon \nabla \cdot  \textbf{E}' \simeq \pm \frac{\varepsilon  \sqrt{\Delta a_*} \alpha^{7/2} \mu^2}{\sqrt{G}}\\ 
&\simeq  \pm 5\cdot 10^7 \textrm{cm}^{-3}
\, \sqrt{\Delta a_*}\bigg(\frac{\varepsilon}{10^{-7}}\bigg)
  \bigg(\frac{\alpha}{0.1}\bigg)^{7/2} \bigg(\frac{\mu}{10^{-12}\,\mathrm{eV}}\bigg)^2,  \nonumber
\end{align}
where the plus and the minus signs correspond to opposite ends of the dipole-like pattern. A crude, non-relativistic estimate indicates 
that the large magnitude of the charge density is consistent with large conductivities: for a non-relativistic collisionless plasma the conductivity is $\sigma \simeq \rho/\mu m_e$, which, using Eq.~\eqref{eq:density}, gives $\sigma/\mu \sim 10^{12}$, where $m_e$ is the electron mass.

Despite the utility of the pulsar analogy, the resemblance with our system is limited. First, in the absence of a plasma, the kinetically mixed superradiance cloud is electrically, instead of magnetically, dominated. Second, while in a pulsar the magnetic field is dipolar and decays away from the neutron star,
in Sec.~\ref{sec:fields} we will show that in our system the resulting visible fields remain strong well outside of the light-cylinder $r_\mathrm{LC}\equiv 1/\mu$ (the radius out to which the plasma can corotate with the BH), up to the Bohr radius 
 $r\gtrsim 1/\mu \alpha \gg r_{\mathrm{LC}}$.
 This means that in the bulk of our system, charges cannot move fast enough to perfectly screen the rotating source field.
Electric dominance and imperfect screening suggest that in our cloud a steady-state force-free solution does not exist, unlike in pulsars where dissipative effects are confined to the current sheets. Instead, our numerical simulations, presented in Sec.~\ref{sec:fields}, show a dynamical interplay between resistive and force-free regions where electric fields have been mostly screened.

Up to now, we have only discussed the electric field dynamics. Complementary insight into the electrodynamics can be gained by  studying instead the  magnetic fields induced by the plasma currents. The magnetic dynamics can be analyzed by combining Ohm's law Eq. \eqref{eq:ohmlaw}, Faraday's, and Ampere's law (derived from Eq.~\eqref{eq:FieldeqInteraction1}) to obtain the magnetic induction equation
\begin{align}
\partial_t \vect{B}=\frac{\varepsilon\mu^2 \vect{B'}}{\sigma}+\frac{1}{\sigma} \nabla^2 \vect{B}+
\nabla \times (\vect{v} \times \vect{B}) ,
\label{eq:inductionequation0}
\end{align}
where the first term on the right hand side accounts for the background superradiant magnetic field. The induction equation is used to study  magnetic fields in a wide variety of astrophysical plasmas, where the electric displacement currents are smaller than plasma currents and can be neglected, an assumption that our simulations show to be valid. The three terms on the right-hand side of the induction equation  describe different characteristic regions of the system. Closest to the BH, the superradiant driving field is large, and the first term on the right-hand side dominates the morphology of the magnetic fields. Away from the neighborhood of the BH, the magnetic field is non-trivially  related to the superradiant driving fields, and the two remaining terms become dynamically relevant in a proportion set by a magnetic Reynolds number $R_m=\sigma |\textbf{v}| \ell$, where $1/\ell$ characterizes the magnetic field gradients. In zones where $R_m \gg 1$, the last term on the right-hand side of Eq.~\eqref{eq:inductionequation0},
which represents pure field advection, is largest. The simulations presented in Sec.~\ref{sec:fields} show that large regions in the bulk of the plasma are dominated by advection, and are characterized by magnetic flux conservation, tight-coupling of the plasma and the magnetic fields, mostly screened electric fields, and some emission of electromagnetic radiation due to the time-dependent plasma charge and current densities. 

Our simulations also show time-dependent regions, especially outside of the light-cylinder, where the plasma cannot corotate with the driving fields as advection would impose. This leads to differential rotation within the plasma and to the twisting and shearing of magnetic field lines, as well as to regions where the second (diffusive) term in Eq.~\eqref{eq:inductionequation0} dominates due to large field gradients and/or small plasma velocities that result in $R_m \ll 1$. In these regions,  we find that the interplay of advection and  diffusion drives turbulent effects, such as breaking and reconnection of field lines, schematically shown in the rightmost panel of Fig. \ref{fig:schematics}. Unscreened electric fields along the direction of plasma currents, expected from the simple kinematic arguments outlined above, are found at these sites. These electric fields lead to significant Ohmic dissipation $\textbf{J} \cdot \textbf{E}$, which in our simulation represent conversion of electromagnetic field energy into particle acceleration and radiation. 

Our simulations thus show that dissipation is associated with magnetic field reconnection and unscreened electric fields, as in the pulsar current sheets. In contrast to  the pulsar system, however, in the kinetically mixed superradiance cloud the resistive effects are realized in dynamically evolving regions throughout the bulk of the plasma. As a consequence, while in pulsars most of the power emitted is due to the time-dependent nature of the currents in the force-free bulk, we find that in our system the comparative preponderance of dissipative effects leads to an emission power that is dominated by Ohmic losses. 

Importantly, the simulations presented in Sec.~\ref{sec:emissionpower} suggest that while increasing the conductivity reduces the size of the dissipative regions, it also increases their number, \textit{i.e.} larger conductivities lead to ``fragmentation'' of the dissipative regions without changing their volumetric fraction. As a result, we find that the dissipative power tends to a $\sigma$-independent value (at large $\sigma$), allowing us to provide a prediction for the emitted power that is set entirely by the dark-photon model parameters and the BH mass.  
The power emitted by our system typically exceeds the  emission power of pulsars by several orders of magnitude, and for  clouds around stellar BHs can be as large as $L\simeq 10^{43}$~erg/s (an exact expression can be found in Eqns. \eqref{eq:powerdissipationfits} or \eqref{eq:luminosity}). We ascribe this difference to the rapid falloff of the dipolar magnetic field of the pulsar away from the neutron star, the large volume of the superradiance cloud when compared with the pulsar's emission regions $V_{\mathrm{cloud}}/V_{\mathrm{light cylinder}}\sim 1/\alpha^3$, and to our system's dissipative features. Given the periodic rotation of the cloud, it is possible that the emitted power will have a pulsating component, and our simulations indeed show some limited evidence that supports this hypothesis (see Sec.~\ref{sec:emissionperiodicity}).
From our simulations we cannot compute the spectral decomposition of the emitted power; however, we can speculate based on results of kinetic treatments of turbulent plasmas (analogous to pulsar current sheet simulations) that charged particles will be highly boosted by the large electric fields resulting in a large component of high-energy radiation in the form of X- and gamma-rays~\cite{Werner:2014spa,Zhdankin:2016lta,Cerutti:2014ysa,Cerutti:2015hvk}; for further discussion see Sec.~\ref{sec:emissionspectrum}. 

In the final stage of our system’s evolution the cloud decays predominantly by gravitational-wave emission, accompanied by the novel electromagnetic emission outlined here, as depicted in the rightmost panel of Fig. \ref{fig:schematics}. These emission channels lead to concrete observational signatures that we describe in Sec.~\ref{sec:observation}, 
such as performing EM follow-up observations of compact binary mergers, searching for a population of same-frequency and/or positive-frequency drift pulsars, and targeting such anomalous pulsars with GW follow-up searches.

With this short summary in hand, we now move on to provide an in-depth discussion of the plasma and field dynamics at each stage of their evolution, starting with the production of the plasma. 

\section{Plasma production}
\label{sec:SFQED}

In this section we describe the production of the conducting plasma within the superradiance cloud and determine the values of the mixing parameter $\varepsilon$ for which a plasma is plausibly created.  We identify two main processes that are crucial for the formation of the plasma, synchrotron radiation emitted by environmental electrons that are accelerated by the superradiance cloud, and subsequent photon-assisted Schwinger pair production in the background electric field. Here we estimate the rates of these two processes and show that they can effectively create the conducting plasma even for kinetic mixing parameters that are several orders of magnitude below current experimental bounds and of the region of interest for the observational prospects discussed later in this work. Several other mechanisms can produce charged particles in background fields and additionally contribute to the formation of the plasma, but for brevity we do not discuss them here (for a comprehensive list we refer the reader to \cite{meszaros1992high}).

\subsection{Synchrotron radiation}
\label{sec:sync}

Any stray charged particle entering the kinetically mixed superradiance cloud will experience strong electromagnetic forces. Since the cloud's magnetic field is subdominant, $|\vect B'| \sim \alpha |\vect E'| < |\vect E'|$, we simply consider the motion of $e^{\pm}$ accelerated by the electric field (equivalently we can perform a boost into a frame with vanishing magnetic field and electric field amplitude reduced by a factor of $1-\alpha \simeq 1$). Inside a fully grown cloud (at distance $r$ such that $r_g \ll r \lesssim 1/\alpha\mu$) the electric field has approximately constant amplitude given by Eq.~\eqref{eq:background} and rotates with angular velocity $\omega \simeq \mu$. The electrons/positrons are then approximately linearly accelerated over a time scale of $1/\mu$, reaching a maximum boost factor of 
\begin{align}
\label{eq:gamma_e}
\nonumber  \gamma_e &\simeq \frac{e\varepsilon |\vect{E}'|}{m_e \mu} \simeq 
e\varepsilon \alpha^{5/2} \sqrt{\Delta a_*} \frac{M_{\rm pl}}{m_e}  \\
   & \simeq 
   10^{12} \left(\frac{\varepsilon}{10^{-7}}\right)\left(\frac{\alpha}{0.1}\right)^{5/2} \left(\frac{\Delta a_*}{0.1}\right)^{1/2},
\end{align}
where in the first line we made use of Eq.~\eqref{eq:background}. As the electric field rotates, the electron/positrons trajectories bend with approximate radius of curvature $r_c \simeq \gamma_e m_e/(e \varepsilon |{\vect E'}|) \simeq 1/\mu$. During this circular motion the charged particles radiate synchrotron photons, predominantly at frequency $\omega_{\rm syn}=\gamma_e^3/r_c \simeq \gamma_e^3\mu$. We can estimate the rate for synchrotron emission at this frequency as $P_{\rm syn}/\omega_{\rm syn}$, which gives
\begin{equation}
 \Gamma_{\rm syn} \simeq \frac{2}{3}\frac{e^2 \gamma_e}{r_c} \simeq \frac{2}{3} \frac{e^3\varepsilon |\vect{E}'|}{m_e} \simeq \frac{2}{3} e^3\varepsilon \, \alpha^{5/2} \sqrt{\Delta a_*} \mu   \frac{M_{\textrm{Pl}}}{m_e},
 \label{eq:syncrate}
\end{equation}
where again, in the last equality we used \eqref{eq:background}. In order for the plasma to be phenomenologically relevant it must be created before the cloud is depleted by gravitational wave emission, on a timescale given by Eq.~\eqref{eq:tgw}. To ensure that this occurs, we impose the sufficient requirement that the synchrotron and photon-assisted pair production rates (discussed in the next section) occur before any particle can escape the cloud, \textit{i.e.}, that the synchrotron and pair-production timescales are shorter than the light-crossing time of the cloud, $1/ \alpha \mu$, which is much smaller than the GW decay time \eqref{eq:tgw}. For the synchrotron emission rate of Eq.~\eqref{eq:syncrate}, this leads to the requirement
\begin{equation}
\label{eq:eps_min_sync}
 \varepsilon  > \frac{1}{e^3 \, \alpha^{5/2} \sqrt{\Delta a_*}}\frac{m_e}{M_{\textrm{Pl}}} \simeq 10^{-18} \left(\frac{0.1}{\alpha} \right)^{5/2} \left(\frac{0.1}{\Delta a_*} \right)^{1/2}
\end{equation}
As evident from Eq.~\eqref{eq:gamma_e}, the above requirement also ensures that the accelerated electrons are highly relativistic ($\gamma_e \gg 1$).

\subsection{Photon assisted Schwinger pair production}
\label{sec:schwinger}

A static electric field can decay to electron-positron pairs through quantum tunneling, a process known as Schwinger pair production. The probability of scalar $e^{\pm}$ pair creation was first computed in Schwinger's seminal work \cite{Schwinger:1951nm}, from vacuum decay in an external, slowly varying electric field $\textbf{E}$. The rate per unit volume $V$ is given by
\begin{equation}
    \label{eq:schwinger}
    \frac{\Gamma_{e^{\pm}}}{V} = \frac{(e |\textbf{E}|)^2}{4\pi^3} \sum_n^{\infty} \frac{1}{n^2} \exp{\left(-\frac{\pi m_e^2}{e |\textbf{E}|}n\right)},
\end{equation}
and is exponentially suppressed for electric fields below the critical value $m_e^2/e \simeq 10^{18}\ \mathrm{V/m}$. Even the large electric field generated by the dark photon superradiance cloud, given in Eq.~\eqref{eq:background}, falls short by a few orders of magnitude, making Schwinger pair production unlikely in our setup. However, pair creation can be greatly enhanced in the presence of highly energetic photons \cite{Schutzhold:2008pz}, such as the synchrotron photons described in the previous section. Photon assisted Schwinger pair production is similar to magnetic pair production~\cite{Erber:1966vv} invoked in Blandford-Znajek processes \cite{Blandford:1977ds}, where radiation with energy above the threshold $2m_e$ can produces electron-positron pairs by scattering off of strong magnetic field. 

Photon-assisted Schwinger pair creation can be viewed as a semiclassical tunneling process and the production rate has been computed with methods similar to the one used for metastable vacuum decay in Ref.~\cite{Coleman:1977py}. The exponential factor in the rate is given by $e^{-S_B}$, where $S_B$ is the Euclidean action evaluated on the bounce solution (the classical trajectory that extremizes the action). In our case \cite{Monin:2010qj, Dunne:2009gi}
\begin{equation}
\label{eq:sbounce}
    S_{\rm B} = -\gamma_{\theta} +\left(\frac{2 m_e^2}{e \varepsilon |\vect E'|}+ \frac{e \varepsilon |\vect E'|}{2m_e^2} \gamma_\theta^2 \right)\arctan \frac{2 m_e^2}{e \varepsilon |\vect E'|\gamma_{\theta}},
\end{equation}
where $\gamma_\theta = \sin\theta\ m_e \omega_\gamma/(e \varepsilon |\vect E'|) $, $\omega_\gamma$ is the photon frequency, and $\theta$ is the angle between the direction of the photon and the background electric field. As the synchrotron photons travel in the cloud, they will encounter electric fields that are almost perpendicular  to their direction of propagation within a time scale of $1/\mu$, when the production rate is maximized. As a result, for a simple estimate of the rate we can take $\sin\theta \simeq 1$. 

There are two limiting cases of Eq.~\eqref{eq:sbounce} depending on the photon frequency. If $\omega_\gamma \ll 2m_e$, $S_B \simeq \pi m_e^2/(e\varepsilon |\vect E'|)$, which reduces to the standard Schwinger result of Eq.~\eqref{eq:schwinger}. Therefore if the photon energy is below the pair production threshold, the electric field still needs to be super-critical for the process not to be exponentially suppressed. On the other hand, if $\omega_\gamma \gg 2m_e$, as is the case for most of the synchrotron photons described in the previous section, $S_B \simeq  2 m_e^3/(e\varepsilon |\vect E'|\omega_\gamma)$. The additional, potentially very small, factor of $2m_e/\omega_\gamma$ significantly enhances the probability of pair production. The full expression of the rate (including the prefactor of the exponential term) is given in Ref.~\cite{Dunne:2009gi} and is larger for photons with polarization perpendicular to the electric field. For the highly energetic synchrotron photons with perpendicular polarization, we have
\begin{align}
    \Gamma^\gamma_{e^\pm} &= \frac{\alpha_{\rm EM}}{2 \pi} \frac{e \varepsilon |\vect E'|}{m_e} \exp \left( -\frac{2 m_e^2}{e\varepsilon |\vect E'|} \frac{2m_e}{\omega_{\rm syn}} \right) \nonumber\\
    &=
    \frac{\alpha_{\rm EM}}{2 \pi} \frac{e \varepsilon |\vect E'|}{m_e} \exp \left[-\frac{4 m_e^6\mu^2}{(e \varepsilon |\vect E'|)^4}\right]. \label{eq:assisted_schw}
\end{align}

\begin{figure*}[t]
    \centering
    \includegraphics[width=\textwidth]{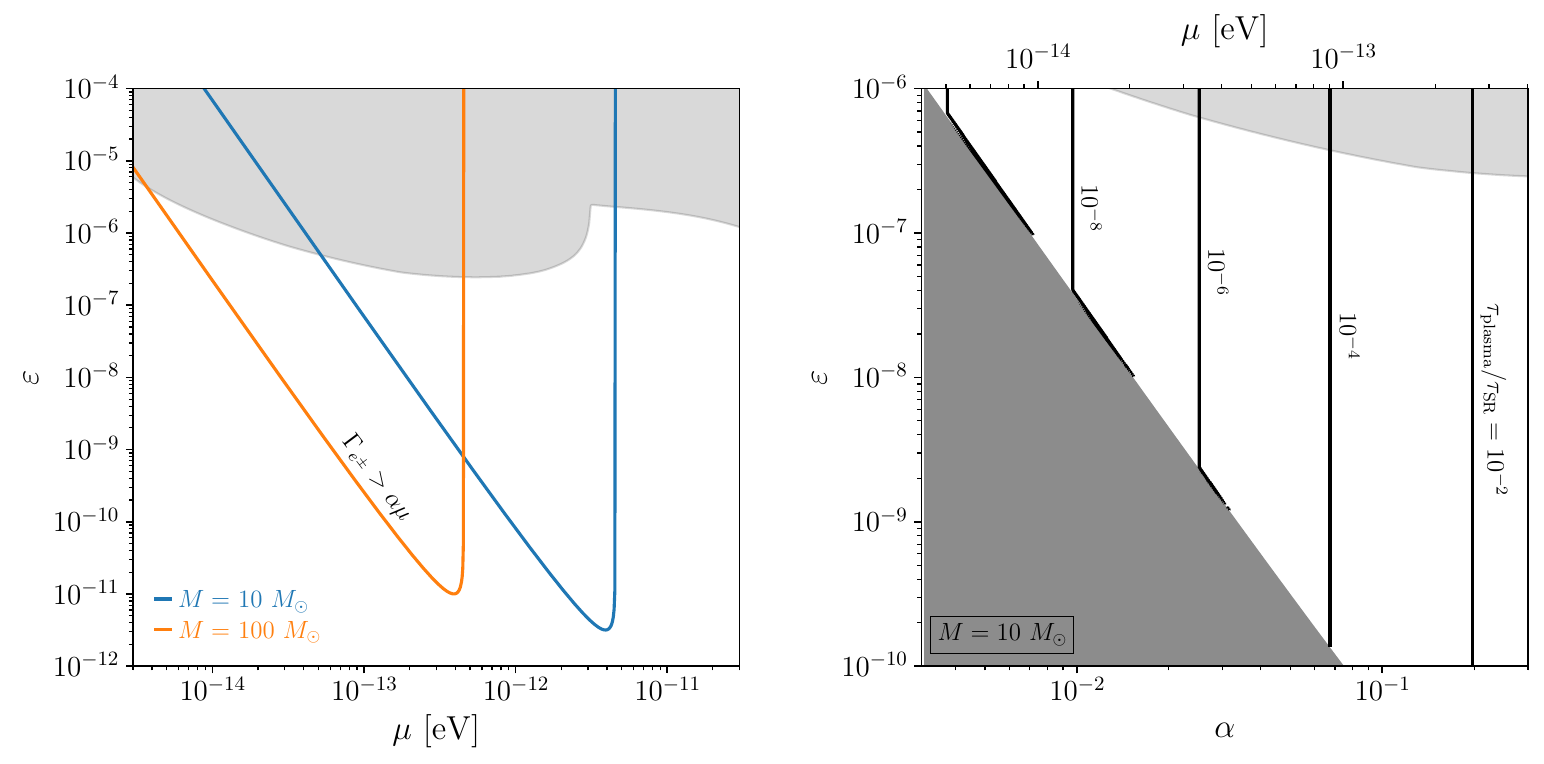}
    \caption{ {\it (left)} Smallest values of the kinetic mixing parameter $\varepsilon$ that allows for efficient $e^{\pm}$ pair production in the superradiance cloud as a function of dark photon mass $\mu$, for BH masses of $10 \, M_\odot$ (blue) and $100\, M_\odot$ (orange), with initial BH spin of $a_* = 0.9$. The rate for photon stimulated Schwinger pair production, given in Eq.~\eqref{eq:assisted_schw}, is required to be greater than the size of the cloud, $\alpha\mu$, when the cloud has fully grown. Smaller dark electric fields at small $\mu$, require larger $\varepsilon$ to initiate the cascade. The sharp cutoff corresponds to the highest dark photon mass that satisfies the superradiance condition for the fastest growing level. {\it (right)} Ratio of the time needed to populate the plasma over the superradiance e-folding time as a function of the kinetic mixing parameter $\varepsilon$ and the gravitational coupling $\alpha$ for a BH mass of $10\ M_{\odot}$ and initial BH spin of $a_{*} = 0.9$ (the ratio is independent of $M $ and only mildly dependent on $a_*$). An estimate of the ratio is given in Eq.~\eqref{eq:tsr_over_tpl}, while in the plot $\tau_{\rm{plasma}}$ is evaluated using the electric field value at the time that the cascade pair production is initiated. In the dark gray shaded region, the electric field is always too small to produce $e^{\pm}$. When the cascade is efficient, the plasma is filled within a small fraction of one superradiance e-folding time ($\tau_{\rm{SR}}$ grows steeply at small $\alpha$). In both panels, the light gray shaded region is excluded by measurements of the CMB spectrum by COBE/FIRAS \cite{Fixsen:1996nj,Caputo:2020bdy}}
    \label{fig:plasma}
\end{figure*}

For cascade production to occur in our system the term in the exponential must reach a magnitude of order unity when the cloud has reached its full size (or before) so that the exponential suppression of pair-production is lifted. This translates into a minimum value for the mixing parameter $\varepsilon \gtrsim m_e^{3/2}\mu^{1/2}/e|\vect{E'}|$. More precisely, in what follows we impose that the pair-production rate for a fully grown cloud is faster than the light crossing time of the cloud $1/\alpha \mu$, so that the synchrotron photons split into $e^{\pm}$ before escaping the superradiance cloud, ensuring a cascade production of the plasma, which translates into a minimal mixing parameter
\begin{widetext}
\begin{align}
 \label{eq:eps_min_schw}
\nonumber    \varepsilon   \gg &  \left(\ln \frac{\alpha_{\rm EM}\gamma_e}{2\pi \alpha} \right)^{-1/4} \frac{m_e^{3/2}\sqrt{2 \mu}}{e|\vect{E'}|}  \\
    \approx &  \left(\ln \frac{\alpha_{\rm EM}\gamma_e}{2\pi \alpha} \right)^{-1/4} \frac{ 1}{e \alpha^{5/2}\sqrt{\Delta a_*} } \frac{m_e}{M_{\rm pl}}\sqrt{\frac{2m_e}{\mu}} 
    \simeq  10^{-10} \left(\frac{0.1}{\alpha}\right)^{5/2}\left(\frac{0.1 }{\Delta a_*} \right)^{1/2} \left(\frac{10^{-12}\ {\rm eV}}{\mu}\right)^{1/2} \left(\frac{\log \frac{\alpha_{\rm EM}\gamma_e}{2\pi\alpha}}{20}\right)^{-1/4},
\end{align}
\end{widetext}
where in the logarithmic term we made use of \eqref{eq:gamma_e}, and in  going to the second line we used Eq. \eqref{eq:background}. Notice that the above requirement also guarantees that $\omega_{\rm syn} \gg 2m_e$. For kinetic mixing parameters saturating the lower bound in Eq.~\eqref{eq:eps_min_schw}, plasma production will be triggered when the cloud has reached a close-to maximal size, while for mixing parameters above this lower bound the plasma will be created before the cloud has fully grown. We show the smallest values of kinetic mixing parameters that allow for cascade pair creation in the superradiance cloud as a function of dark photon mass in the left panel of Fig.~\ref{fig:plasma}.

\subsection{Dynamics leading to a quasi-steady state}
\label{sec:quasi-steady}

The plasma begins to be populated once the pair-production cascade initiates,
which as noted previously happens when the superradiance cloud has grown to a size such that the pair-production rate becomes of the order of the cloud's Bohr radius, {\it i.e.} $\Gamma^\gamma_{e^\pm }\simeq \alpha\mu$. For this to occur and up to a logarithmic correction, 
the superradiant field must reach a critical value $ e \varepsilon |\vect{E}'_{\mathrm{crit}}|\simeq \sqrt{2}m_e^{3/2}\mu^{1/2}$. The superradiant field grows to $\vect{E}'_{\mathrm{crit}}$ in a few superradiance times $\tau_{\mathrm{SR}}\simeq 1/4 \alpha^6 a_* \mu$. After that, the plasma is created by cascade production on the much shorter light-crossing timescale, over which the superradiant field and the cascade production rates are approximately fixed to $\vect{E'}\simeq \vect{E}'_{\mathrm{crit}}$ and $\Gamma^\gamma_{e^\pm }\simeq \alpha \mu$ respectively\footnote{For simplicity, throughout Secs.~\ref{sec:sync} and \ref{sec:schwinger}, we use the superradiant field value at saturation, Eq.~\eqref{eq:background}, instead of the critical field, $\vect{E}'_{\mathrm{crit}}$. The latter will be as large as the saturation field only for the smallest values of $\varepsilon$ allowed, given numerically in Eqs.~\eqref{eq:eps_min_sync} and \eqref{eq:eps_min_schw}.}.

During cascade production, the charge density grows exponentially as $n_e = n^0_e \exp \left( 2\Gamma^\gamma_{e^\pm } t \right)$, where $n_e = n^-_e + n^+_e$ is the total number density of electrons and positrons.
Pair production stops when the charged plasma effectively screens the critical electric field due to charge separation, which happens when the electron number density reaches $n^f_e \simeq  \varepsilon \nabla \cdot \vect{E}'_\mathrm{crit}/e \simeq\sqrt{2} \alpha (m_e \mu)^{3/2}/e^2$. 
Assuming that when the cascade begins we start from one single electron in the cloud, $n^0_e \simeq (\alpha\mu)^3$, the plasma grows $n^f_e/n^0_e$ e-folds before the cascade stops, so that the plasma formation time can be estimated as
\begin{align}
  \nonumber  \tau_{\rm plasma} \simeq &\frac{1}{2 \Gamma^\gamma_{e^\pm }} \ln\bigg(\frac{n^f_e}{n^0_e}\bigg)
  \simeq \frac{1}{2 \alpha \mu}\ln \frac{m_e^{3/2}}{e^2 \alpha^2\mu^{3/2}}  \quad .
\end{align}
The plasma production time is thus parametrically shorter than the superradiance timescale by a factor 
\begin{align}
    \frac{\tau_{\rm plasma} }{\tau_{\rm SR}} &\simeq {2 a_* \alpha^5  } \ln \frac{m_e^{3/2}}{e^2 \alpha^2\mu^{3/2}} \label{eq:tsr_over_tpl}
\end{align}
In the right panel of Fig.~\ref{fig:plasma} we show the above ratio of timescales for the values of $\varepsilon$ and $\alpha$ that satisfy the plasma pair production requirement from Eq.~\ref{eq:eps_min_schw}.

As the dark photon cloud continues to grow and the electric field increases, more charged particles will be created and the plasma will rearrange itself in the screening configuration, until $\vect E'$ has reached its maximum value Eq. \eqref{eq:background} after $\approx 180\ \tau_{\rm SR}$, at which point the charge density achieves its maximal value Eq. \eqref{eq:density}.
The numerical simulation presented in the next sections will show how charge separation in the cloud and electric screening are indeed good approximations. Note that the formation of the plasma induces a plasma mass for the SM photon ($\omega_p = e\sqrt{n_e/m_e}$), but does not significantly affect the dark photon mass. In fact, in the limit $\mu\rightarrow 0$, a massless mode must remain in the theory even in the presence of the plasma, which indicates that the leading contribution to the dark photon mass is simply $\mu$, up to $\varepsilon^2$ corrections.
The plasma frequency does not affect the mixing between the dark and visible photons either, nor the propagation of visible fields in the plasma, the reason being that in our system the energy density in the visible electromagnetic fields greatly exceeds the energy density in the charged $e^{\pm}$ plasma (by a factor $\sqrt{m_e/\mu}$), so the tenuous plasma cannot impede the propagation of the comparatively larger EM fields. This is different from the case in \cite{Dubovsky:2015cca}, where the EM fields are a small perturbation on top of a comparatively dense charged plasma. A more detailed discussion of  plasma effects on the dark photon is presented in App.~\ref{appendix:darkphoton}. 
Note also that the total mass of the plasma,
\begin{widetext}
\begin{align}
    M_{\rm plasma} \approx \frac{m_e \rho}{e (\alpha \mu)^3} \simeq \varepsilon \alpha^{1/2} (\Delta a_*)^{1/2} \frac{m_e M_{\rm pl}}{\mu} 
    \simeq 10^{-29} M_\odot \left(\frac{\epsilon}{10^{-7}}\right)\left(\frac{\alpha}{0.1}\right)^{1/2}\left(\frac{\Delta a_*}{0.1}\right)^{1/2}\left(\frac{10^{-12}\ {\rm eV}}{\mu}\right)
    \label{eq:mplasma}
\end{align}
\end{widetext}
is much smaller than the mass of the cloud, $M_c =  \alpha^2 \Delta a_* M_{\rm pl}^2/\mu$, leaving the gravitational potential unaltered. We can then safely assume that the growth and dynamics of the superradiance cloud is not affected by the presence of the standard model plasma.

In the left panel of Fig.~\ref{fig:plasma}, we show the range of $\varepsilon$ and $\mu$ where fast cascade production occurs and the plasma is populated, $\Gamma_{e^\pm}^\gamma > \alpha\mu$. Much slower processes can also populate this plasma in the parameter spaces where the fast cascade production is inactive. Assuming, for example, Bondi accretion and $\mathcal{O}(1)$ sound speeds $c_s$, it takes roughly
\begin{widetext}
\begin{align}
    \tau_{\rm acc} = \frac{ M_{\rm plasma}}{\dot{M}_{\rm Bondi}}\approx \varepsilon \alpha^{-3/2} (\Delta a_*)^{1/2} \frac{c_s^3 M_{\rm pl}\mu}{\pi n_{M}}
    \simeq 10\ {\rm years} \left(\frac{\varepsilon}{10^{-7}}\right)\left(\frac{\alpha}{0.1}\right)^{-3/2}\left(\frac{\Delta a_*}{0.1}\right)^{1/2}\left(\frac{\mu}{10^{-12}\ {\rm eV}}\right)\left(\frac{1/{\rm cm^3}}{n_{M}}\right)\left(\frac{c_s}{1}\right)^3
\end{align}
\end{widetext}
to populate enough charged particles inside the superradiance cloud, where $n_{M}$ is the matter number density in the interstellar medium. Such a time scale suggests that for parameters where the cascade production is active, accretion from interstellar medium can be safely ignored during dark photon superradiance and plasma generation. On the other hand, in the parameter space where the cascade production is inactive, such processes can be fast enough to populate a plasma inside the superradiance cloud around isolated BHs inside our galaxy, which are most likely more than thousands if not millions of years old.

Finally, it is worth pointing out that the transient process of cascade particle production discussed in this section only produces a small and unobservable amount of emission. The transient effects discussed in~\cite{Caputo:2021efm}, for example, occur when $|\vect{E}|^2 \sim m_e^3\mu$, which amounts to a total energy of $10^{32}(10^{-12}~\mathrm{eV}/\mu)^2(0.1/\alpha)^3$~ergs (independent of $\varepsilon$). As we will demonstrate in section \ref{sec:emissionpower}, this is
about 20 orders of magnitude smaller than the total electromagnetically dissipated power from the superradiance cloud. Similarly, photon superradiance \cite{Pani:2012vp, Cardoso:2020nst,Blas:2020kaa,Cannizzaro:2020uap,Cannizzaro:2021zbp} will saturate long before the field strengths and energy densities in the photon superradiance cloud reaches sizes relevant for observation.

\section{Field configurations} 
\label{sec:fields}

We have shown that a pair production cascade ensues on short timescales, once the superradiance dark photon cloud surpasses a critical electric field strength during the exponential growth of the cloud. The generated charges form a highly conducting plasma that is subject to the electromagnetic fields of the dark photon cloud. In this section, we study the macroscopic state this plasma equilibrates into. To that end, we consider, numerically, the superradiance cloud of a kinetically mixed dark photon on a fixed Kerr spacetime of mass $M$ and dimensionless spin parameter $a_*$. As outlined throughout secs.~\ref{sec:cloudsummary} and~\ref{sec:SFQED}, it is crucial to work with an Ohm's law that accounts for the energy dissipation into the plasma inside the cloud. Building on the analytical discussions in the previous section, we first examine the structure of the visible electromagnetic fields of the cloud-plasma system on large scales. On small scales, we demonstrate that, in the high conductivity limit, turbulent dynamics emerges accompanied by efficient magnetic field line reconnection in the bulk of the dark photon cloud. Throughout the section, we study the system considering all relativistic effects of the background spacetime in the interaction basis. 

We begin by introducing the plasma model in the context of a kinetically mixed massive vector field with the SM photon in Sec.~\ref{ssec:RFFE}. We establish that within this model, a quasi-stationary end state of the pair production cascade is reached and characterize its large scale behavior in Sec.~\ref{sec:largescalesolution}, and small-scale dynamics in Sec.~\ref{sec:Highlyconductinglimit}. In Sec.~\ref{sec:turbulentscalesummary}, we briefly summarize the main findings.

\subsection{Plasma model}
\label{ssec:RFFE}

We study the kinetically mixed field equations \eqref{eq:FieldeqInteraction1} of the visible field ${A}_\mu$ in the interaction basis \eqref{eq:interlagrangian}. A strongly magnetized, highly-conducting pair plasma, is well-described by the force-free limit of ideal magnetohydrodynamics (see, e.g., Ref.~\cite{Font:2008fka} for a review). Specifically, force-free electrodynamics is applicable
when \textit{(i)} resistive effects are negligible, \textit{(ii)} the magnetic
field strength is larger than the electric field's, i.e.,
${B}^2>{E}^2$, and \textit{(iii)} the plasma mass density
$\rho_p$ and pressure $P_p$ are much smaller than the electromagnetic
energy density, i.e., ${B}^2\gg\rho_p,P_p$ (see e.g., \cite{1969ApJ...157..869G,Blandford:2002bp}). 
In force-free electrodynamics, one can numerically evolve only the electromagnetic fields, without keeping track of the fluid quantities, with the current being
uniquely determined from the electromagnetic fields by the requirement that the Lorentz force vanishes (see, e.g., Refs.~\cite{Paschalidis:2013gma,Komissarov:2004ms} for details on the force-free limit). 
In essence, $F_{\alpha \beta} I^{\alpha}=0$ allows $\nabla_\beta F^{\alpha \beta}=I^{\alpha}$ to be rewritten as 
$F_{\alpha \gamma}\nabla_\beta F^{\alpha \beta}=0$. (More details, include the equations of force-free electrodynamics
in terms of electric and magnetic fields can be found in App.~\ref{app:resistivitytest}.)

As discussed in the previous 
sections, the superradiant system considered here satisfies condition
\textit{(iii)} [see \eqref{eq:mplasma}], but \textit{a priori} violates conditions \textit{(i)} and
\textit{(ii)}. Exactly how and why these violations persist, even at large
conductivities is the subject of the following subsections. However, one can already
anticipate that in the limit of vanishing backreaction of the pair plasma onto the
visible electromagnetic fields, the magnetic dominance is lost by virtue of the
superradiant solution being electrically dominated: ${B}^2=\varepsilon^2 B'^2 < \varepsilon^2
E'^2={E}^2$. Since electric dominance implies that there is no frame where
the electric field, and hence the acceleration on any charges, vanishes, it is furthermore not surprising that
dissipative processes also become important.  
Therefore, in order to relax
assumptions \textit{(i)} and \textit{(ii)}, we modify the force-free equations to explicitly allow for dissipation by 
introducing an Ohm's law with finite conductivity $\sigma$, while still requiring the current to be a function of the 
electromagnetic fields. 
This is done in such a way that, under a certain set of assumptions, the force free limit can be recovered as $\sigma \rightarrow \infty$.
Slightly abusing terminology, we shall refer to this as resistive force-free electrodynamics.
Several variations of this approach have been applied to simulating pulsar magnetospheres~\cite{Komissarov:2005xc,Gruzinov:2007se,Li:2011zh,Parfrey:2016caq,Mahlmann:2020yxn}, and 
in particular the electrically dominated current sheet~\cite{Li:2011zh,Mahlmann:2020yxn}, and here we generalize these to the kinetically mixed case (see also  App.~\ref{app:resistivitytest} for more details).
In the following, we will describe how we set up the coordinate system, the Maxwell equations, the fluid of charged particles and the Ohm's law in a fully relativistic simulation of the plasma of electrons and positrons inside the superradiance cloud.

\paragraph{Space-time decomposition:} We begin by discussing how the kinetically-mixed
Maxwell's equations for the visible fields can be decomposed  
according to a given choice of time slicing of a spacetime,
which for this study will be given by the Cartesian Kerr-Schild coordinate time $t$.
Using the (future-pointing) unit normal to slices of constant time $n^\mu$,
the visible electric and magnetic fields defined 
with respect to this slicing are
\begin{align}
E^i \equiv n_\nu F^{i\nu}, && B^i \equiv n_\nu (*F)^{i\nu} = \frac{1}{2} n_\nu \varepsilon^{i \nu \alpha \beta} F_{\alpha \beta} . 
\label{eq:em_euler}
\end{align} 
where $\varepsilon^{\alpha\beta\gamma\delta}$ is the Levi-Civita tensor. 
These are the quantities that we evolve on the BH spacetime. 
Projecting the kinetically mixed Maxwell equations~\eqref{eq:FieldeqInteraction1} into components
orthogonal and parallel to the time slice, one obtains the evolution equations in terms of three-dimensional spatial quantities.
We give the explicit form of these equations that we use to carry out the numerical evolution in App.~\ref{app:numericalsetup}.

In order to evolve the electric and magnetic fields, we also need to specify the electromagnetic current. The four dimensional current $I^\mu$ can be decomposed into a spatial component $J^i$, and a component perpendicular to the slices of constant time 
\begin{align}
J^i= I^i- \rho_q n^i,  & & \rho_q=-n_{\mu} I^\mu,
\label{eq:currentdeceuler}
\end{align}
where $\rho_q$ is the Eulerian frame (i.e., with respect to the slices
of constant time) charge density.
We directly calculate the charge density from the divergence of 
the electric field,
\begin{align}
\rho_q =& \ D_i E^i-\varepsilon\mu^2 n_\mu A'^\mu,
\label{eq:chargeconstraint}
\end{align}
using the Gauss's law constraint equation obtained from projecting the
kinetically mixed Maxwell equations for the visible fields \eqref{eq:FieldeqInteraction1} onto the time slice.
However, we still need to specify the spatial part of the current $J^i$.

\paragraph{Ohm's law:} 
As discussed above, we introduce the effect 
of finite conductivity using a simple Ohm's law in the frame of the fluid (plasma)
\begin{align}
    j^\mu = \sigma e^\mu,
    \label{eq:ohmslaw}
\end{align}
where $j^\mu$ and and $e^\mu$ are, respectively, the electromagnetic current and visible electric field in the fluid frame. The latter is defined in terms of the fluid velocity $u^\mu$,
\begin{align}
e^\mu \equiv u_\nu F^{\mu \nu}, && j^{\mu} \equiv I^\mu+(u^\nu I_\nu) u^\mu,
\label{eq:em_fluid}
\end{align} 
in an analogous way to the Eulerian frame quantities.
With this prescription, we are neglecting anisotropic magnetic field effects. However, as the superradiance cloud system is characterized by strong electric fields, this is the dominant contribution away from the force-free limit. 
Note that the Ohm's law~\eqref{eq:ohmslaw} instantaneously relates currents and electric fields, a prescription that is valid for low-inertia plasmas such as ours. In eq. \eqref{eq:ohmslaw}, the conductivity $\sigma$ is a phenomenological parameter that allows for energy dissipation via mechanisms that are set by the microphysics, which is left unspecified. While without a microphysical description it is not possible to determine the value of the conductivity, we expect that the conductivity of the pair plasma considered here is large when measured in terms of the system's natural length scale (the inverse dark-photon mass), i.e., $\sigma/ \mu \rightarrow \infty$ [see also a brief discussion below eq.~\eqref{eq:density}]. 
Note that in this limit, other charge transport mechanism such as diffusion due to charge gradients can be safely neglected. 

The resistive relation \eqref{eq:ohmslaw} allows us to compute the fluid-frame
currents from the visible electromagnetic fields. However, since we are not directly evolving the
fluid, and in particular its velocity, this is not sufficient to give the
Eulerian current entering into the evolution equations. In ideal
magnetohydrodynamics (including the limiting case of force-free), given the
electromagnetic fields, one can reconstruct one component of the velocity,
referred to as the drift velocity
\begin{align}
v^i_{d,{\rm ideal}}=\frac{\varepsilon^{ijk}E_j B_k}{B^2} ,
\label{eq:ffdrift}
\end{align} 
where here we use $v^i$ to refer to the Eulerian spatial velocity of the fluid $v_i=-u_i/(n_\mu
u^\mu)$, though not the component of the velocity parallel to the magnetic field. In our resistive extension, 
following \cite{Gruzinov:2008um} (see also Ref.~\cite{Komissarov:2005xc}),
we use a drift velocity that is augmented with an electric component to allow for electrically dominated regions to be treated
self-consistently, 
\begin{align}
v^i_d &=\frac{\varepsilon^{ijk}E_j B_k}{B^2+E_0^2}, \quad
E_0^2=B_0^2+E^2-B^2, \nonumber\\
B_0^2& = \frac{1}{2}\left[B^2-E^2+\sqrt{(B^2-E^2)^2+4(E_iB^i)^2}\right].
\label{eq:Augmetneddriftvel} 
\end{align} 
This ensures that the drift velocity is bounded by the speed of light, even in electrically dominated regions, and is further
reduced in regions with non-vanishing resistive effects, i.e.,
$E_i B^i\neq 0$. 
We note that, while the quantities above are written in terms of $E$ and $B$ fields, 
$B^2-E^2=F_{\mu\nu}F^{\mu\nu}/2$ and $E_i B^i=F_{\mu\nu}(*F)^{\mu\nu}/4$ are spacetime scalars.

In general, the full fluid velocity cannot be determined from the
electromagnetic fields alone, without extra conditions. Here,
following~\cite{Li:2011zh}, we identify $v^i=v_d^i$,
i.e., we set the non-drift velocity component to zero in the BH frame
defined by $n^\mu$. With the fluid velocity specified, we can transform~\eqref{eq:ohmslaw} into the
Eulerian frame, giving
\begin{align} 
J^i= \rho_q v^i_d +\sigma E_0
    \sqrt{\frac{B^2+E_0^2}{B_0^2+E_0^2}}\left (\frac{E_0 E^i+B_0
B^i}{B^2+E_0^2} \right ).  
\label{eq:resistivecurrent} 
\end{align} 
This is the kinetically mixed extension of the current considered in
\cite{Li:2011zh}. In App.~\ref{app:resistivitytest}, we further discuss the
advantages and limitations of this choice of current, and contrast it with
other currents developed in the literature. Notice, however, this ansatz is merely a prescription to extend the regime of validity of the force-free paradigm to resistive regions (such as current sheets), which has physical meaning only in the high-conductivity limit, where $B^2>E^2$, $E_iB^i=0$, and $\sigma E_0<\infty$, such that $v_d^i\rightarrow v_{d,\text{ideal}}^i$. Hence, while different choices of currents may result in different physics at moderate conductivity, for $\sigma\rightarrow\infty$ all prescripts converge towards the well-defined force-free limit. Therefore, we primarily focus on the trend towards this limit.

\paragraph{Numerical setup:}

The visible fields $E^i$ and $B^i$ are numerically evolved on a fixed Kerr spacetime in
Cartesian Kerr-Schild coordinates. The massive vector field solutions on this
background spacetime, which enter as source terms in the evolution equations,
are constructed numerically (this is
discussed in App.~\ref{app:procaconstruction}). We restrict our attention
to solutions where the superradiant instability has been saturated,
which occurs when the cloud oscillation frequency is synchronized with the 
horizon frequency
$\omega_R=\Omega_{\rm BH}$. The computational domain extends from the BH horizon to
spatial infinity through the use of compactified coordinates, 
and mesh refinement is used to concentrate resolution in the central region,
enabling us to resolve the
BH-cloud system sufficiently up to $r=10r_c$. The conductivity $\sigma$
is set to be constant in space and time, as is serves merely as a proxy for the
local resistivity present in the cloud. In light of the lack of a microphysically
motivated conductivity, and the success of analogous choices in the case of the
pulsar magnetosphere (see e.g., \cite{Li:2011zh}), this choice, while unphysical, is a 
first step towards a more complete analysis. We also note that, any choice of spatially dependent conductivity that varies on a macroscopic length scale likely results in similar qualitative and quantitative behavior of the system. This follows provided the system's state is independent of conductivity for sufficiently high values, because resistive features (magnetic reconnection, energy dissipation, etc.) are active on scales below the macroscopic scales of the system.
Finally, we evolve the system of equations forward in
time using a higher-order explicit Runge-Kutta algorithm. The limitations of
this choice in the context of stiff equations in the large conductivity limit,
as well as further details of the numerical methods, are discussed in App.~\ref{app:numericalsetup}. 
We evolve the system, starting from suitably chosen initial data, for a sufficiently
long period of time ($\sim 200/\mu$ or longer) such that it relaxes towards an approximate 
equilibrium, as measured by the Poynting flux at large radii becoming nearly constant.
Details are given in App.~\ref{app:numericalsetup}.

\subsection{Field solutions: Large scale behaviors} \label{sec:largescalesolution}

\begin{figure*}[t]
    \centering
    \includegraphics[width=0.95\textwidth]{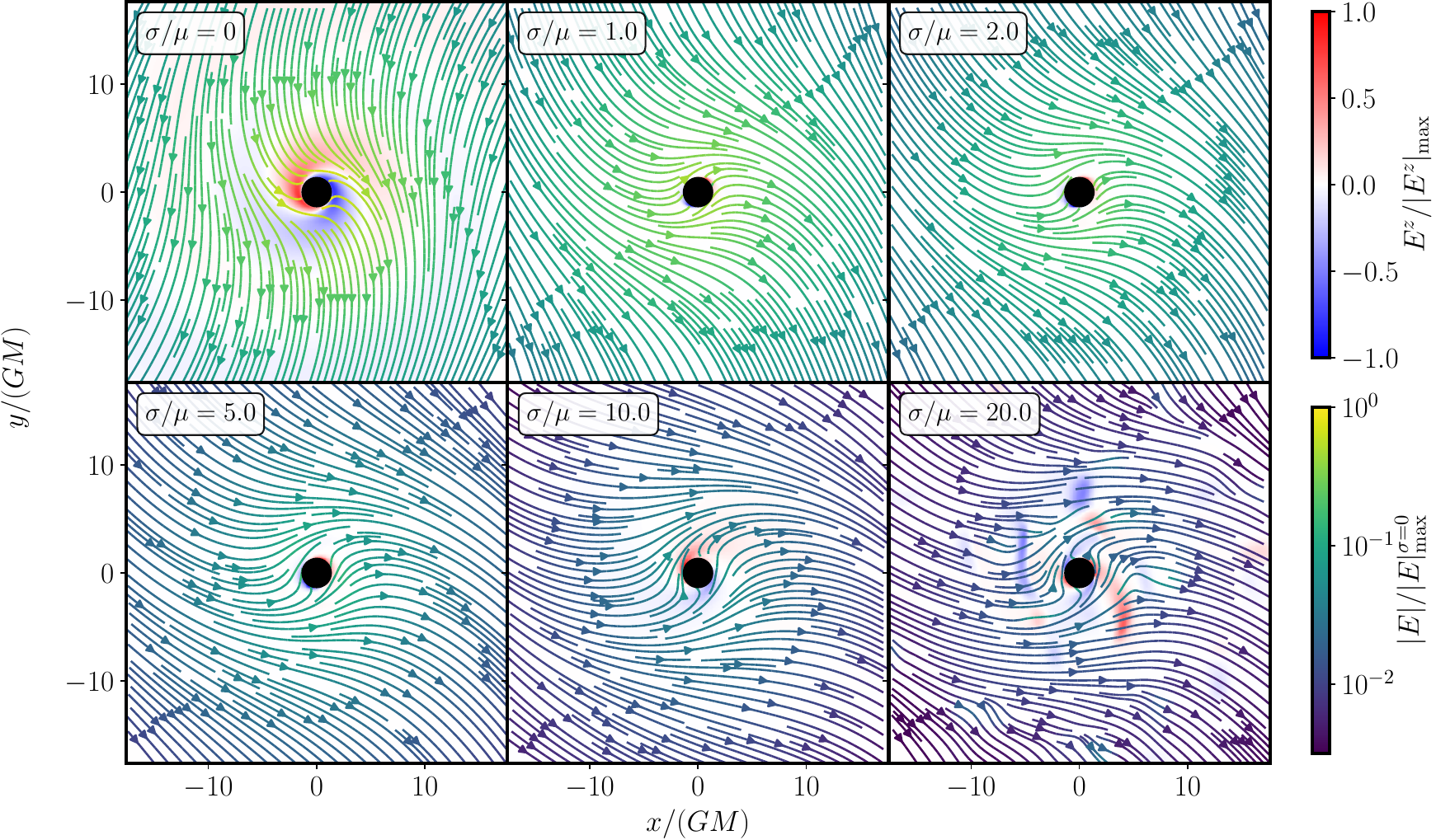}
    \caption{We show the visible electric field geometry, $E^i$, and magnitude as a function of the plasma conductivity $\sigma$ in the equatorial plane of the BH. The superradiance cloud's phase is the same in each panel. Color (red/blue) correspond to the magnitude of the component along the spin axis, i.e., in the $z$-direction, normalized by the maximal magnitude of that component at the given conductivity. Field lines are projections of the electric field onto the equatorial plane, while the color of the field lines (yellow/green) indicates the magnitude of the visible electric field normalized by the maximal magnitude at $\sigma=0$. The BH and cloud parameters are as in \figurename{ \ref{fig:vaccumFieldlines}}, i.e., $\alpha=0.3$ and $a_*=0.86$. With increasing conductivity, the electric field magnitude decreases compared with the vacuum limit, and $E^z/E\sim\mathcal{O}(1)$ for $\sigma/\mu=20$. The field geometry undergoes a phase-shift of $\pi/2$ between vacuum and large conductivity limits.}
    \label{fig:condEFieldlines}
\end{figure*}

In the following, we demonstrate how in the quasi-steady state solution, the
strong electric field of the superradiance cloud is mostly 
screened by locally produced charges at large conductivities, and the visible
magnetic field begins to play an important role in the system.  As noted above, the
physical value of the plasma conductivity $\sigma$ will be set by a
microphysical scale (due to scattering, synchrotron radiation, pair production and annihilation, etc.)  
that we expect to be much smaller than the other physical
length or time scales in the system that we consider (e.g., $1/\mu$). 
However, due to numerical
limitations, we only consider values up to $\sigma/\mu=20$. Therefore,
in order to make qualitative and quantitative statements about the properties
of the quasi-endstate, we proceed by discussing the behavior of the system as a
function of conductivity, focusing primarily on $\sigma/\mu\geq 1$ (though we
include the low-conductivity limit in App.~\ref{app:lowconductivityregime}
for completeness), and extrapolate trends towards $\sigma\rightarrow\infty$ if
possible. This is the approach typically used to study resistive effects in
pulsar magnetospheres \cite{Li:2011zh,Parfrey:2016caq,Mahlmann:2020yxn}; we
discuss possible shortcomings of these methods below.

We begin by considering the behavior of the visible electric field $E^i$
of the quasi-stationary endstate of the
pair cascade as function of conductivity in \figurename{
    \ref{fig:condEFieldlines}}.  In the vacuum limit, $\sigma/\mu=0$, there is no electric field generated by a charged plasma, and the depicted field lines are just
an equatorial slice of the electric field $E^i=\varepsilon E'^i$
shown in \figurename{ \ref{fig:vaccumFieldlines}}. For $\sigma/\mu\geq 1$, the
main qualitative feature, as summarized in Sec.~\ref{sec:cloudsummary}, is the buildup
of the dipolar screening charge density leading to a significant reduction of
the visible electric field compared to the vacuum case. As the conductivity
increases, this suppression of $E^2$ grows and the component of $E^i$ along the
BH spin-axis becomes more important, as $E^z/E\sim\mathcal{O}(1)$ in the
$\sigma/\mu=20$ panel of \figurename{ \ref{fig:condEFieldlines}}. 

Another important qualitative feature we find is that the visible electric field exhibits a
global de-phasing of $\pi/2$ with respect to the dark electric
field at large conductivities. 
This de-phasing can be understood analytically
in the non-relativistic limit, $\alpha\ll 1$. In this
limit, a spatial derivative, which is set by the inverse cloud size $(\nabla
\sim \alpha\mu)$, is much smaller than the time derivative $(\partial_t \sim
\mu)$. For electric fields with similar or larger strengths compared to the magnetic field, as indicated by our numerical simulations at small and
moderate conductivities (see \figurename{ \ref{fig:condBsqoverEsq}} below), the Maxwell 
equations in the interaction basis~\eqref{eq:evolutionequations} reduce to 
\begin{align}
    \partial_t {E}^i \approx -\sigma E^i + \varepsilon \mu^2 A'^i,
    \label{eq:nonrelelectricfieldeq}
\end{align}
where we assume the same Ohm's law as before in \eqref{eq:ohmslaw}. With the non-relativistic superradiant field solutions $A'_i$ in \eqref{eq:srprofiles}, we see that the visible electric field is driven towards the inhomogeneous solution 
\begin{align}
    \textbf{E}\propto\begin{pmatrix} \sigma +i\mu \\ \mu -i \sigma\\ 0 \end{pmatrix} e^{-i\omega t}+c.c.
    \label{eq:analyticEfield}
\end{align}
over timescales of $1/\sigma$. Hence, as
conductivity increases towards $\sigma\sim \mu$, the visible electric field direction rotates with
respect to a fixed superradiance cloud phase, such that at large conductivities,
$\sigma/\mu \gg 1$, the visible and superradiant electric fields exhibit a
phase-offset of $\pi/2$. Notice, however, at very large conductivity, this
analytic approximation, in principle, is no longer valid, as it neglects the
magnetic field effects, which become important as the electric field is
screened, for $\sigma/\mu\gg 1$. These effects induce the appearance of small
scale structures visible in the last panel of \figurename{
    \ref{fig:condEFieldlines}}. We will elaborate on the break down of this
approximation and the emergence of small scale features in more detail in
Sec.~\ref{sec:Highlyconductinglimit}.

\begin{figure*}[t]
    \centering
    \includegraphics[width=.95\textwidth]{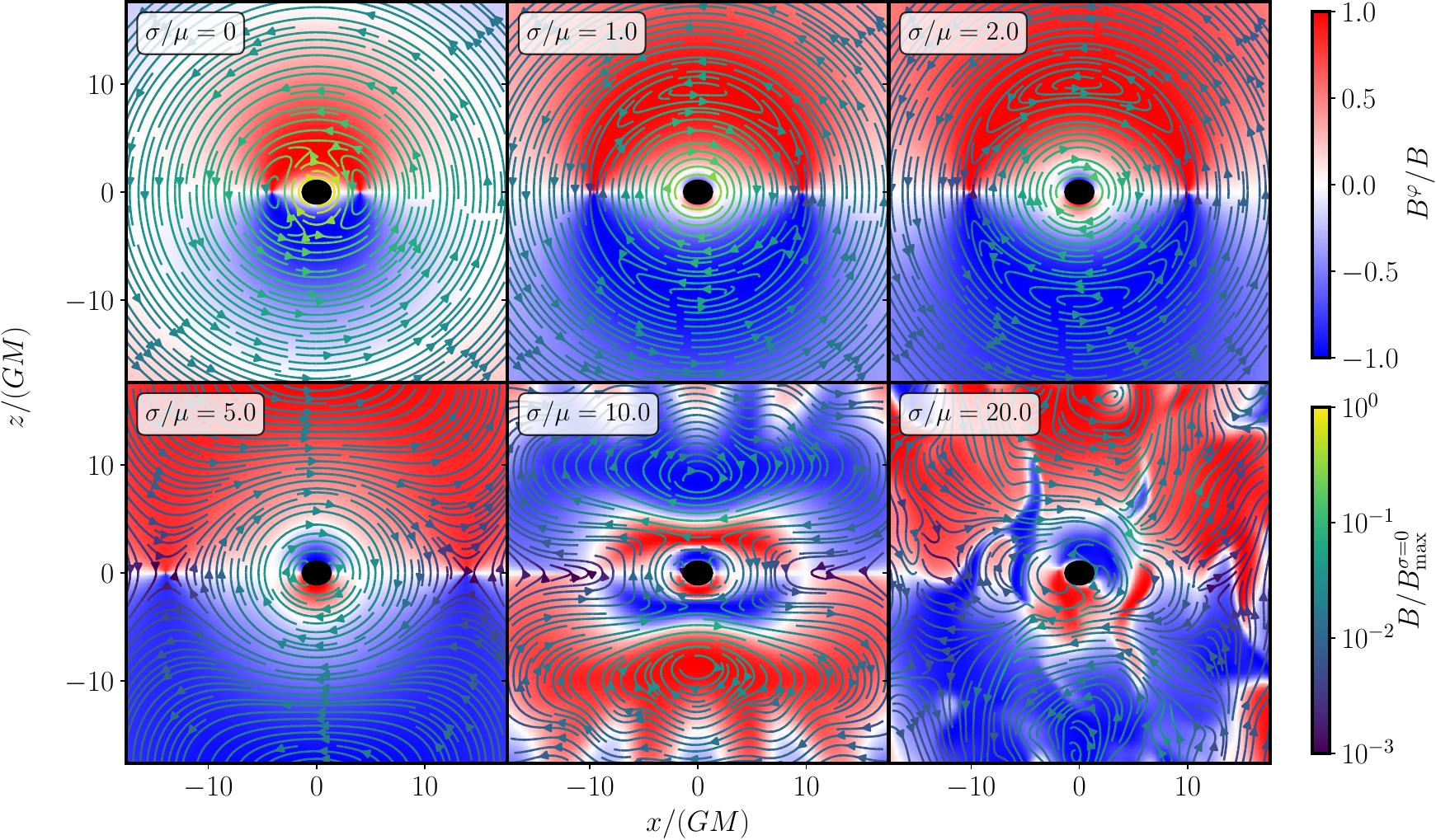}
    \caption{Magnetic field lines $B^i$ and magnitudes in a coordinate slice spanned by the BH spin (pointing in the $z$-direction), and an arbitrarily chosen superradiance cloud phase.
    The six panels show the field configurations in the same slice for successively larger conductivities $\sigma$. The background colors (red/blue) indicate the magnitude of the component perpendicular to the slice $B^\varphi$ in the $\varphi$-direction around the BH normalized by the magnitude of the visible magnetic field. The colors of the field lines (yellow/green) indicate the magnitude of the visible magnetic field along the field lines normalized by the maximal magnitude in the vacuum case $B_{\rm max}^{\sigma=0}$. The BH and cloud parameters in all panels are as in \figurename{ \ref{fig:vaccumFieldlines}}, i.e., $\alpha=0.3$ and $a_*=0.86$. The magnitude of the magnetic field, while exponentially decaying in the vacuum limit, is roughly uniform at large conductivities $\sigma/\mu=20$. The small-scale features are discussed in detail in Sec.~\ref{sec:Highlyconductinglimit}}
    \label{fig:condBFieldlines}
\end{figure*}

As the conductivity grows, the electric field decreases in amplitude and the
magnetic field plays a more important role. The field line geometry and
magnitude of the visible magnetic fields are shown as a function of
conductivity in \figurename{ \ref{fig:condBFieldlines}}. At vanishing $\sigma$,
the solutions are identical to the vacuum solution, and the snapshot in
\figurename{ \ref{fig:condBFieldlines}} simply represents a slice of the
geometry shown in \figurename{ \ref{fig:vaccumFieldlines}}. The magnetic null
line (i.e., where $B^i=0$) crosses this slice once on either side of the BH as
they spiral away from the BH. The field lines close around this null line, and
the magnetic field strength is largest close to the BH and decays exponentially
towards spatial infinity. In the vacuum limit, the dark photon field
exhibits an exact helical symmetry about the BH spin-axis\footnote{The helical
Killing field is $k^\mu=\xi^\mu-\Omega_{\rm BH}\varphi^\mu$ in terms of the stationary
and axisymmetric Killing vectors. The superradiant field $A'_\mu$
strictly retains this symmetry, at leading order in the kinetic mixing, i.e.
$\mathcal{L}_\textbf{k}A'_\mu \sim\mathcal{O}(\varepsilon^2)$.}.  For
$\sigma= \mu$, the magnetic field pattern still exhibits this helical symmetry
approximately on the spatial scales depicted in \figurename{
    \ref{fig:condBFieldlines}}. This symmetry is broken for $\sigma/\mu>1$.
Therefore, the last two snapshots of \figurename{ \ref{fig:condBFieldlines}},
while representing the magnetic field geometry qualitatively, are not
indicative of the full three-dimensional field geometry. Qualitatively, at large conductivities, the plasma turns
into a highly conducting pair plasma attaching to the visible magnetic field
lines. This implies a differential rotation of the magnetic field lines at 
large and small distances from the BH,  which breaks the helical geometry into a more complex configuration presenting small-scale features, that are further discussed in the next section.
For both the electric and magnetic field, the presence of the plasma leads to
field strengths that are relatively uniform in magnitude within the Bohr radius
(as can be seen in the $\sigma/\mu=20$ panel of \figurename{
    \ref{fig:condEFieldlines}}). The small-scale features, crucial for the
high-conductivity dynamics of the system, are discussed in detail in the next
section.

\begin{figure*}[t]
    \centering
    \includegraphics[width=.99\textwidth]{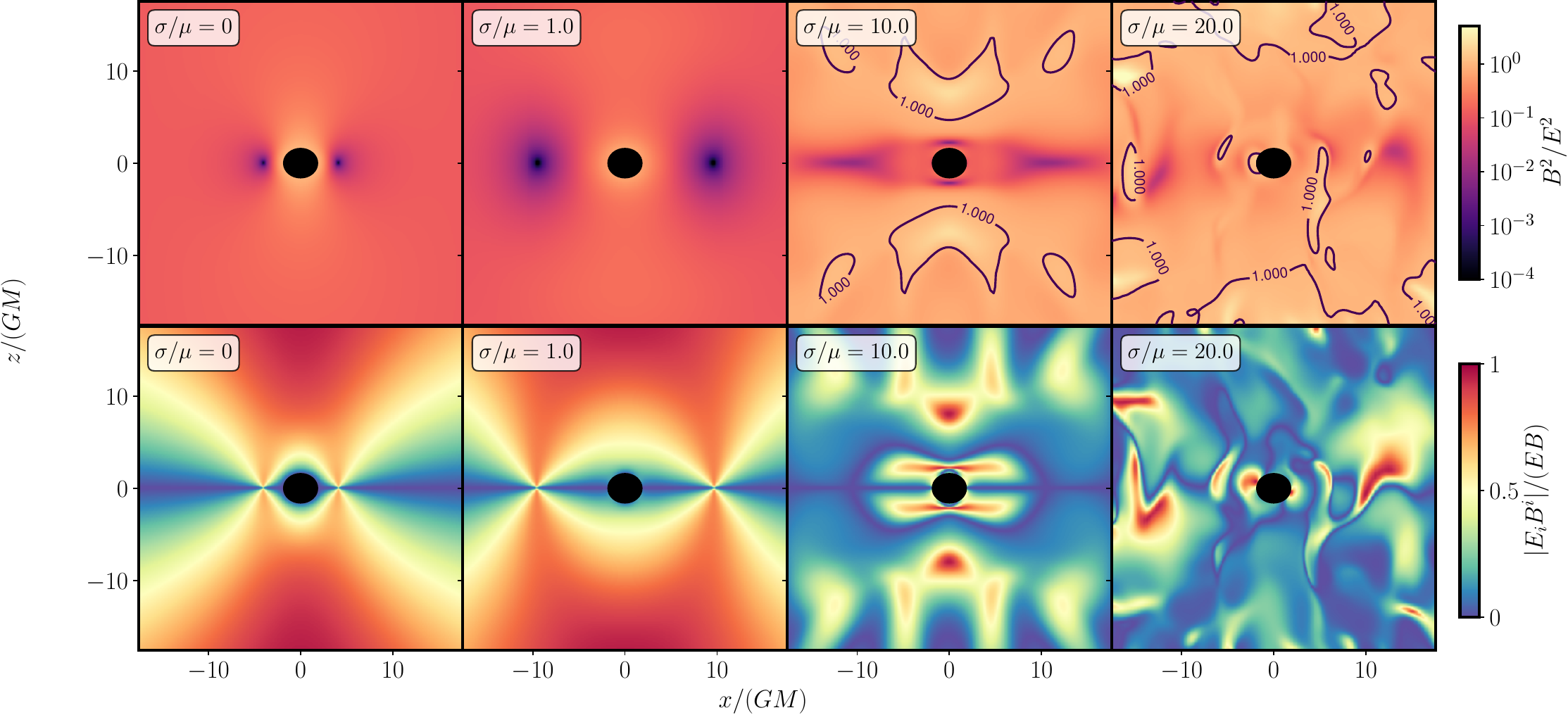}
    \caption{\textit{(top row)} The ratio between visible magnetic and electric
field strengths, $B^2$ and $E^2$, respectively, as function of conductivity
$\sigma$ in the same slices as in \figurename{ \ref{fig:condBFieldlines}}
(i.e., spanned by the BH spin-axis in the z-direction and an arbitrary
superradiant phase). Contour lines indicate, where $B^2/E^2=1$. \textit{(bottom
row)} The magnitude of the visible electric field component in the direction of
the visible magnetic field, $|E_iB^i|$, normalized by both magnitudes. The
slices of the top and bottom rows are identical. The BH and cloud parameters in
all panels are as in \figurename{ \ref{fig:vaccumFieldlines}}, i.e.,
$\alpha=0.3$ and $a_*=0.86$. For $\sigma/\mu\lesssim 1$, the electric field is
dominant everywhere and the violations of $|E_iB^i|=0$ is strong, while for
$\sigma/\mu\gtrsim 1$, the magnetic field begins to dominate in some regimes
and $|E_iB^i|=0$ is violated only in isolated regions.}
    \label{fig:condBsqoverEsq}
\end{figure*}

\begin{figure*}[t]
    \centering
    \includegraphics[width=1\textwidth]{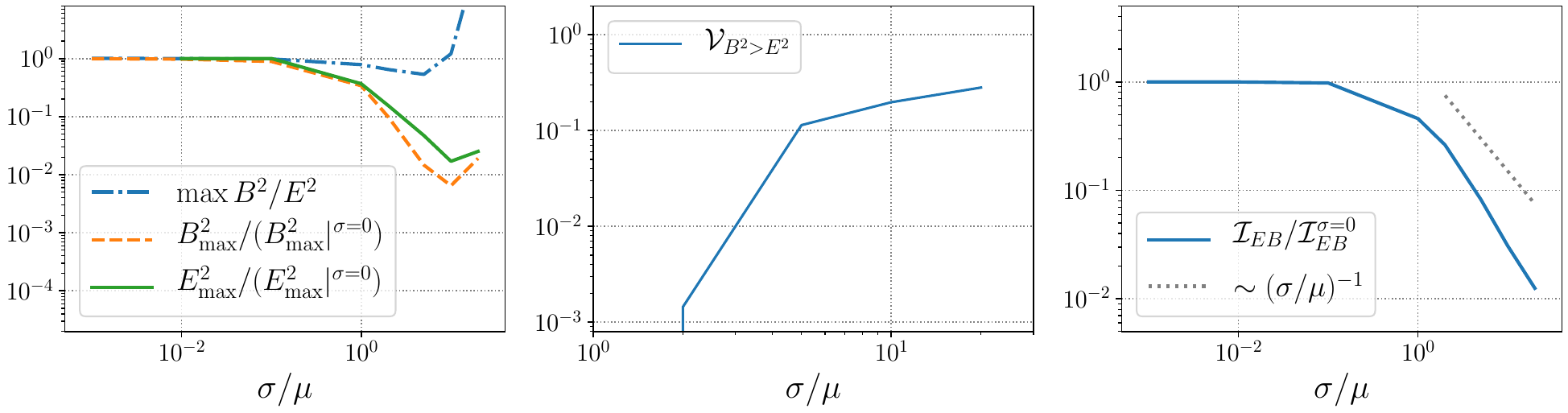}
    \caption{\textit{(left)} We plot the maximal ratio of visible magnetic to
     electric field magnitudes $\max B^2/E^2$, and the maximal magnetic and electric
     field magnitudes, $B_{\rm max}^2$ and $E_{\rm max}^2$,
     normalized by their maximal vacuum values, 
     as a function of
     plasma conductivity $\sigma/\mu\in\{0.001,0.01,0.1,1,2,5,10,20\}$.
     \textit{(middle)} The fractional coordinate volume $\mathcal{V}_{B^2>E^2}$ of
     magnetically dominated regions inside a coordinate sphere of radius
     $4/(\alpha\mu)$ around the central BH as function of conductivity.
     \textit{(right)} We show the behavior of the volume integral of $|E^iB_i|$ over a coordinate sphere of radius $10r_c$,
     $\mathcal{I}_{EB}$, 
     as a function of conductivity, normatlized to its 
     vacuum value $\mathcal{I}_{EB}^{\sigma=0}$. As above, we consider here a
     BH-cloud system with $\alpha=0.3$ and spin $a_*=0.86$ in all panels.}
    \label{fig:limitandcurrent}
\end{figure*}

In order to understand the degree to which our solutions approach a force-free
solution in the $\sigma\rightarrow \infty$ limit, we consider how violations of
the force-free conditions $E_iB^i=0$ and $B^2>E^2$ (respectively, conditions
\textit{(i)} and \textit{(ii)} discussed in the beginning of subsection~\ref{ssec:RFFE}),
change with increasing conductivity.  In \figurename{ \ref{fig:condBsqoverEsq}},
we show, in representative slices, pointwise measures of the violations of these conditions, 
while in \figurename{ \ref{fig:limitandcurrent}} we show how volume
integrated measures of these violations decrease with increasing conductivity.

Examining a volume integral of $E_iB^i$ as a function of conductivity, shown in
rightmost panel of \figurename{ \ref{fig:limitandcurrent}}, we find that it
begins to decrease like $1/\sigma$ for $\sigma/\mu \geq 1$. From the bottom
panels of \figurename{ \ref{fig:condBsqoverEsq}}, we can see that,
in contrast to low and moderate value of $\sigma/\mu$, at high
conductivity, large values of $E_iB^i$ (relative to the magnitude of the
fields) occur only in isolated, smaller-scale regions.

We also find that the fraction of the volume that is magnetically dominated
increases with increasing conductivity, as shown in the middle panel of
\figurename{ \ref{fig:limitandcurrent}}, in particular, for a coordinate sphere
of radius $4/(\alpha\mu)$. For $\sigma=0$, none of this volume is magnetically
dominated (as expected, since $B'\sim\alpha E'$, $B = \varepsilon B'$, and $E = \varepsilon E'$), while for
$\sigma/\mu=20$, approximately one-third of the volume is. The spatial extent
of these magnetically dominated regions can be seen in the top panels
of~\figurename{ \ref{fig:condBsqoverEsq}}.  Similarly, as shown in the left
panel of \figurename{ \ref{fig:limitandcurrent}}, the global maximum visible
electric and magnetic field strengths are comparable, while the value of $\max
B^2/E^2$ increases to large values with large conductivities.

Na\"ively, one might expect that in the infinite conductivity limit, the
visible electric field in the fluid frame is completely shorted out by large
scale charge separation, leading to $e^\mu\rightarrow 0$, and hence a
magnetically dominated solution everywhere (recalling that magnetic dominance
is equivalent to the existence of a frame where the electric field vanishes, i.e., $B^2/E^2>1$).
The results above suggest that a force-free solution might exist for
$\sigma\rightarrow\infty$, at least in a significant fraction of the total
volume taken up by the plasma and superradiance cloud.  However, given the slow
increase in the magnetically dominated fraction of the volume with increasing
conductivity, it seems plausible that electrically dominated regions with
non-zero volume may
persist as $\sigma\rightarrow\infty$. Electric dominance implies an unscreened
electric field in the fluid frame, allowing for strong particle
acceleration, and the dissipation of field energy.  This is consistent with our
force-free simulations (without a guide field), which always evolve towards
developing electrically dominated regions (see \figurename{
    \ref{fig:forcefreequantities}} in App.~\ref{app:resistivitytest} for a
discussion on these force-free simulations), and will be discussed further in
the following section.  In the next section, we will also discuss how the
violations of the force-free conditions are connected to the turblent behavior
the plasma.

\subsection{Field solutions: Small scale turbulence} \label{sec:Highlyconductinglimit}

In the previous section, we found that a large-scale charge separation
screens the superradiant electric field, lifting the importance of the
visible electric field and leading to magnetically dominated regions. In
what follows, we focus on the magnetic field dynamics in the
large-conductivity regime. We show that, in this regime, the dark photon
superradiance cloud-plasma system is characterized by turbulent plasma
dynamics in the bulk of the system.  A trend towards small-scale features
can already be seen in \figurename{ \ref{fig:condBFieldlines}}, and
becomes more apparent in \figurename{ \ref{fig:condBsqoverEsq}} in the
previous section. The turbulent regions emerge not in isolated and clearly
structured lower-dimensional regions, but rather across the bulk of the
cloud.  This is in contrast to the pulsar magnetosphere, where, at least
in the high conductivity limit, small scale features and dissipation are
expected to be contained mostly in a two-dimensional current sheet.

In order to understand the turbulent behavior of the visible magnetic field dynamics at 
moderate and large conductivities, it is instructive to consider the visible magnetic induction equation in the presence of a finite kinetic mixing with the massive vector field. For simplicity, we focus on the flat spacetime limit only, noting that the
following qualitative arguments are unchanged on curved backgrounds.
Furthermore, for clarity, we assume that the fluid (i.e., the plasma) is mostly
non-relativistic, and that at large $\sigma$, the plasma is
conduction (as opposed to advection) dominated, as explicitly shown in App.~\ref{app:lowconductivityregime}. 
Finally, we can neglect the displacement current $\partial_t E^i\sim \mu
E^i$, as it is suppressed compared to the conduction term $\sim\sigma
E^i$, if $\sigma\gg \mu$. All qualitative arguments outlined below translate to the fully-relativistic case. Making these assumptions, the evolution
equations of the (kinetically mixed) Maxwell equations
\eqref{eq:evolutionequations} together with the current \eqref{eq:ohmslaw}
reduce to the visible magnetic induction equation
\begin{align}
\partial_t {B}^i=\frac{\varepsilon\mu^2 B'^i}{\sigma}+\frac{1}{\sigma} \partial_j\partial^j {B}^i+\varepsilon^{ijk}\varepsilon_{klm}\partial_jv^l{B}^m,
\label{eq:inductionequation}
\end{align}
where $B'^i=\varepsilon^{ijk}\partial_j A'_k$ and $v^i$ is the fluid velocity that we identify with the drift velocity $v^i=v^i_d$, such that $v^i{E}_i=0$. For later convenience, we define the Cartesian Kerr-Schild coordinate radius $\hat{\rho}=(x^2+y^2+z^2)^{1/2}$ (see App.~\ref{app:numericalsetup} for details).
With a characteristic length scale $\ell=1/\mu$ of the
system, we are able to define an effective \textit{magnetic Reynolds
number}, $R_m=\ell v_d \sigma = \sigma v_d/\mu$, of the effective plasma
defined in \eqref{eq:resistivecurrent}\footnote{One could instead chose $\ell=r_c$. However, the
qualitative arguments are unaffected by the precise choice of magnetic Reynolds number.}.  For $R_m\gg 1$,  the magnetic field dynamics is dominated by the third term in Eq. \eqref{eq:inductionequation}, which represents field advection. In this regime the magnetic field dynamics is entirely determined by the plasma, as both are strongly coupled (i.e., the magnetic field is comoving with the plasma). For $R_m\ll 1$, the second term in the induction equation is most important, which accounts for magnetic field diffusion. In this case the magnetic field decouples from the plasma motion and relaxes to a diffusive state. The effects of the superradiant driving fields are included explicitly in the first term of Eq. \eqref{eq:inductionequation} and implicitly in the plasma velocity, which depends on the driving electric fields, and in the visible magnetic field themselves, which are sourced by electrically induced currents.

\begin{figure*}[t]
    \centering
    \includegraphics[width=1\textwidth]{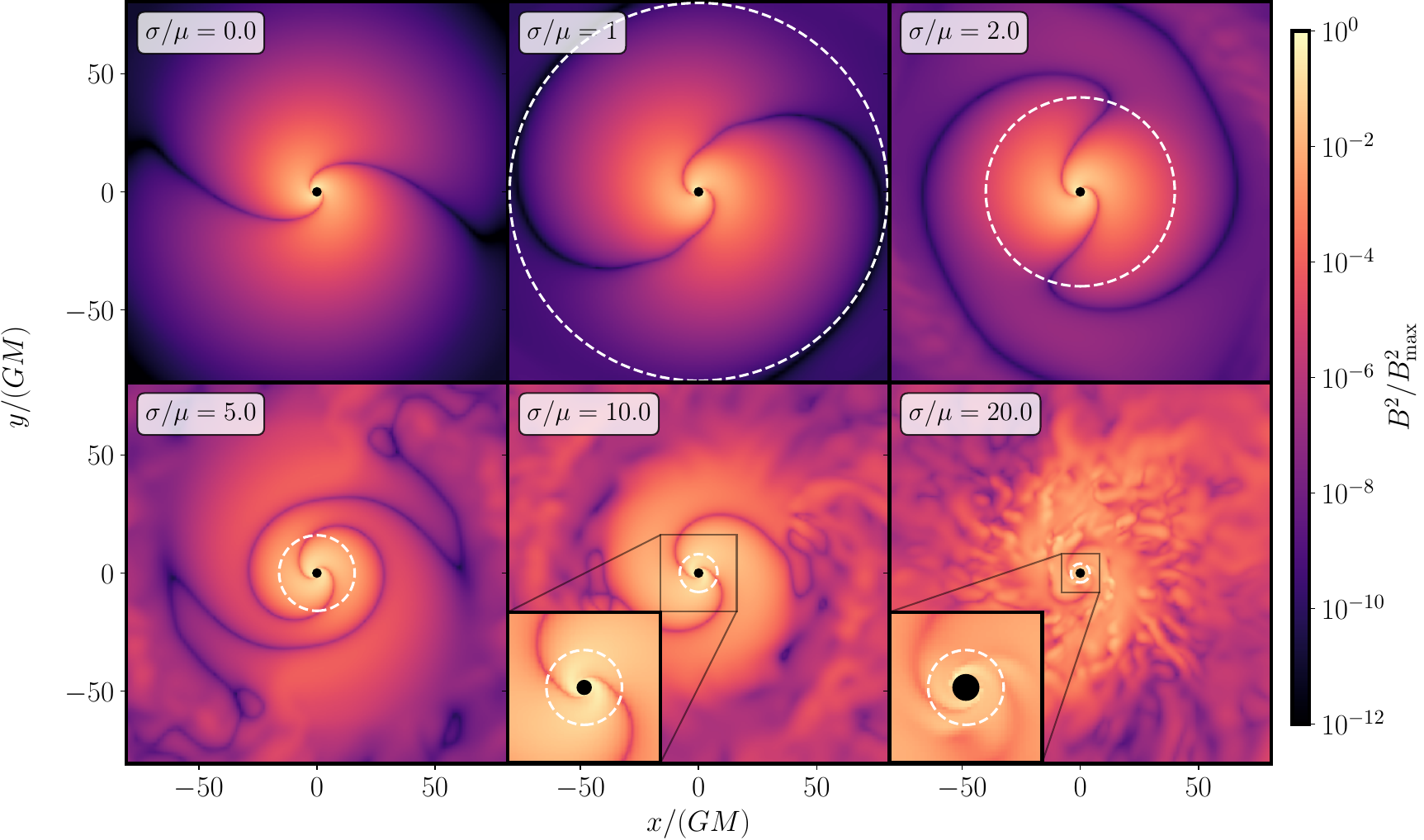}
    \caption{We plot the visible magnetic field strength $B^2$ in the equatorial plane of the system in the vacuum limit, $\sigma=0$, as well as at moderate to high plasma conductivities, i.e., $\sigma/\mu\gtrsim 1$. The BH and cloud parameters are as in \figurename{ \ref{fig:vaccumFieldlines}}, i.e., $\alpha=0.3$ and $a_*=0.86$. The superradiance cloud phase is identical in each of the panels. The color is normalized by the maximal visible magnetic field strength at each conductivity. The white dashed line indicates the critical coordinate radius $r_*=80\mu GM/\sigma$, discussed in the main text. The region $\hat{\rho}<r_*$ is dominated by superradiant driving, while the regions with $\hat{\rho}>r_*$ are characterized by an interplay of advective and diffusive regions. The flat spacetime light cylinder for this system is roughly $R_{\rm LC}=GM/\alpha\approx 3.33 GM$. Notice, the resolution of our numerical methods decreases with increasing coordinate distances $|x|$ and $|y|$, resulting in, for instance, a suppression of small-scale features in the $\sigma/\mu>2$ cases for $|x|,|y|>50GM$.}
    \label{fig:Bsqlargesigma}
\end{figure*}

To illustrate the different domains of the magnetic field dynamics, in \figurename{ \ref{fig:Bsqlargesigma}}  we show the magnetic field magnitude inside the equatorial plane as a function of conductivity. In the vacuum limit, $\sigma/\mu=0$, the magnetic field coincides with the superradiant magnetic driving field shown in \figurename{ \ref{fig:vaccumFieldlines}}. With increasing conductivity, i.e., $\sigma/\mu\gtrsim 1$, from \figurename{ \ref{fig:Bsqlargesigma}} we see that the regions where the morphology of the magnetic field resembles the superradiant magnetic fields are confined to distances from the BH that are smaller than a characteristic radius $r_*$, which we heuristically find to be 
\begin{equation}
 r_*\approx 80\mu GM/\sigma.
 \label{eq:rstar}
\end{equation}
Inside this critical radius, the superradiant driving field $B'^i$ is exponentially large, and the first term of \eqref{eq:inductionequation} dominates, compared with the terms it induces in the diffusion and advection contributions\footnote{Note, the superradiant components of the visible magnetic fields in the second and third term in the induction equation \eqref{eq:inductionequation}, are $\alpha^2$ and $\alpha$ suppressed, respectively, to the leading contribution at intermediate conductivities.}.
Note also that at large conductivities an overall phase-offset of $\pi/2$ between the superradiant and visible magnetic fields appears,  similar to the behavior of the electric field case (see
\figurename{ \ref{fig:condEFieldlines}}); at the level of the magnetic field, this phase-offset emerges from the first term in the induction equation \eqref{eq:inductionequation}.

For radii larger than $r_*$ we see that the magnetic fields are non-trivially related to the driving superradiant electric and magnetic fields. At large distances from the BH and on  scales of order $1/\mu$, the
magnetic field strength increases with growing conductivity. On smaller scales, on the other hand, and especially at large conductivities, we see that a series of small scale features appear. These features, which arise on scales of order $1/\sigma$, emerge from the interplay of the diffusion and advection
components of the magnetic induction equation. At large conductivities and due to the large plasma velocities, most
regions with $\hat{\rho}>r_*$ are advection dominated. As a result, the oscillating superradiant driving fields source
visible magnetic fields that couple strongly to the plasma, while conversely,
the plasma cannot back react onto the superradiant driving fields. Our results show that for radii $\hat{\rho}>r_*$, and on spatial scales of the entire cloud (i.e., $>1/\mu$), the plasma is unable to corotate with the driving fields; hence, differential rotation between plasma cells at different radii occurs. 
This shear velocity results in twisting of the magnetic field lines on scales of
the cloud. The twist builds up magnetic
energy that is dissipated through turbulent magnetic reconnection in regions of
large magnetic diffusion, i.e., $R_m\ll 1$ (as we illustrate in detail below). 
Thus, the small scale features in
\figurename{ \ref{fig:Bsqlargesigma}} are a result of this turbulent
reconnection. The radius
$r_*$ may saturate at the light cylinder of the system for
$\sigma\rightarrow\infty$ as inside it the plasma could rotate rigidly
with the superradiant driving fields. However, we do not find the light
cylinder to be a location of special importance for the largest conductivity that we considered: $\sigma/\mu= 20$.

Another qualitative feature is the disappearance of the magnetic null line of
the superradiant magnetic field for moderate conductivities, $\sigma/\mu>5$,
outside the critical radius $r_*$, as can be seen in \figurename{
\ref{fig:Bsqlargesigma}}. Vanishing magnetic field strength implies vanishing
plasma bulk velocity, i.e., $R_m\approx 0$, and equivalently the presence of
strong magnetic diffusion. Hence, we find that the magnetic null line is
quickly filled by magnetic field lines diffusing into the null line from
surrounding areas with finite magnetic field strength. This process efficiently
removes the null line outside the critical radius $r_*$. 

In addition to the large scale differential rotation about the BH, in 
advection dominated regions (where the plasma and
magnetic field are co-moving) we
observe localized roughly uniform \textit{oscillatory} motion of the plasma
within the equatorial plane (with oscillation radius given by $1/\mu$), 
as well as periodic \textit{longitudinal} motion
of the plasma along the BH spin-axis\footnote{Even in the force-free
simulations, this periodic oscillatory motion of the magnetic field strength in
the equatorial plane can be observed.}. This periodic motion is likely driven
by the large scale superradiant electric field throughout the plasma, in conjunction
with large scale charge separation of the pair plasma for $\sigma/\mu\gtrsim
1$. Charges are accelerated along the large scale superradiant electric field. However, the
orbital frequency $\mu$ of the field's direction forces the charges into a
circular trajectory with radius of $1/\mu$. This is reflected in
the circular motion of features in the magnetic field of scale $1/\mu$ inside the
equatorial plane. The circular motion of negative
and positive charges is exactly out of phase by $\pi$, resulting in out-of-phase
oscillatory motion of the plasma on either side of the BH due to the large scale
charge dipole screening the superradiant electric field.
The longitudinal periodic motion along the spin axis
is more complex, and likely a result of the electric field driving within the
equatorial plane. We will discuss the observational consequence of this motion
of charge densities in Sec.~\ref{sec:emissionperiodicity}.

\begin{figure*}[ht]
    \centering
    \includegraphics[width=0.8\textwidth]{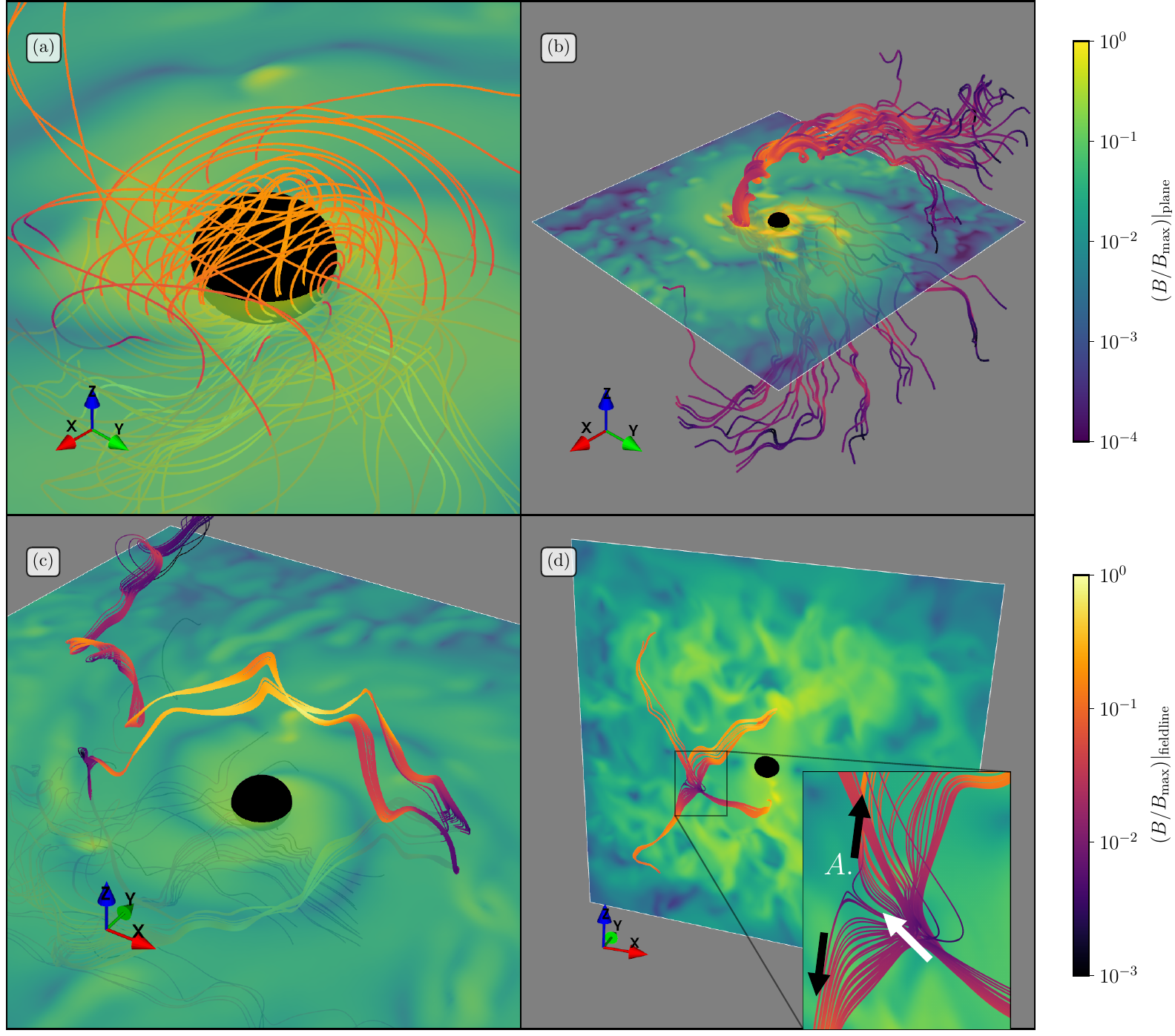}
    \caption{We plot a selection of visible magnetic field lines of the
	superradiance cloud-plasma system with conductivity $\sigma/\mu=20$,
	$\alpha=0.3$, and $a_*=0.86$. In panels (a), (b) and (c) we also plot the
	visible magnetic field strength within the equatorial plane, while in panel (d)
	we show the visible magnetic field strength in the plane spanned by the BH spin
	and an arbitrary superradiance cloud phase. We discuss this plot in detail in
	the main text. The main macroscopic scales involved are the BH-scale, set by
	the mass $M$, the superradiance cloud's oscillation timescale $1/\mu\approx 3.33
	G M$, and the cloud's Bohr radius $r_c\approx 11.1 G M$.}
    \label{fig:Magfieldlines}
\end{figure*}

Let us illustrate some of these observations explicitly in \figurename{
\ref{fig:Magfieldlines}}. In panel (a), we show the three-dimensional geometry of the 
visible magnetic field lines in the vicinity of the central BH.
The field lines are mostly closed around the BH, and only occasionally thread
the event horizon. This confirms the discussion above, as these field lines are
roughly inside $\hat{\rho}<r_*\approx 4G M$ for the choice of parameters in the figure,
and hence, are approximately set by the superradiant driving field (compare
also to \figurename{ \ref{fig:vaccumFieldlines}}). 

The large-scale differential rotation and twisting of magnetic field lines on scales of the entire cloud is
shown in panel (b) of \figurename{ \ref{fig:Magfieldlines}}. There, we focus on
a few representative field lines crossing the equatorial plane. Both above and
below the equatorial plane, we find that the azimuthal angular velocity of the
magnetic field lines decreases with increasing distance from the BH's spin
axis, leading to a lag of portions of the field lines far away from the BH
compared to those closer to the center. This lag causes twisting of the field
lines roughly around the spin-axis\footnote{This is only roughly true, since
the plasma motion is more complex as pointed out above.} on scales of the
superradiance cloud, which, ultimately, results in the opening of the visible
magnetic field lines at large distances. 

In addition to this large scale feature, the field line geometry also exhibits
features on scales of roughly $1/\mu$, as can be seen in panel (c) of
\figurename{ \ref{fig:Magfieldlines}} (notice, this scale compared with the BH
mass is $1/(\mu GM)\approx 3.33$). There we show two collections of field
lines exhibiting variations on spatial scales set by the superradiance oscillation frequency.
These arise likely as a result of the periodic motion of the plasma in the
large scale superradiant electric field, twisting the field lines on scales of $1/\mu\ll
r_c$ as well. Both the twisting on scales of the cloud, and small
scales, $1/\mu$, builds up magnetic energy that is dissipated by means of
magnetic reconnection. 

In panel (d) of \figurename{ \ref{fig:Magfieldlines}}, we isolate one such
reconnection sight, representative for a class of reconnection processes active
throughout the bulk of the cloud.
We leave the details to the discussion in Sec.~\ref{sec:dissipationmechanism},
and just point out here that the visible magnetic field lines entering the
reconnection region along the white arrow, diverge away into \textit{two}
distinct directions along the black arrows. This indicates that the
connectivity of the field lines is discontinuously changed at the location
labelled as A. Another indication of magnetic reconnection at location A in
panel (d) of \figurename{ \ref{fig:Magfieldlines}}, is the appearance of
\textit{four} bundles of field lines from the region around A. In the highly
conducting and highly magnetized plasma limit, (at least one of) the dimensions
of the reconnection regions are expected to scale as $1/\sigma$. Speculating
about the $\sigma\rightarrow\infty$ limit, we hypothesize that these
reconnection sites turn into one- or two-dimensional highly fragmented
localized filaments and current sheets, where potentially large amounts of
electromagnetic energy is injected into the plasma (as typically observed in 
turbulent highly magnetized plasmas, e.g., \cite{Zhdankin:2016lta,Comisso:2018kuh,2011A&A...525A..57P}). 
We discuss the connection
between significant energy dissipation into the plasma and magnetic
reconnection in Sec.~\ref{sec:dissipationmechanism} in detail.

\subsection{Summary of turbulent plasma scales} \label{sec:turbulentscalesummary}

Before concluding this section, we provide a summary of the main features of the quasi-equilibrium endstate of the pair production cascade. We begin with the largest scales first, and work towards small scales:
\begin{itemize}
\item On spatial scales of the superradiance cloud, $\sim r_c=1/(\mu\alpha)$, the
superradiant electric field is efficiently screened by a roughly dipolar charge
distribution. A significant fraction of the cloud's volume is magnetically
dominated at $\sigma/\mu=20$ with trend towards larger fractional volumes for
larger plasma conductivities. This increases the importance of the magnetic field
dynamics for the cloud-plasma system. For sufficiently high conductivity, the
magnetic field and the plasma become strongly coupled except in isolated
diffusion regions. Hence, at large distances from the central BH's spin axis,
the plasma rotates around the BH with period much longer than the superradiant
cloud's period, inducing differential rotation on the scale of the entire
cloud. The resulting shearing magnetic field lines reconnect inside the bulk of the cloud.

\item On spatial and temporal scales set by the Compton wavelength of the dark photon, $1/\mu$, a variety of features appear. The plasma orbits with the superradiance cloud in circular motion with radius given by roughly $1/\mu$. This is likely due to the large scale electric field set by the superradiance cloud. Hence, negative and positive components of the local charge density orbit exactly out of phase due to the large charge dipole. In the large conductivity regime, this circular motion implies circular motion of the visible magnetic field due to the strong coupling in advection dominated regions. Features of size $1/\mu$ in the global visible magnetic field geometry appear due to the built-up magnetic field twisting, which is released in magnetic field line reconnection sites.

\item Besides the two macroscopic scales discussed above, the conductivity
$\sigma$ sets the size of non-ideal features, which are expected to be of
microscopic size. We found that at moderately large conductivities
$\sigma/\mu=20$, these non-ideal regions begin to form filaments inside the
superradiance cloud's plasma, setting the scale of the turbulent behavior. 
Speculating, for very large conductivities, $\sigma/\mu\gg 20$, the non-ideal 
regions may fragment into a large number of current sheet-like structures 
filling the turbulent plasma\footnote{This is typically found in treatments of turbulent magnetized plasmas, e.g., \cite{Comisso:2018kuh,Zhdankin:2016lta,2011A&A...525A..57P}.}. 
Below, in Sec.~\ref{sec:emissionpower},
we elaborate on this and identify these regions as sites of enhanced energy dissipation.
\end{itemize}
Therefore, the superradiance cloud-plasma system is characterized by
differential rotation, as well as periodic plasma motion with period given by
the boson mass scale $1/\mu$, leading the magnetic field lines to be twisted
both on cloud size scales $1/(\mu\alpha)$, and on scales set by the dark photon mass 
$1/\mu$. This twisting is relaxed through magnetic reconnection in the bulk
of the superradiance cloud in features with size set by $1/\sigma$. 
These processes likely lead to strong
electromagnetic transients with periodicity set by the dark photon mass $\mu$.
Characterizing the power and periodicity of these transients is the subject of
the next section.

\section{Electromagnetic emission} \label{sec:luminosity}

The pair production cascade within the superradiance cloud saturates in a
turbulent, differentially-rotating plasma surrounding the central BH with
a partially screened electric field and magnetic field line reconnection in the
bulk. In highly-magnetized astrophysical plasmas, particles are
efficiently accelerated at reconnection sites, leading to high-energy
electromagnetic emission. Therefore, we expect strong electromagnetic
signatures from the superradiance cloud system. 

In the following, we
illustrate the radiation and dissipation channels in our setup in 
Sec.~\ref{sec:radchannel}, quantify the emitted electromagnetic luminosity in Sec.~\ref{sec:emissionpower}, identify the dominant emission mechanism in Sec.~\ref{sec:dissipationmechanism},
discuss the periodicity of the emission pattern in Sec.~\ref{sec:emissionperiodicity}, and comment on the possible emission
spectra in Sec.~\ref{sec:emissionspectrum}. 

\subsection{Radiation and dissipation channels} \label{sec:radchannel}

The effective description of the pair plasma that we use, introduced in the
previous sections, includes only the leading-order resistive correction to the
force-free limit of ideal magnetohydrodynamics. In the context of this
formalism, any microphysical processes (e.g. pair production, scattering, photon emission,
or other QED effects) are averaged over,
or only roughly approximated by the macroscopic conductivity $\sigma$
that characterizes the local dissipation in the plasma, and not included from first
principles\footnote{To some degree, this
could be achieved within the context of kinetic theory and particle-in-cell
simulations. However, we leave exploring this avenue to future work.}. However,
our approximation is sufficient to reliably estimate the total electromagnetic
power output of the system through the outgoing Poynting flux, as well as
through dissipation due to macroscopic spatial currents along
the visible electric field. The Poynting flux is typically invoked,
within the force-free paradigm to estimate the rotational energy extraction
rates of pulsars \cite{1969ApJ...157..869G,Contopoulos_1999,Komissarov:2005xc,Spitkovsky:2006np,Ruiz:2014zta,Petri:2015rbe,Carrasco:2018kdv} (which have been confirmed within kinetic theory in
\cite{Cerutti:2015hvk,Philippov:2014mqa,Philippov:2017ikm}). For typical pulsars, this macroscopic coherent energy flux is expected to be
emitted from the system in the form of lower-energy radio waves, as well as
dissipate in particle acceleration processes in current sheets resulting in X-rays and gamma-rays.
While in the pulsar magnetosphere, dissipation is mainly confined to roughly two-dimensional
current sheets, the
superradiance cloud exhibits reconnection in the bulk, enabling efficient energy
transfer into high-energy emission. The dissipation of electromagnetic energy 
can be interpreted as sourcing local particle acceleration, synchrotron and curvature
radiation, and plasma heating. Therefore, estimating the total emitted Poynting
flux and dissipative energy losses of the superradiant system is crucial to
understand the overall electromagnetic signatures. In the following, we briefly outline how these quantities are computed in our setup.

\paragraph{Modified Poynting theorem:} The conservation of energy in the
interaction basis, \eqref{eq:energymomentumconservation}, can be used to
identify the macroscopic sources and types of energy flows present in the
system. The background Kerr spacetime has an asymptotically-timelike Killing
field $\xi^\mu$, endowing the system with a local energy conservation law.
Therefore, we define the total energy $\mathcal{E}$ of $A_\mu$, with respect to $\xi^\mu$,
within the (coordinate) domain $D$ as 
\begin{align}
    \mathcal{E}=\int_Dd^3x\sqrt{\gamma}T^{\alpha}{}_\mu n_\alpha \xi^\mu,
\end{align}
with the volume form $d^3x\sqrt{\gamma}$ of a $t=$const. slice of Kerr spacetime (see App.~\ref{app:numericalsetup} for details), and the energy-momentum tensor $T_{\mu\nu}$ of the \textit{visible} fields
\begin{align}
    T_{\mu\nu}= F_{\mu}{}^\lambda F_{\lambda\nu}-\frac{1}{4}g_{\mu\nu}F^{\alpha\beta}F_{\alpha\beta}.
\end{align}
Throughout this work, the domain $D$ of consideration is the exterior of the BH
up to a coordinate sphere $S_{\hat{\rho}}^2$ at coordinate radius
$\hat{\rho}=\sqrt{x^2+y^2+z^2}$ in Kerr-Schild coordinates (defined in App.~\ref{app:numericalsetup}). 
In the following, we focus entirely on the \textit{visible} electromagnetic field. Intuitively, this visible field is a superposition of the massless (i.e., the SM photon) and the massive vector fields. The former is propagating freely, and sourced only by the plasma, while the latter is gravitationally bound to the BH, and non-radiative. Given this, and the energy-momentum conservation \eqref{eq:energymomentumconservation}, we can relate the change of the total energy of the visible electromagnetic fields within $D$, $\partial_t\mathcal{E}$, to the energy fluxes across the boundary of the domain, $\partial D$, as well as the work done on the plasma within $D$, by the modified Poynting theorem, written as
\begin{align}
    \partial_t \mathcal{E}=-P_{\rm EM}-\dot{\mathcal{E}}_{\rm BH}-L_{\rm diss}+\dot{\mathcal{E}}_{\rm A'}.
    \label{eq:poyntingtheorem}
\end{align}
The first two terms on the right-hand side correspond to the Poynting flux emitted towards infinity,
$P_{\rm EM}$, and the visible electromagnetic field flux across the event
horizon of the BH, $\dot{\mathcal{E}}_{\rm BH}$, respectively. The third term describes the
energy loss of the visible fields to the pair plasma through resistive
processes, $L_{\rm diss}$. Lastly, the source of energy of the superradiant
system is the energy injection of the massive vector field
$\dot{\mathcal{E}}_{A'}$. Notice, we assumed that the energy of the visible electromagnetic
fields is much larger than the energy contained in the pair plasma,
i.e., $T_{\mu\nu}\gg T^{\rm plasma}_{\mu\nu}$. Hence, any finite mass loss due
to the accretion or emission of fermions is not contained in the above
analysis, which was shown to be a good approximation in \eqref{eq:mplasma}.
In the following, we consider each
component on the right hand side of Eq.~\eqref{eq:poyntingtheorem} and take the
flat spacetime limit to connect to familiar expressions. 

\paragraph{Poynting fluxes:} The electromagnetic luminosity---the Poynting flux---through a coordinate sphere at radius $\hat{\rho}$, is
\begin{align}
    P_{\rm EM}= -\oint_{S^2_{\hat{\rho}}}d\Omega_\mu T^{\mu}{}_\nu\xi^\nu \overset{\rm flat}{=} \oint_{S^2_{\hat{\rho}}}d\Omega \ \boldsymbol{\hat{\rho}}\cdot(\boldsymbol{E}\times\boldsymbol{B}).
    \label{eq:poyntingflux}
\end{align}
Here, $d\Omega_\mu$ is the oriented area element of $S^2_{\hat{\rho}}$ pointing outwards, $\boldsymbol{\hat{\rho}}$ the radial unit vector, and $d\Omega$ the solid angle. Hence, positive $P_{\rm EM}$ implies visible electromagnetic energy leaving the domain $D$. Since the massive linear combination of $A_\mu$ and $A'_\mu$ is gravitationally bound to the BH and decays as $\sim\exp(-r/r_c)$, at large distances, the visible Poynting flux, $P_{\rm EM}$, will receive a contribution only from the massless linear combination (corresponding to the SM photon) for sufficiently large $\hat{\rho}$. Analogously, the energy flux across the BH's event horizon is
\begin{align}
    \dot{\mathcal{E}}_{\rm BH}= -\oint_{S^2_{\rm BH}}d\Omega_\mu T^{\mu}{}_\nu\xi^\nu,
    \label{eq:eventhorizonflux}
\end{align}
where $S^2_{\rm BH}$ is the event horizon, and $d\Omega_\mu$ the oriented area
element pointing outwards. Hence, negative $\dot{\mathcal{E}}_{\rm BH}$ implies
visible electromagnetic energy accreting onto the BH. This, of course, vanishes
in the $\alpha\ll 1$ limit (i.e., in the flat spacetime limit).
At the saturation point, i.e., if
$\omega=\Omega_{\rm BH}$, the massive field has vanishing flux across the
horizon, such that $\dot{\mathcal{E}}_{\rm BH}$ contains only massless fluxes.
We can therefore interpret \eqref{eq:eventhorizonflux} as a measure of the
amount of accretion, or energy
extraction from the BH (e.g. the Blandford-Znajek process \cite{Blandford:1977ds}), triggered by the plasma and superradiance cloud.

\paragraph{Dissipative energy losses:} Besides these fluxes of energy across the boundary of the domain, the resistive pair plasma is able to dissipate energy in the bulk of $D$. This macroscopic dissipation is captured by the dissipative losses\footnote{Again, due to the kinetic mixing, both the massive and massless linear combinations of $A_\mu$ and $A'_\mu$ dissipate energy.}
\begin{align}
    L_{\rm diss}=- \int_Dd^3x\sqrt{-g}F^{\alpha\beta}\xi_\alpha I_\beta \overset{\rm flat}{=} \int_Dd^3x\boldsymbol{E}\cdot \boldsymbol{J}.
    \label{eq:dissipativelosses}
\end{align}
Here $g$ is the metric determinant of Kerr spacetime. In the flat spacetime limit, this expression reduces to the Joule heating within $D$. For later convenience, we define the local dissipation density
\begin{align}
    \rho_{\rm diss} = N F^{\alpha\beta}\xi_\alpha I_\beta,
    \label{eq:dissipationdensity}
\end{align}
where $N=\sqrt{-g/\gamma}$ is the lapse providing a macroscopic measure of the local rate at which energy is lost to heating, particle
acceleration, etc. 

\paragraph{Energy transfer from superradiance cloud:} 
The main energy source, driving the radiation and dissipation, is the superradiance cloud, which extracted a non-negligible amount of rotational energy from the BH. The rate of
replenishment of $\mathcal{E}$ from the superradiance cloud $A'_\mu$ is given
by the last term in \eqref{eq:poyntingtheorem}:
\begin{align}
    \dot{\mathcal{E}}_{\rm A'}= -\varepsilon\mu^2\int_Dd^3x\sqrt{-g}F^{\alpha\beta}\xi_\alpha A'_\beta \overset{\rm flat}{=} \varepsilon\mu^2\int_Dd^3x\boldsymbol{E}\cdot \boldsymbol{A}'.
\end{align}
This describes the energy transfer from superradiant to the visible fields. \\

\noindent Most important for our discussion in the following sections are the
Poynting flux $P_{\rm EM}$ and the dissipation rate $L_{\rm diss}$. These determine
the total electromagnetic power output of the system, and provide insights into
the characteristics of the emission, such as the primary emission mechanism,
the time-dependence, and the emission spectrum.

\subsection{Power output} \label{sec:emissionpower}

We find that the cloud-plasma system settles into a driven
turbulent state with a large electric dipole screening the superradiant
electric field and bulk magnetic field reconnection. We demonstrate below that
the dipole results in coherent electromagnetic Poynting flux,
while the magnetic reconnection is associated with strong energy dissipation from the visible
electromagnetic fields into the plasma. As we show in the following, due to the
turbulent nature of the dissipation in the bulk of the cloud, the dissipative
losses dominate over the Poynting flux from the system by orders of magnitude. 

In the vacuum limit, i.e., for $\sigma=0$, there is no dissipation and 
electromagnetic modes propagate freely. However, both the dark and visible
electromagnetic fields fall off exponentially at large distances
away from the BH, and there is no flux to infinity. At non-zero conductivities of the medium, on the other hand,
any Poynting flux is re-absorbed by the plasma on scales set by the
skin-depth of the effective fluid, which is a complex function of
propagating mode frequencies, conductivity, background electromagnetic field
strengths, and local Ohmic losses. As we show below, this leads
the Poynting flux to go to zero in the intermediate regime
$\sigma/\mu\sim\mathcal{O}(1)$. In this regime,
the energy dissipation is expected to be
largest, while any freely propagating electromagnetic modes are re-absorbed on
the skin depth length scales. 
In the limit where $\sigma/\mu\gg 1$,
the regime in which the cloud-plasma system is expected to reside, the Poynting
flux is expected to mostly decouple from the plasma, except in locations of
large dissipation into the plasma, and propagate freely.

\begin{figure*}[t]
    \centering
    \includegraphics[width=0.45\textwidth]{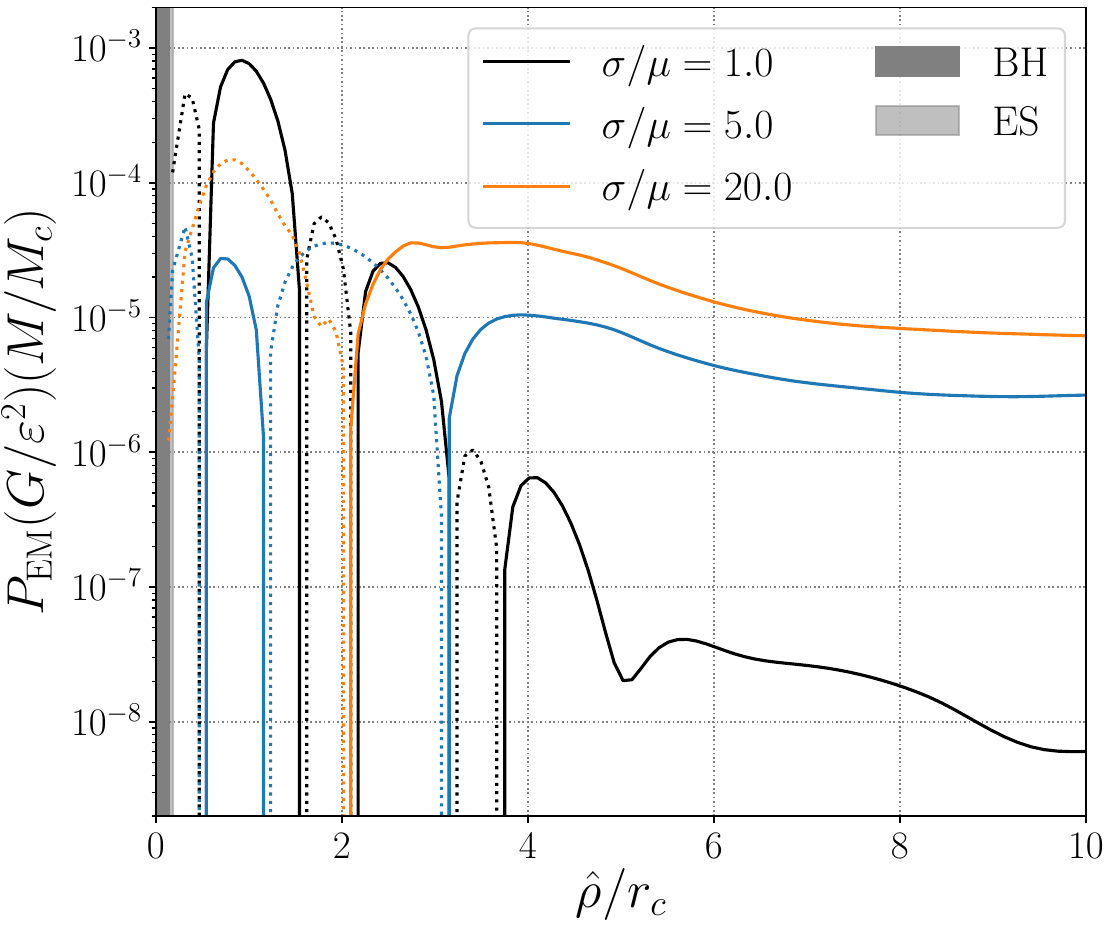}
    \hfill
    \includegraphics[width=0.45\textwidth]{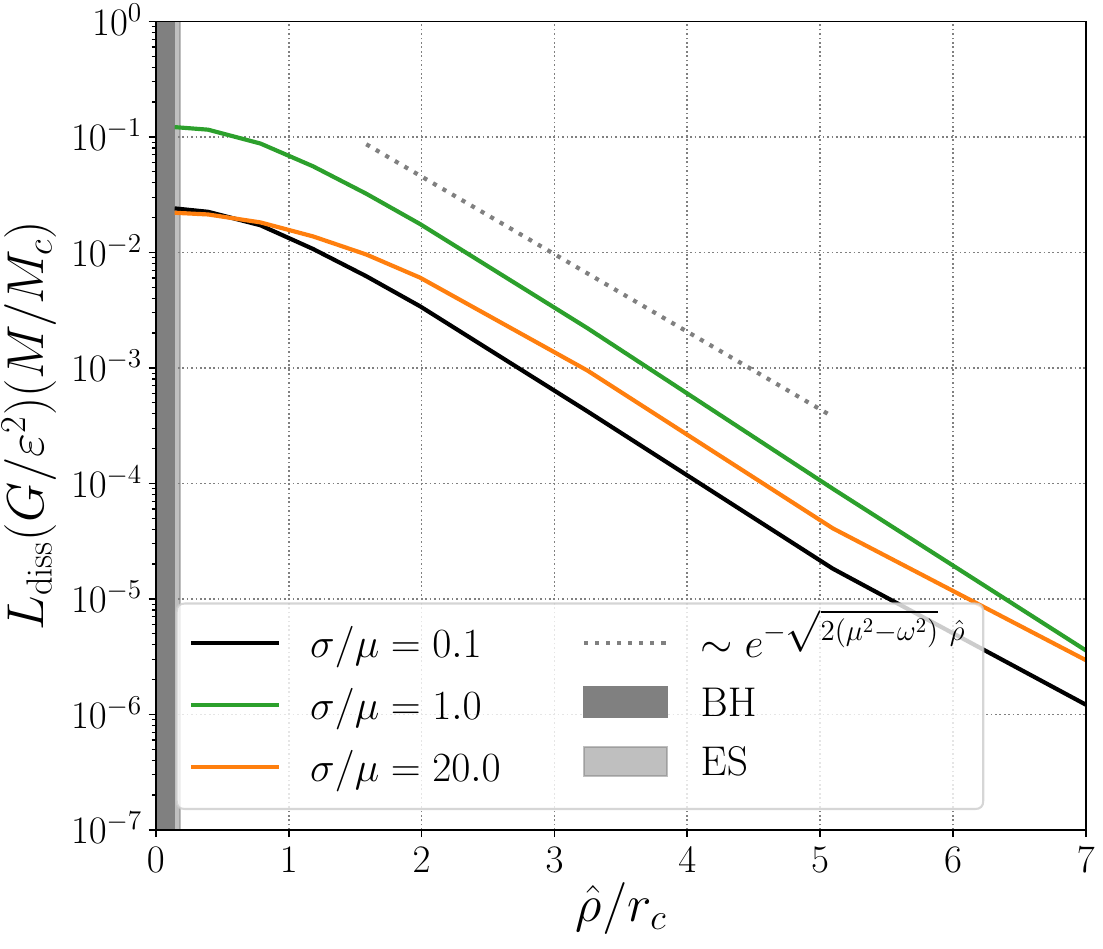}
    \caption{\textit{(left)} We plot the total visible, time-averaged, Poynting
    flux $P_{\rm EM}$, defined in \eqref{eq:poyntingflux}, through a coordinate
    sphere at radius $\hat{\rho}$ around the BH, for various conductivities
    $\sigma$. Solid lines are positive (locally outgoing) fluxes, whereas
    dotted lines are negative (locally ingoing) fluxes. The interior of the
    BH and the ergosphere (ES) in the equatorial plane are
    indicated by shaded regions; the smallest radius value indicates the flux through the event
    horizon $\dot{\mathcal{E}}_{\rm BH}$. 
    \textit{(right)} We plot the total
    energy dissipation rate due to Ohmic losses $L_{\rm diss}$, defined in
    \eqref{eq:dissipativelosses}, everywhere \textit{outside} a coordinate
    radius $\hat{\rho}$ for various conductivities. In both panels, we focus on
    an $\alpha=0.3$ cloud with a BH of spin $a_*=0.86$, and Bohr radius of the
    superradiance cloud of $r_c=1/(\mu\alpha)$; notice, $\sqrt{2(\mu^2-\omega^2)}\rightarrow\alpha\mu$ for $\alpha\ll 1$. Note, our simulations assume a
    conductivity constant everywhere in space. At intermediate conductivities,
    $\sigma\sim\mu$, the Poynting flux is efficiently absorbed by the effective
    plasma, while towards large conductivity, the electromagnetic modes
    propagate freely. The energy injection into the plasma $L_{\rm diss}$
    follows the profile of the cloud for all but the highest conductivities
    considered here.}
    \label{fig:FluxJouleradius}
\end{figure*}

To understand the high-conductivity regime, we compute the quantities $P_{\rm
EM}$, $L_{\rm diss}$, and  $\dot{\mathcal{E}}_{\rm BH}$ 
in our numerical simulations for
$\alpha=0.3$ and various different conductivities $\sigma/\mu\gtrsim 1$ (note,
in App.~\ref{app:lowconductivityregime}, we discuss the small-$\sigma$
regime for completeness). In \figurename{ \ref{fig:FluxJouleradius}}, we
present the visible Poynting flux and dissipative losses as functions of
conductivity and coordinate radius for the superradiant
cloud-plasma system. We postpone a discussion on time-dependence of the
electromagnetic emission to Sec.~\ref{sec:emissionperiodicity}, and focus here
on quantities time-averaged over one period of the superradiance cloud. As
anticipated, the energy dissipation into the plasma is largest for intermediate
conductivities, $\sigma\sim\mu$. As a result, the visible Poynting flux is
efficiently re-absorbed by the fluid and decays exponentially as it propagates
away from the BH. The sinusoidal features of $P_{\rm EM}$ for
$\sigma/\mu\lesssim 1$ are discussed in detail in App.~\ref{app:fluxdiscussion}. The local energy
dissipation follows the radial profile of the superradiant
cloud $\sim \exp(-\sqrt{2(\mu^2-\omega^2)}\hat{\rho})$, driving the energy
injection into the plasma, at large distances from the BH. Moving towards
larger conductivities, $\sigma/\mu> 1$, the weakening of the dissipative losses
roughly follows a $\sim 1/\sigma$ scaling, however, with important corrections
at $\sigma/\mu\gtrsim 5$ discussed below. 
As can seen in the left panel of \figurename{ \ref{fig:FluxJouleradius}},
energy flows into the BH for $\sigma/\mu\geq 1$ (though the horizon 
is actually a source of energy for lower conductivies, see App.~\ref{app:lowconductivityregime})
at rate comparable to the total Poynting flux to the wavezone.
The radial re-absorption length scale of the
electromagnetic flux increases significantly with conductivity for
$\sigma/\mu>1$, enabling efficient transfer of propagating modes from the
center of the superradiance cloud to the emission zone far away from the BH.
Focusing on the outgoing Poynting fluxes at large distances, $\hat{\rho}=10 r_c$,
a trend emerges, from small fluxes at intermediate conductivities to large
power at large conductivity, saturating at a conductivity-independent outgoing
electromagnetic emission.

\begin{figure*}[t]
    \centering
    \includegraphics[width=0.45\textwidth]{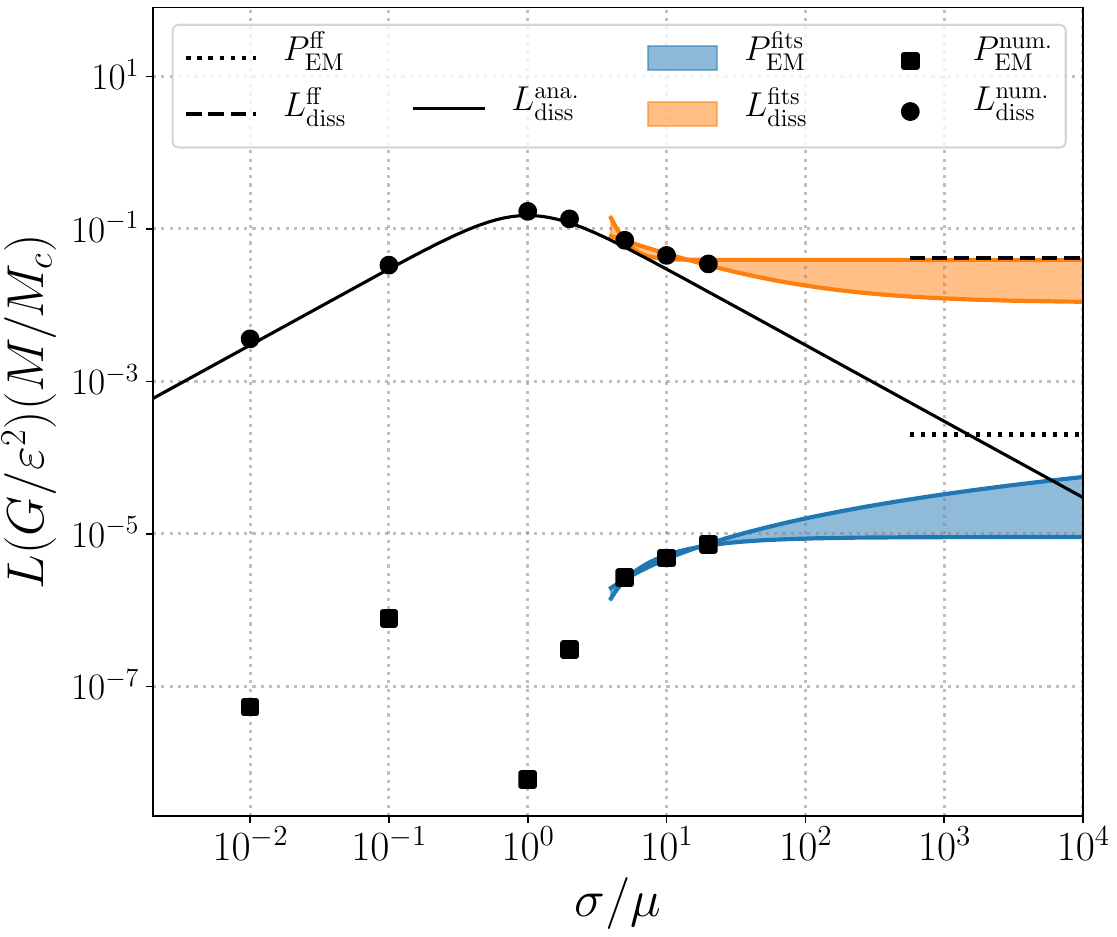}
    \hspace{1cm}
    \includegraphics[width=0.45\textwidth]{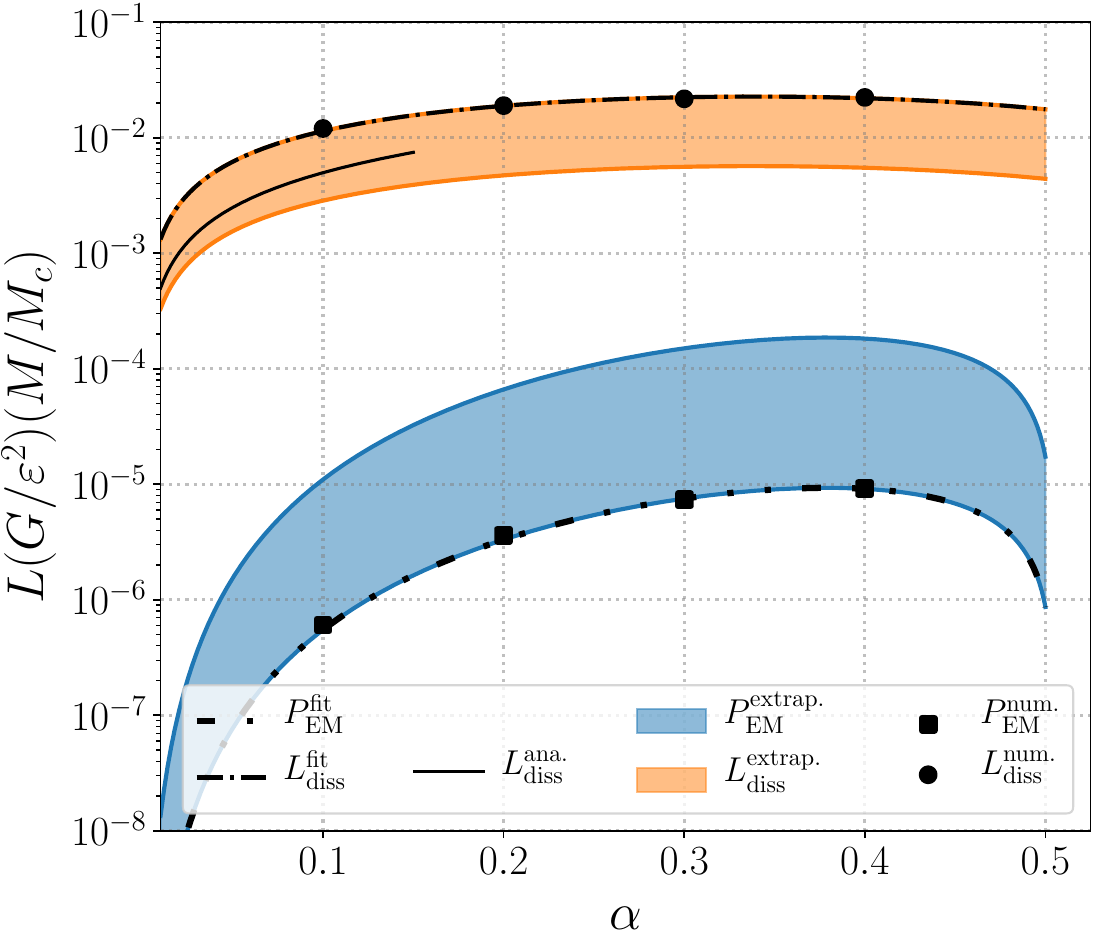}
    \caption{
    We show the energy dissipation rate, integrated over the entire cloud
    $L_{\rm diss}$, and the total (time-averaged) Poynting flux $P_{\rm EM}$ extracted at
    $\hat{\rho}=10 r_c=10/(\mu\alpha)$. 
    \textit{(left)} Focusing on
    $\alpha=0.3$ and $a_*=0.86$, we plot the quantities obtained from our
    resistive force-free simulations, black circles representing $L_{\rm diss}^{\rm num.}$, and black squares
    representing 
    $P_{\rm EM}^{\rm num.}$, as functions of conductivity. 
    The orange and blue
    bands, labelled $L_{\rm diss}^{\rm fits}$ and $P_{\rm EM}^{\rm fits}$, respectively, are a
    series of fits of the form $a_1+a_2(\mu/\sigma)^p$ to the simulation results
    with the three largest conductivities.
    The fits are motivated by the discussion in the
    main text that the energy dissipation remains finite at infinite
    conductivities. The bands are bounded by the most optimistic and
    pessimistic fits to the data. The black dotted and dashed lines show
    the force-free estimates for the emitted Poynting flux and total
    dissipation, labelled $P_{\rm EM}^{\rm ff}$ and $L_{\rm diss}^{\rm ff}$,
    valid formally at $\sigma\rightarrow\infty$ (how these are obtained is
    discussed in the main text). Lastly, we show the analytical
    approximations \eqref{eq:analyticdissipativelosses}, labelled as $L_{\rm
    diss}^{\rm ana.}$, for comparison. \textit{(right)} 
    We also show $L_{\rm diss}^{\rm num.}$ and $P_{\rm EM}^{\rm num.}$
    from the simulations, but now fixing
    $\sigma/\mu=20$ and varying $\alpha$, assuming $\omega=\Omega_{\rm BH}$. 
    The two fits in
    \eqref{eq:powerdissipationfits} to the numerical data $L_{\rm diss}^{\rm
    num.}$ (dash-dotted line) and $P_{\rm EM}^{\rm num.}$ (sparse-dashed line) are labelled
    as $P_{\rm EM}^{\rm fit}$ and $L_{\rm diss}^{\rm fit}$ (and use the ansatz
    $a_1\alpha^3+a_2\alpha^4$ and $a_1\alpha^1+a_2\alpha^2$, respectively). The
    orange and blue bands, labelled $L_{\rm diss}^{\rm extrap.}$ and $P_{\rm
    EM}^{\rm extrap.}$, are the $\sigma/\mu\rightarrow\infty$ extrapolations of
    the corresponding bands in the plot on the left (there for $\alpha=0.3$)
    applied to the two fits $P_{\rm EM}^{\rm fit}$ and $L_{\rm diss}^{\rm
    fit}$. Lastly, the analytic estimate \eqref{eq:analyticdissipativelosses}
    is indicated as $L_{\rm diss}^{\rm ana.}$. A discussion of both plots can
    be found in the main text. }
    \label{fig:FluxJouletotal}
\end{figure*}

The conductivity of the pair plasma within the superradiance cloud is expected
to be set by a micro-physical scale far smaller than any macroscopic scale of the
system, $\sigma\gg\mu$. Therefore, in the left panel of \figurename{
    \ref{fig:FluxJouletotal}}, we consider the trends of total dissipation
$L_{\rm diss}$ and Poynting flux at large radii $P_{\rm EM}$ towards the
large-$\sigma$ limit. As pointed out above, the coherent electromagnetic flux
emitted from the system increases rapidly from $\sigma\sim\mu$ towards a non-zero 
value for $\sigma\gg \mu$. The blue band in the left panel of \figurename{
    \ref{fig:FluxJouletotal}}, indicates possible fits with various
$\sigma$-scalings of the trend, extrapolating to physically relevant regimes,
$\sigma\gg\mu$. We discuss the behavior of $P_{\rm EM}$ for $\sigma/\mu <1$ in App.~\ref{app:fluxdiscussion}. Turning to the
dissipation of energy into the plasma, the behavior in the low to medium
conductivity regime, $\sigma/\mu\lesssim 1$, is as expected. The energy
injection at the macroscopic level increases as $\sim \sigma$ from the vacuum
limit towards intermediate resistivity levels. Beyond the peak dissipation
power around $\sigma\sim \mu$, a decay following $L_{\rm diss}\propto 1/\sigma$ 
is predicted by simple arguments of the bulk
dissipation inside the cloud (outlined below). However, instead of following
this behavior, the energy dissipation rate deviates 
from this trend. In order to extrapolate to large $\sigma$, 
we can therefore split the dissipation into two different components:
\begin{align}
    L_{\rm diss}=L_{\rm diss}^{\rm bulk}+L_{\rm diss}^{\rm turb},
\end{align}
heuristically representing the bulk and the turbulent dissipation,
respectively. Based on analytic estimates (discussed below) the bulk
dissipation $L_{\rm diss}^{\rm bulk}$ is expected to decrease as $\sim
1/\sigma$ towards small resistivity, while the turbulent dissipation
component $L_{\rm diss}^{\rm turb}$ will have a non-zero value in the infinite
conductivity limit. 
To capture this, in \figurename{ \ref{fig:FluxJouletotal}},
we fit the results from the simulations with $L_{\rm diss}\sim a_1+a_2(\mu/\sigma)^p$, considering $p \in [1,0.025]$, finding $L_{\rm diss}=L_{\rm diss}^{\rm turb}\approx
5\times 10^{-2}(\varepsilon^2/G)(M_c/M)$ for $\alpha=0.3$ and $\sigma\rightarrow \infty$.
The orange band in
\figurename{ \ref{fig:FluxJouletotal}} represents the range of values for the
conductivity dependence, and is bounded by the most optimistic and pessimistic
fits considered, to illustrate the uncertainty of this extrapolation. 

We compare these extrapolations with the results from force-free
simulations, valid formally at $\sigma\rightarrow\infty$, (see App.~\ref{app:resistivitytest} for
details on the numerical implementation and setup; in particular the current
is given by \eqref{eq:forcefreecurrent}). 
In the force-free simulations, the continually development of electrically dominated
regions (where the evolution equations breakdown) must be handled
in an ad hoc manner, by reducing the magnitude of the electric field by hand,
which gives rise to an artificial type of dissipation in regions where 
current sheets might develop in a more complete description of the plasma dynamics.
Nevertheless, we can determine the effective dissipation rate by assuming energy conservation
\eqref{eq:poyntingtheorem}, as is common in the literature (e.g.~\cite{Most:2020ami}),
and also compute the Poynting flux given by the force-free simulations at large
distances and across the BH horizon.
Encouragingly, these infinite conductivity results are in good agreement 
with the extrapolation of the resistive plasma simulations
towards large conductivity, shown as the orange bands in the left panel of 
\figurename{ \ref{fig:FluxJouletotal}}. As we discuss in below, this turbulent
dissipation component is associated with magnetic reconnection and other small scale features of the
solution, and hence, is expected to remain finite even at very large bulk
conductivities, $\sigma\rightarrow\infty$. The Poynting flux at large distances from the central BH in the force-free setting, shown in \figurename{ \ref{fig:FluxJouletotal}}, is consistent with the large-$\sigma$ extrapolations of $P_{\rm EM}$ (i.e., is within the blue band in the left panel of \figurename{ \ref{fig:FluxJouletotal}}).

The bulk dissipation is linked to the large-scale visible electric field induced by the superradiance cloud, while the turbulent dissipation emerges from higher-order magnetic field corrections. The former can be understood analytically in the non-relativistic limit, i.e., for $\alpha\ll 1$, by means of the solutions derived in \eqref{eq:analyticEfield}. Given this solution and neglecting magnetic field effects, we can determine the electromagnetic current density of the plasma to be
\begin{align}
\begin{aligned}
    \textbf{J} & \ =\sigma \textbf{E} \\
     & \ =-\frac{e^{-r/r_c}\sqrt{M_c\mu}\alpha^{3/2}\varepsilon \sigma\omega^2}{2\sqrt{\pi}(\sigma^2+\omega^2)}\begin{pmatrix} \sigma+i\omega \\ -\omega+i\sigma \\ 0\end{pmatrix} e^{-i\omega t}+c.c.
    \label{eq:nonrelcurrentdensity}
\end{aligned}
\end{align}
In the non-relativistic limit, the spatial extend $r_c$ of the superradiance cloud
is large compared with the oscillation frequency $1/\mu$, implying local charge
neutrality $\rho_q=0$. Hence, the spatial dependence of $J^i$ is a large scale
modulation of the locally neutral plasma for $\alpha\ll 1$. Using the flat
spacetime limit of \eqref{eq:dissipativelosses}, together with the above
current density and the visible electric field solution
\eqref{eq:analyticEfield}, the bulk dissipation rate of the cloud in
the $\alpha\ll 1$ limit is 
\begin{align}
    L_\text{diss}^{\rm bulk}=\frac{\sigma\alpha\varepsilon^2}{\mu(1+(\sigma/\mu)^2)}\frac{M_c}{G M} \ .
    \label{eq:analyticdissipativelosses}
\end{align}
As expected, this quantity goes to zero in both the insulating and highly
conducting limits, leading to free propagation of Poynting flux away from the
system. From \figurename{ \ref{fig:FluxJouletotal}}, we can see that this
expression provides a good approximation for the total dissipation for
$\sigma\lesssim 2\mu$ for $\alpha=0.3$, while for higher conductivities, the
turbulent contributions, i.e., the magnetic field driven component, to $J^i$
are more important. 

Most relevant for determining the potentially observable electromagnetic signatures of superradiant systems are the
$\sigma\rightarrow\infty$ estimates for $L_{\rm diss}$ and $P_{\rm EM}$ as
functions of $\alpha$. In the right panel of \figurename{
    \ref{fig:FluxJouletotal}}, we show our results for $L_{\rm diss}$ and
$P_{\rm EM}$ in simulations with $\sigma/\mu=20$ as functions of $\alpha$.
Focusing on the numerical results first, it is evident from the right panel
of \figurename{ \ref{fig:FluxJouletotal}} that the Poynting flux and the total
dissipation power have different scalings with $\alpha$. The analytic estimate
\eqref{eq:analyticdissipativelosses} for $L_{\rm diss}^{\rm bulk}\sim \alpha
M_c/M$ suggests a leading order $\alpha$-scaling of $L_{\rm diss}^{\rm
num.}\sim \alpha M_c/M$, which we find in \figurename{
    \ref{fig:FluxJouletotal}} to provide the best fit to the data. For the
Poynting flux, we find a leading-order scaling of $P_{\rm EM}\sim\alpha^3$ to fit
best. The two fits shown in the right panel of \figurename{
    \ref{fig:FluxJouletotal}}, are\footnote{Note, fits of the form $L^{\rm fit}_{\rm diss}\sim \alpha^p+\alpha^{2p}$ with $p\in(0.7,1)$ are plausible based on the numerical data and result in louder signals for $\alpha<0.1$. Fits with $p\in(1,1.2)$ are equally plausible, however, are entirely consistent with the large-$\sigma$ extrapolation uncertainty of \eqref{eq:lumdissipationfits} down to $\alpha\sim 10^{-4}$.}
\begin{align}
    L_{\rm diss}^{\rm fit} &=  \ \varepsilon^2 F(\alpha)\frac{M_c}{GM},
    \label{eq:lumdissipationfits}\\
    P_{\rm EM}^{\rm fit} &=  \ \varepsilon^2 G(\alpha)\frac{M_c}{GM},
    \label{eq:powerdissipationfits}
\end{align}
with
\begin{align}
\begin{aligned}
    F(\alpha)&= 1.31\times 10^{-1}\alpha -1.88\times 10^{-1}\alpha^2, \\
    G(\alpha) &= 6.86\times 10^{-4}\alpha^3 - 1.36\times 10^{-3}\alpha^4.
\end{aligned}
\end{align}
Determining whether these scalings are also valid in the $\sigma\rightarrow\infty$ limit
requires simulations with larger conductivities across a larger range of values for
$\alpha$. Hence, we estimate the theoretical uncertainties of the fits \eqref{eq:lumdissipationfits} and \eqref{eq:powerdissipationfits}, indicated as orange and blue bands in the
right panel of \figurename{ \ref{fig:FluxJouletotal}}, as follows. For
$\alpha=0.3$ (i.e., the left panel of \figurename{ \ref{fig:FluxJouletotal}}),
we obtain a series of different $\sigma\rightarrow\infty$ extrapolations for
both the total dissipation and the Poynting flux (blue and orange bands in the
left panel of \figurename{ \ref{fig:FluxJouletotal}}). The spread of these
$\sigma\rightarrow\infty$ extrapolating fits in the left panel of \figurename{
    \ref{fig:FluxJouletotal}}, corresponds to the width of the orange/blue
bands at $\alpha=0.3$ in the right panel of \figurename{
    \ref{fig:FluxJouletotal}}. We then use this relative uncertainty of the
large-$\sigma$ extrapolation at $\alpha=0.3$ and apply it to the fits for total
dissipation and Poynting flux, i.e., \eqref{eq:lumdissipationfits} and \eqref{eq:powerdissipationfits}, for all
$\alpha$, hence, obtaining the orange and blue bands in the right panel of
\figurename{ \ref{fig:FluxJouletotal}}. Therefore, the fit in
\eqref{eq:lumdissipationfits} for $L_{\rm diss}$ is likely an over-estimate
of the $\sigma\rightarrow\infty$ result (the lower bound of this extrapolation
uncertainty is given by $L_{\rm diss}^{\rm fit}/4$), while the fit in
\eqref{eq:powerdissipationfits} for $P_{\rm EM}$, is likely under-estimating
the true flux at $\sigma\rightarrow\infty$. It is clear from \figurename{
    \ref{fig:FluxJouletotal}}, that the turbulent dissipation power dominates
over the total Poynting flux across the
entire parameter space of the $m=1$ superradiant state. Furthermore, the
relatively flat $\alpha$-scaling of the total power output is in stark contrast
to the dependence of the total emitted gravitational wave energy flux $P_{\rm
GW} \propto \alpha^{10} {M_c^2}/{M^2}$ from the oscillating dark photon cloud
in the $\alpha\ll 1$ limit. Lastly, for comparison, the superradiant instability
growth rate of the $m=1$ state scales as $\Gamma\sim a_*\alpha^6\mu$. 

\subsection{Dissipation mechanism} \label{sec:dissipationmechanism}

\begin{figure*}[t]
    \centering
    \includegraphics[width=1\textwidth]{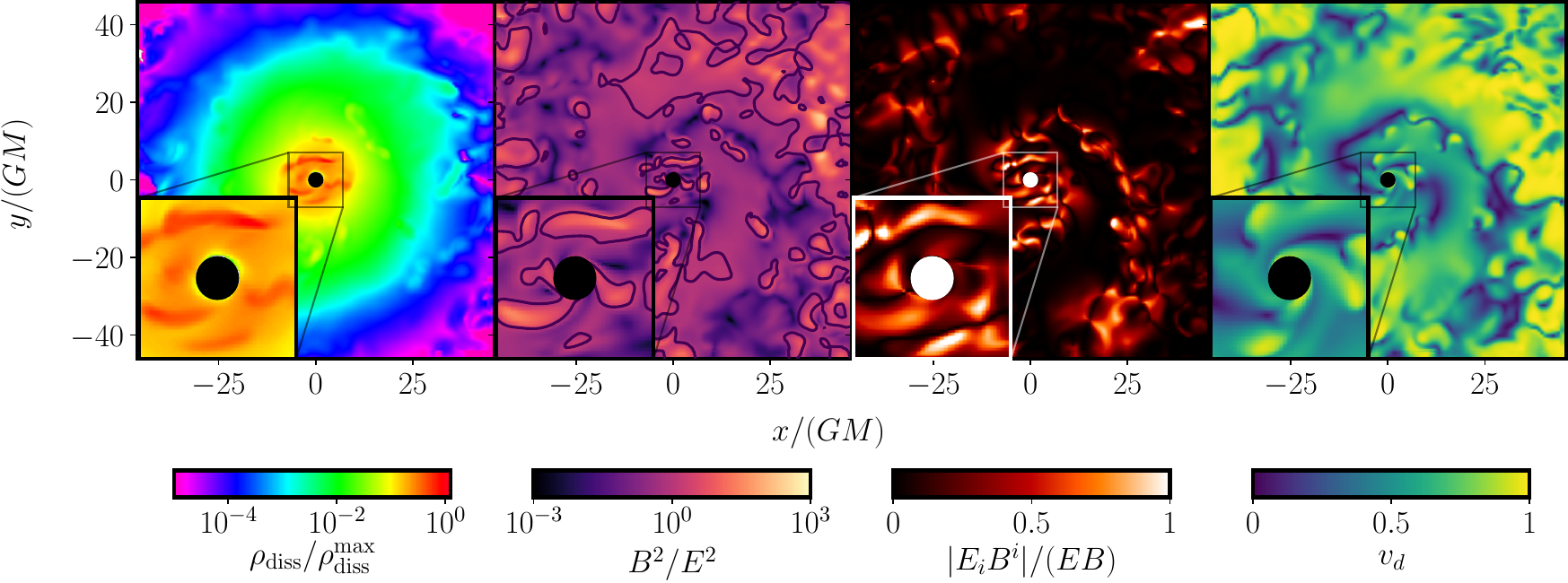}
    \caption{We plot the dissipation density $\rho_{\rm diss}$ (defined in \eqref{eq:dissipationdensity}) normalized by the maximal density $\rho_{\rm diss}^{\rm max}=\max\rho_{\rm diss}$, the ratio of visible electromagnetic fields $B^2/E^2$, the violations of the force-free condition $E_iB^i=0$ normalized by the magnitude of the visible electric and magnetic fields, and the magntiude of the plasma velocity $v_d=|\boldsymbol{v}_d|$ in the equatorial plane of the central BH. All panels correspond to the same coordinate time and a BH of spin $a_*=0.86$, cloud with $\alpha=0.3$, and plasma conductivity of $\sigma/\mu=20$. We indicate where $B^2/E^2=1$ by a contour line. Regions of small plasma velocities, i.e., large magnetic diffusion, are also sites of large $E_iB^i\neq 0$ and locally enhanced energy injection density $\rho_{\rm diss}$. This implies that magnetic reconnection sites are locations of enhanced energy injection into the plasma.}
    \label{fig:jdotedetail}
\end{figure*}

Given the importance of the turbulent dissipation power, even at large
conductivity, we next discuss the spatial dependence of the energy dissipation density $\rho_{\rm diss}$, defined in \eqref{eq:dissipationdensity},
and demonstrate that regions of high dissipation are associated with magnetic
reconnection. This density captures the
macroscopic energy injection of the electromagnetic fields into the plasma,
driving dissipative processes at the microscopic level. 
Recall, magnetic reconnection sites are regions, where the connectivity of the 
otherwise frozen-in magnetic field is changed, resulting in large spatial current 
along the visible electric fields dissipating energy.
In the following, we focus on the $\alpha=0.3$, $a_*=0.86$ and $\sigma/\mu=20$
case, while commenting on how these results extrapolate to the physically relevant 
limit of high conductivity.

In the left panel of \figurename{ \ref{fig:jdotedetail}}, we show the local
dissipation rate per volume $\rho_{\rm diss}$ [defined in
\eqref{eq:dissipationdensity}]. On large scales, this quantity follows the same
exponential fall-off in the radial direction as the superradiance cloud (at $\sigma/\mu=20$), while
on smaller scales, strong variation associated with turbulent features is
apparent\footnote{These two
spatial components are naturally associated with the bulk and turbulent
dissipation powers $L_{\rm diss}^{\rm bulk}$ and $L_{\rm diss}^{\rm turb}$,
driven by electric and magnetic fields, respectively (as discussed in the previous
section).}. We focus on the latter since, as argued in the previous subsection,
we expect these to persist (though develop smaller scales) in the
$\sigma\rightarrow\infty$ limit. From \figurename{ \ref{fig:jdotedetail}}, it
is clear that the regions of local enhancement in the dissipation are
associated with magnetic dominance (second panel) combined with $|E_i B^i|\sim E B$
(third panel), or with electric dominance. From
Eq.~\eqref{eq:resistivecurrent}, the plasma either allows for a significant
component of the current parallel to the electric field.  Focusing on the
insets showing the neighborhood of the BH, locations of the locally enhanced
dissipation density, in addition to having large $|E_iB^i|$ and being
magnetically dominated, are also bordered by zones of low drift velocity $v_d$ (or
equivalently, small magnetic Reynolds number $R_m$; fourth panel), which are associated
with efficient magnetic reconnection.  
Farther away from the BH, large regions (at $\sigma/\mu=20$)
are strongly electrically dominated, and associated with enhanced dissipation
and low drift velocity (in contrast to the magnetically dominated regions
which have $v_d\sim 1$), again indicating strong magnetic diffusion and
reconnection.

The magnetic field geometry within the plasma is set by the three different
spatial scales discussed in Sec.~\ref{sec:turbulentscalesummary}. 
The differential rotation on scales of the cloud induces a shearing of
the magnetic field lines on scales of the cloud $r_c$ around the spin-axis of the BH,
while the intermediate scale oscillations of the plasma (both in the equatorial plane 
and longitudinally along the BH spin axis) drive twisting of the field lines on
scales of $1/\mu$. This combined macroscopic build-up of magnetic field
twisting is released in magnetic diffusion regions associated with currents along 
the visible electric fields of thickness $1/\sigma$.  As discussed in more detail in
Sec.~\ref{sec:Highlyconductinglimit}, regions of small magnetic Reynolds number
$R_m=\sigma v_d/\mu$ are characterized by efficient magnetic diffusion. In
these diffusive regions, the connectivity of the magnetic field lines changes,
i.e. reconnection occurs, which drives enhanced dissipation and accounts 
for the dominant channel for the loss of
electromagnetic energy in the superradiance cloud-plasma system.

\begin{figure*}[t]
    \centering
    \includegraphics[width=0.9\textwidth]{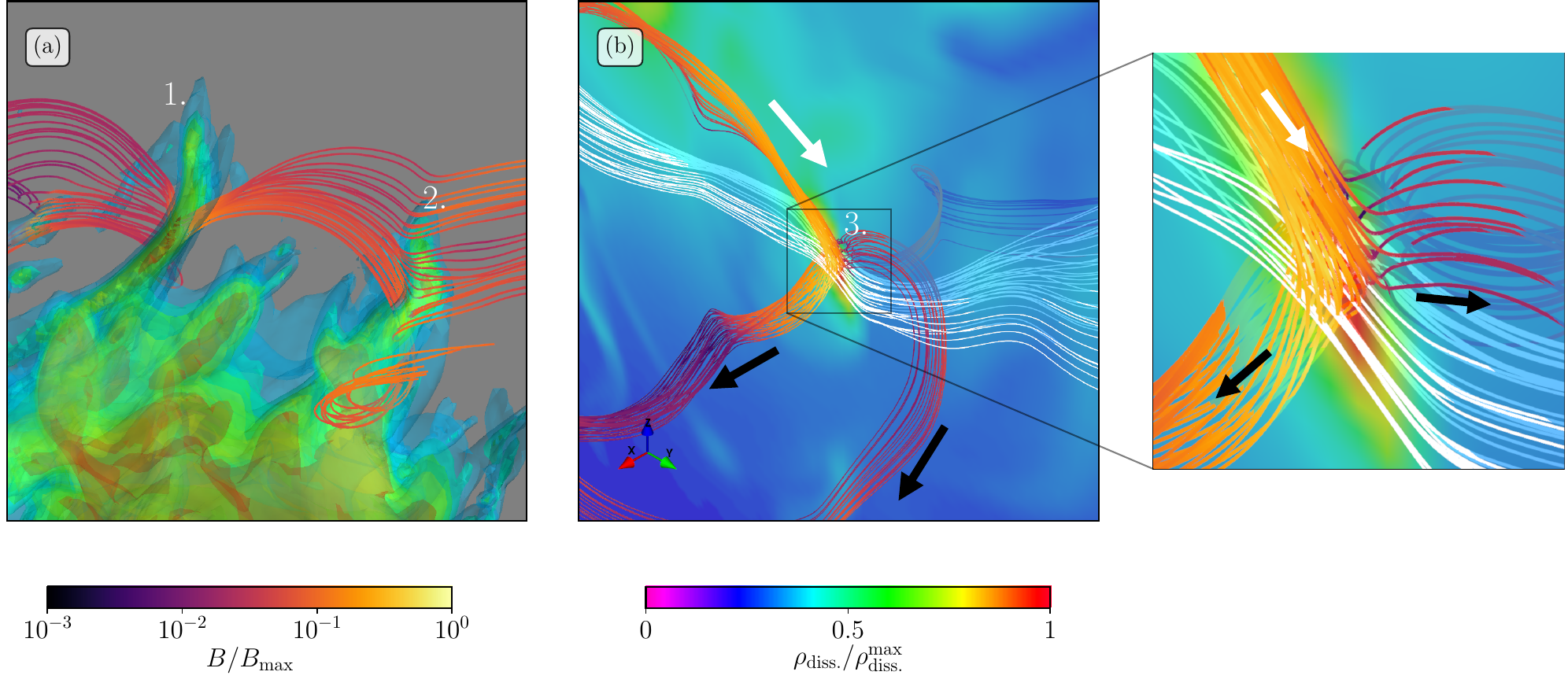}
    \caption{We show the visible magnetic field lines (in dark/yellow, color indicating the visible magnetic field strength normalized by the global maximum $B/B_{\rm max}$) and electric field lines (in white), as well as the local dissipation density $\rho_{\rm diss}$ (all colors) in two different contexts. Both panels show a close-up of the plasma roughly $15 GM$ away from the central BH of spin $a_*=0.86$, as well as $\alpha=0.3$ and $\sigma/\mu=20$. In \textit{(a)}, the dissipation density is shown as semi-transparent isosurfaces. The BH is located towards the bottom of the plot. In \textit{(b)}, the dissipation density is plotted on a semi-transparent plane spanned by the $z$ and $y$ directions. Here the BH is located towards the top left of the plot. The numbers in both panels indicate regions of large energy injection into the plasma. The arrows show the divergence of magnetic field lines away from the reconnection site. A detailed discussion can be found in the main text.}
    \label{fig:magneticreconnection}
\end{figure*}

To illustrate this connection explicitly, we show two example magnetic
reconnection events in \figurename{ \ref{fig:magneticreconnection}}.  In panel
(b) of \figurename{ \ref{fig:magneticreconnection}}, magnetic field lines enter
the reconnection region (labelled ``3.") from the top along the white arrow
(and from behind the semi-transparent plane on which $\rho_{\rm diss}$ is
plotted). These same field lines exit the region in \textit{two} directions
along the two black arrows (similarly for the lines entering from behind the
semi-transparent plane). The point where the field lines diverge is associated
with the magnetic field magnitude dropping to near zero (indicated by the color
of the magnetic field lines) and locally enhanced dissipation density $\rho_{\rm diss}$
(indicated by the color in the semi-transparent plane). This is characteristic
of discontinuous reconnection, where the magnetic field lines change
connectivity discontinuously along a line or plane where the magnetic field goes to 
zero. In two dimensions, X-point reconnection is the canoncial example
of discontinuous reconnection, and most prominent in current sheets of the pulsar 
magnetosphere powering the high-energy component of the electromagnetic emissions.
In particular, the field line geometry shown in panel (b) of 
\figurename{ \ref{fig:magneticreconnection}} resembles spine-fan type
magnetic reconnection~\cite{2021RSPSA.47700949L,2012RSPTA.370.3169P}.

We show a second example of reconnection in panel (a) of \figurename{
    \ref{fig:magneticreconnection}}, where we isolate two sites of large
magnetic field gradients. There, we show a set of visible magnetic field lines
that are strongly twisted as they
connect two regions of locally enhanced $\rho_{\rm diss}$ (labelled 
as ``1." and ``2."), separated by a distance of $\sim 1/\mu$.
Within each dissipation region, the field lines are curved on smaller scales
(plausibly set by $1/\sigma$). The strong field gradients in these regions,
as well as the fact that the magnetic field magnitude does not
go to zero, make this example more consistent with continuous reconnection,
where magnetic field lines pass through each other in a diffusion dominated 
region of small plasma velocity\footnote{Continuous type reconnection typically occurs at quasi-separatrix layers with large, but bounded, squashing degree \cite{2021RSPSA.47700949L,2012RSPTA.370.3169P,DEMOULIN20061269}. We do not attempt to identify quasi-separatrix layers, instead resort to identifying reconnection zones based on field diffusivity, magnetic field curvature and dissipation density.} (see also \figurename{ \ref{fig:jdotedetail}}). 

Therefore, we find that the visible magnetic field line connectivity changes
discontinuously (and we find evidence for continuous reconnection) at various
places in the bulk of the cloud. Both types are accompanied by enhanced energy
injection into the plasma.  The reconnection is fundamentally driven by the
fixed orbital frequency of the superradiant magnetic field, suggesting that, for
even larger conductivities, $\sigma/\mu\gg 20$, the qualitative picture is
unchanged. We expect that, in this limit, though the size of the diffusion regions may
decrease as $\sim 1/\sigma$ down to micro-physical scales, the rate of
energy dissipation, which is driven by reconnection, remains roughly
constant.  Identifying the changing connectivity of the magnetic field as the
driver of dissipation allows us to make a connection with existing kinetic
analyses to roughly estimate the particle and emission spectrum of the system,
which is the subject of Sec.~\ref{sec:emissionspectrum}.

\subsection{Periodicity of emission} \label{sec:emissionperiodicity}

\begin{figure*}
    \centering
    \includegraphics[width=0.33\textwidth]{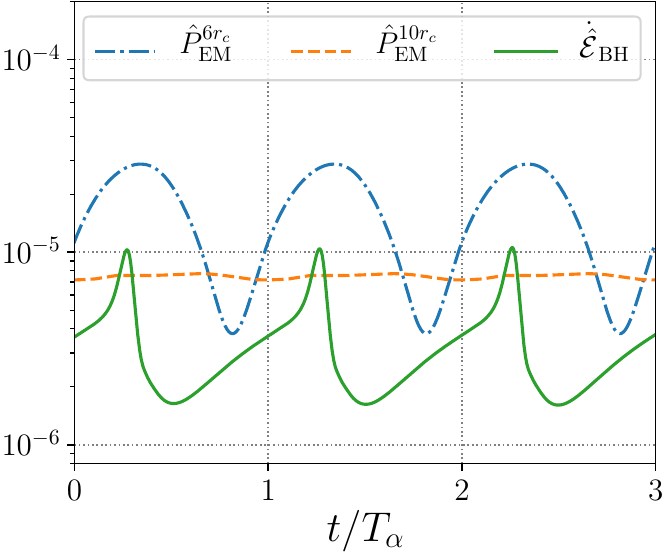}
    \hfill
    \includegraphics[width=0.64\textwidth]{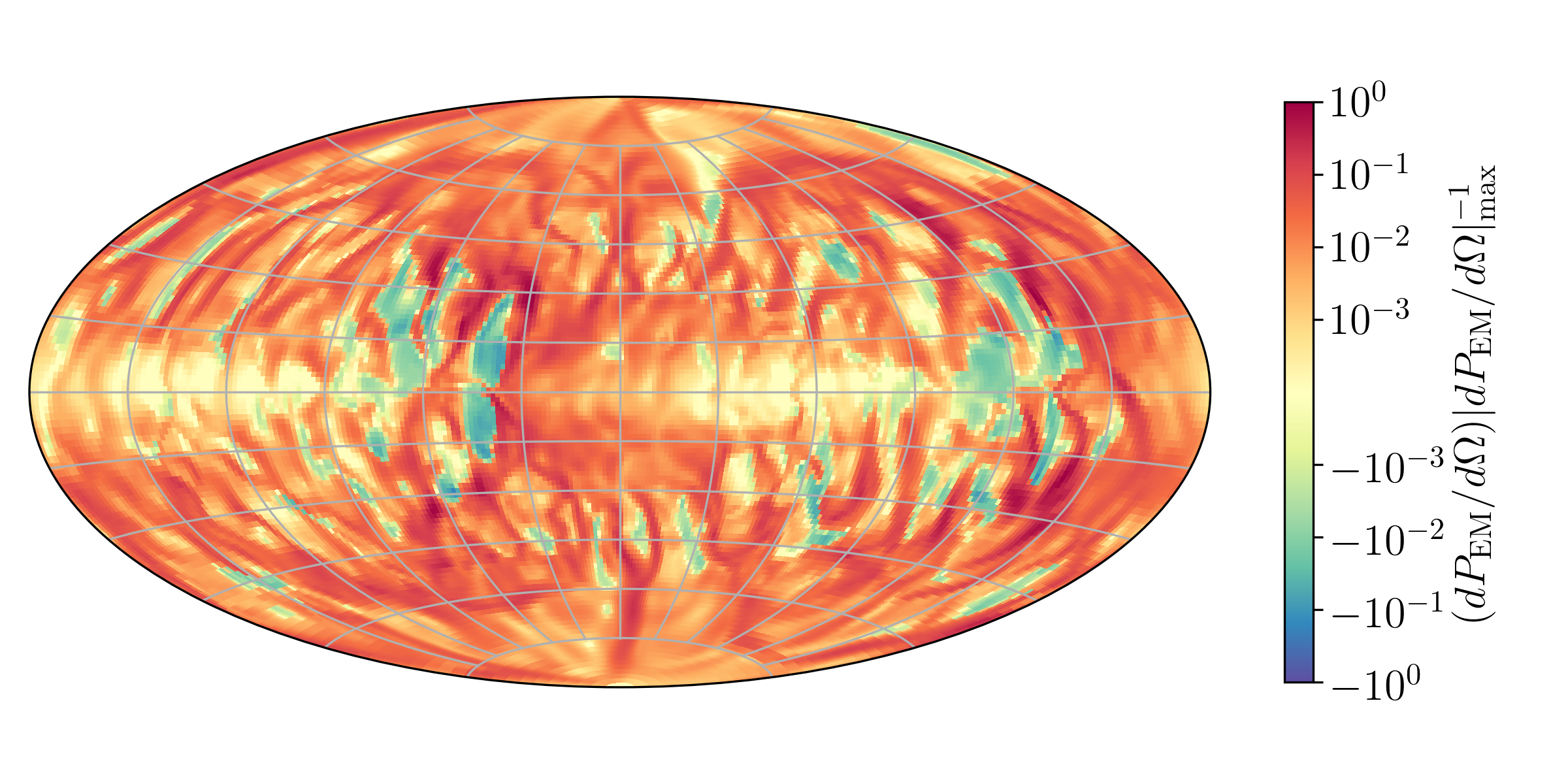}
    \caption{Here we consider a system with a BH of spin $a_*=0.86$, a superradiant
cloud with $\alpha=0.3$, and a plasma conductivity of $\sigma/\mu=20$.
    \textit{(left)} We show the time dependence of the total visible Poynting
    flux entering the BH $\dot{\hat{\mathcal{E}}}_{\rm EH}$ and the outward flux
    $\hat{P}_{\rm EM}$ through spheres of coordinate radii of $6r_c$ and $10r_c$. Hats indicates the rescaling $\hat{P}=P(G/\varepsilon^2)(M/M_c)$.
    Time is normalized by the period of the superradiance cloud
    $T_\alpha=2\pi/\omega$.  \textit{(right)} We show a snapshot of the visible
    Poynting flux per solid-angle (centered on the BH) through a coordinate
    sphere at $6r_c$, normalized by the maximum value.  Due to the differential
    rotation of the turbulent plasma, this pattern only rotates slowly along
    the azimuthal direction, i.e., with period $T\gg T_\alpha$.  At a
    coordinate radii of $10r_c$, the periodic modulation of the amplitude of
    the Poynting flux is mostly gone, indicating that the dissipation in the
    interior region is periodic.  The small-scale features in the angular
    distribution $dP_{\rm EM}/d\Omega$ is a result of the formation of current
    sheets and turbulence in the plasma.
    }
    \label{fig:poyntingperiodicity}
\end{figure*}

So far, we have discussed the total time-averaged Poynting flux and
turbulent energy injection into the plasma, ignoring the time-dependence
of the emission. In the case of a pulsar, the beamed radio emission, as well as
the pulsed high-energy component of the spectrum are characteristics that
emerge from the oscillation of the magnetic dipole field around the spin-axis
of the star. Since even in the well-studied pulsar case, the radio emission mechanism is a topic of debate, we
ignore it in the following, returning to a brief discussion of this
low-frequency component in Sec.~\ref{sec:emissionspectrum}. The pulsed X-ray
component of the pulsar spectrum requires a large-scale, coherent
magnetic field geometry deep inside the light cylinder, along which charges
are accelerated and radiate, or an oscillating current sheet outside the light
cylinder \cite{Bai1,Bai2,Cerutti:2015hvk,Philippov:2017ikm,Kalapotharakos:2017bpx}. In the context of the
superradiant system considered here, we do not find such coherent and
persistent field or current sheet structures, at least at the conductivities
we consider in this study. Instead, we find the plasma surrounding the BH to be in a turbulent
state without persistent large-scale magnetic or electric fields. However, 
since this is driven periodically by the superradiant fields at a frequency
$\omega$ one may na\"ively expect the electromagnetic emission to still be periodic as well.
Because modeling the light curve, as is done in the 
pulsar case (see e.g., Refs.~\cite{Bai2,Cerutti:2015hvk}), is challenging
for the superradiant system, we
consider the time-dependence of the macroscopic Poynting flux at large
distances, as well as the energy dissipation density throughout the bulk of the
cloud in order to understand the temporal evolution of the electromagnetic emission. We
find that the (sub-dominant) Poynting flux shows signs of periodicity, and the
dissipation density locally exhibits weak evidence of periodicity.  We close
this section by discussing techniques which could improve our understanding of
the temporal and viewing angle dependence of the electromagnetic emission.

We begin by discussing the time-dependence of the amplitude and angular
distribution of the total Poynting flux. In the left panel of \figurename{
    \ref{fig:poyntingperiodicity}}, we demonstrate that both the total Poynting
flux entering the BH, and the flux passing through a coordinate sphere at distance $6r_c$
from the central BH, vary periodically on timescales set by the cloud's
frequency\footnote{Notice, the \textit{visible} Poynting flux at finite
distances from the BH contains propagating massive dark photon states that are
bound to the BH. This component is exponentially suppressed at large distances.
Hence, we have made sure that the contribution to $P_{\rm EM}^{6r_c}$ in
\figurename{ \ref{fig:poyntingperiodicity}} from the massive states is
negligible.}. The periodic absorption of electromagnetic energy by the BH is
driven by the longitudinal periodic plasma motion in the vicinity of the
event horizon elaborated on below. The plasma density is expected to roughly
follow the profile of the superradiance cloud (compare
\eqref{eq:density}), and thus the conductivity should decrease exponentially away
from the central BH. This, paired with the decreasing contribution of the periodic component of $P_{\rm
EM}$ with increasing distance\footnote{Recall, the conductivity is spatially
constant in our simulations. At $\hat{\rho}=10r_c$, the conductivity is several
orders of magnitude smaller compared with \eqref{eq:density}; hence, the
physical relevance of the Poynting flux at those large distances should be
interpreted with caution.} (as shown in \figurename{
    \ref{fig:poyntingperiodicity}}), suggest that the Poynting flux periodically injects
energy into the plasma $\sim\mathcal{O}(r_c)$ away from the BH. 
Hence, the periodicity of the total emitted Poynting flux is
suggestive of periodic electromagnetic emission. 

\begin{figure*}
    \centering
    \includegraphics[width=1\textwidth]{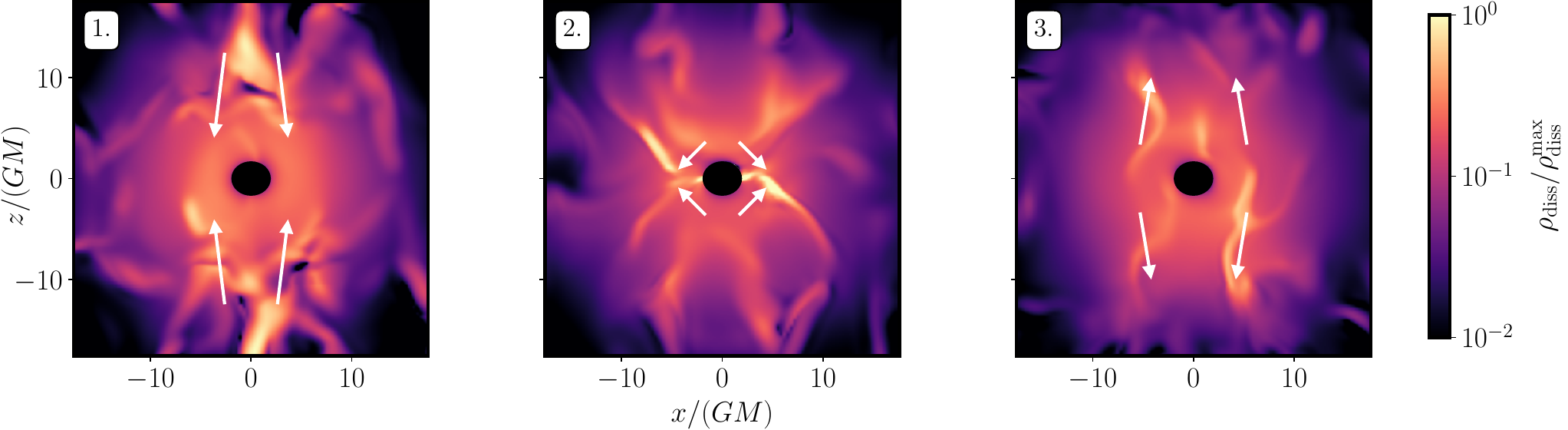}
    \includegraphics[width=0.7\textwidth]{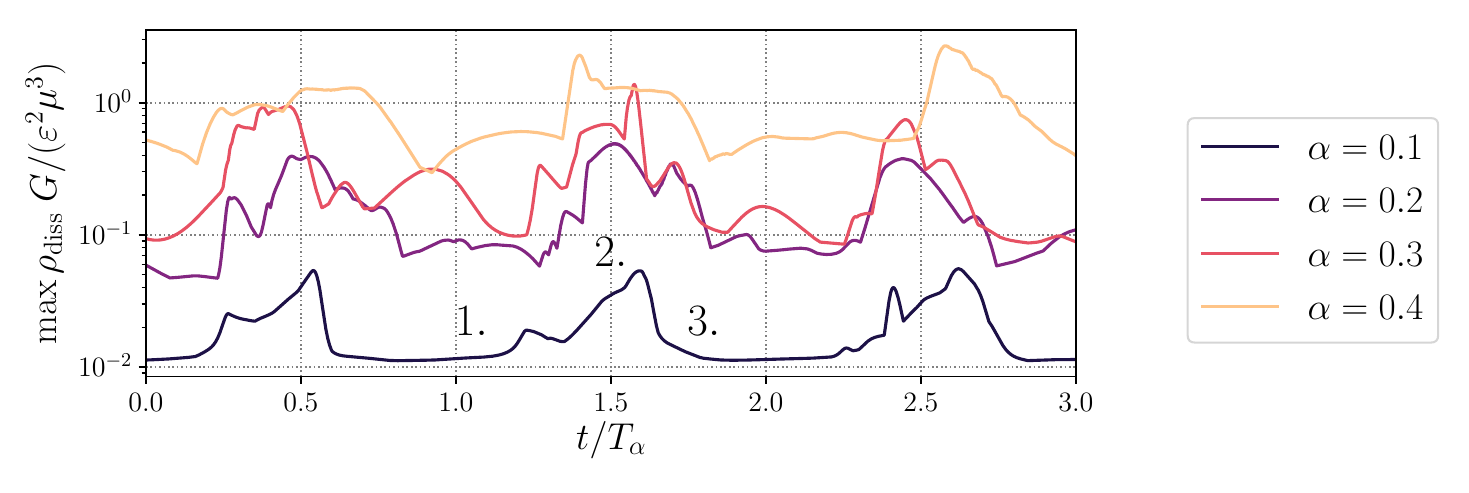}
    \caption{\textit{(top row)} We show the dissipation density $\rho_{\rm
diss}$ in the plane spanned by the BH spin and a fixed direction in the
equatorial plane at three instances during a single superradiance cloud period.
Here we focus on a BH with spin $a_*=0.86$, $\alpha=0.3$, and plasma
conductivity $\sigma/\mu=20$. The arrows indicate the direction of the motion
of the features, and are discussed in detail in the main text.
\textit{(bottom)} We plot the maximum of the dissipation
density $\max \rho_{\rm diss}$ as a function to time (normalized by the cloud's period
$T_\alpha=2\pi/\omega$) for different values of $\alpha\in\{0.1,0.2,0.3,0.4\}$
(and associated saturated BH spins satisfying $\omega=\Omega_{\rm BH}$, see
Tab.~\ref{tab:clouds}). 
In the bottom panel, we also indicate the times of the snapshots
in the top panels by their corresponding number labels.
Over the course of a single period of the superradiance cloud, the plasma
undergoes periodic motion along the BH's spin axis (as indicated by the arrows
in 1. and 3. in the top row), leading to peaks in the local dissipation
density, when the plasma from below and above the BH collide in the equatorial
plane (corresponding to snapshot 2.).}
    \label{fig:jdoteevo}
\end{figure*}

We now turn to the temporal variation of the energy injection density
$\rho_{\rm diss}$ into the plasma. Generally, the dissipation
density follows the motion of the plasma on scales of $1/\mu$ and $r_c$. Close
to the BH, the most relevant periodic motion of the plasma is the longitudinal
motion (discussed in Sec.~\ref{sec:Highlyconductinglimit}) along the spin-axis
of the BH. In the first panel of \figurename{ \ref{fig:jdoteevo}}, the
dissipation density is peaked in pockets above and below the BH, moving along
the spin-axis towards the equatorial plane, as indicated by the arrows.
This corresponds to the time when the overall maximum of the dissipation density is at its
lowest value per period, as shown in the bottom plot in \figurename{
    \ref{fig:jdoteevo}}. 
Subsequently, two regions of enhanced dissipation density (and the
associated plasma) collide within the equatorial plane, 
as show in the middle panel of the top row, leading to locally and
temporally large amplitudes of the energy injection rate $\rho_{\rm diss}$.
Finally, the regions of enhanced dissipation density begin to move away from
the equatorial plane along the spin-axis in the last panel in the top row of
\figurename{ \ref{fig:jdoteevo}}, and the associated maximum of the dissipation
density decreases. This process repeats on timescales of the cloud's period. It
is non-trivial to translate this behavior directly into observable variations
of the electromagnetic signature. We may speculate, however, that this periodic
enhancement of the dissipation density could lead to a periodic flaring of the
superradiance cloud (analogous to e.g.,~\cite{Most:2020ami,Nathanail:2020wap,Ripperda:2020bpz}). It should be noted though, that 
the total integrated turbulent energy dissipation does not show significant
temporal modulations.

In summary, both the Poynting flux and
the local dissipation rate exhibit weak evidence of periodicity on
timescales set by the dark photon mass $1/\mu$. Ultimately, our large-scale
macroscopic description of the system is insufficient to resolve and understand
the microscopic particle acceleration processes active in the turbulent plasma.
As well, with our current treatment, we are unable to determine how the local
dissipation rate translates into observing-angle-dependent radiation, though
na\"ively  one expects this to be strongly modulated by the oscillation of the
superradiance cloud.  This could be improved by consider a small domain near the
BH and using higher resolution to resolve higher values of conductivity than
considered in this study, e.g., using local resistive force-free techniques.
This could determine whether coherent magnetic and electric field geometries
remain inside the light cylinder, even at very large conductivities. These
sufficiently large scale field geometries may then be used to perform light
curve modeling paralleling the advances made in understanding high-energy
pulsar light curves.  Another avenue could be to utilize particle-in-cell (PIC)
simulations of the turbulent regions of the plasma. This would recover the
particle acceleration and non-thermal heating within the plasma, and could
therefore be utilized to understand the time- and angular-dependence of the
X-ray/gamma-ray sky map of the superradiant system.

\subsection{Emission spectra} \label{sec:emissionspectrum}

The plasma is characterized by differential rotation on scales of the entire
cloud, superradiant driving on scales of $1/\mu$, and turbulence down to
microscopic scales accounted for in our setup by the inverse conductivity
$1/\sigma$. Visible electromagnetic energy is dissipated into the pair plasma by
resistive processes. Likely this dissipation occurs primarily through particle acceleration, with the subsequent
synchrotron emission of highly boosted particles leading to high energy photons
that escape the system. Besides this non-thermal component at the high-energy end
of the emission spectrum, various coherent low-energy radio emission processes
may be active in regions of the superradiance cloud. In the following, we
briefly review existing kinetic theory results for the spectra of 
turbulent pair plasma and possible low-frequency radio emission mechanisms
that may be relevant to the emission spectrum for the system considered here.

In the pulsar magnetosphere, resistive processes occur mainly in
current sheets outside the light cylinder. There, magnetic dominance is
lost, and electromagnetic energy is efficiently dissipated by accelerating and heating the plasma. In order to gain
insight into the microphysical processes in these
accelerating regions, kinetic approaches based on numerical PIC
methods are typically utilized~\cite{Nishikawa:2020rwe}. Within this
framework, 
the distributions of charged particles are evolved in time 
according to the Lorentz force of the local electromagnetic field,
while back-reacting on the ambient fields through the charge and current they source.
Local simulations resolve microphysical scales such as the Lamour radius
$r_L=m_e\gamma/(e B)$ of an electron with mass $m_e$ and boost factor
$\gamma=E_e/m_e$ in an ambient magnetic field of strength $B$. Therefore, these
methods are powerful tools to determine the classical particle kinetic spectrum
self-consistently from first principles. On the other hand, radiative
corrections to the particle motion, pair production, are
neglected, or added in an ad-hoc fashion, and the expensive nature of these simulations
make it difficult to apply in a global setting, while still achieving sufficient resolution to accurately approximating the microphysics. Nonetheless, PIC methods have
successfully recovered the global pulsar magnetosphere, the expected non-thermal
particle spectrum, and have played a central role in studies of the
radio emission mechanism of pulsars
\cite{Philippov:2014mqa,Cerutti:2015hvk,Zhang:2020qgp}.

In the case of magnetic reconnection, PIC approaches have found that the local
electron acceleration results in a particle distribution $N_e(\gamma)$ with a
high-energy, power-law tail below a cutoff $\gamma_c$, of the form
$dN_e/d\gamma\propto \gamma^{-p}e^{-\gamma/\gamma_c}$
\cite{Werner:2014spa,Sironi:2014jfa,Guo:2014via}, for $\gamma\gtrsim 1$ (see
also
Refs.~\cite{Kagan:2014hea,Werner:2016fxe,Cerutti:2014ysa,Cerutti:2015hvk,Philippov:2017ikm}).
The size of the resistive region $\ell$ sets the high-energy cutoff $\gamma_c$
as the boost factor where $\ell$ equals the Lamour radius. 
Most applicable to the superradiant
system at hand are studies focusing on three-dimensional
turbulent pair plasmas \cite{Zhdankin:2016lta,Comisso:2018kuh},
determining the power-law to be roughly $p=2.8$, for large plasma magnetizations. 
We leave a
detailed investigation of the kinetic spectrum in the context of a kinetically
mixed superradiance cloud to future work, and
in the following make a crude estimate of the high-energy component of the emission associated with this electron kinetic spectrum
based on the characteristic length scales and field strengths.
The
high-energy cutoff $\gamma_c=e\ell \langle B^2\rangle^{1/2}/m_e$ is set by the
average ambient magnetic field strength $\langle B^2\rangle^{1/2}$, defined by 
$\langle B^2\rangle = 1/S_{2r_c} \int_{S_{2r_c}}dV B^2$, where $S_{2r_c}$ is a 
coordinate volume of a sphere of radius $2r_c$ centered on the BH. From the
resistive force-free simulations with $\sigma/\mu=20$, we extract this 
root-mean-square 
magnetic energy for each value of $\alpha$ we consider. Fitting
the $\alpha$-dependence by\footnote{Functions of the form $\sim \alpha^2$ or
$\sim\alpha^3$ provide worse fits, but are plausible given the numerical and
theoretical uncertainty.} $\sim \alpha^{5/2}$, the average magnetic field is
\begin{align}
    \langle B^2\rangle^{1/2}=2.5\times 10^{8} \ \text{Gauss}\left(\frac{\varepsilon}{10^{-7}}\right)\left(\frac{M_\odot}{M}\right)\left(\frac{\alpha}{0.1} \right)^{5/2}.
    \label{eq:ambientmagneticfield}
\end{align}
With this, the cutoff electron and positron boost factor, with $\ell=1/\mu$, is given by
\begin{align}
    \gamma_c\approx 2.2 \times 10^{7}\left(\frac{\varepsilon}{10^{-7}}\right)\left(\frac{\alpha}{0.1}\right)^{3/2}.
    \label{eq:highenergycutoff}
\end{align}
The size of the resistive region could be set by smaller length scales than the $1/\mu$
value used above. However, for most the parameter space of interest,
the effects of the radiation reaction will become important for much lower boost
factors than in
\eqref{eq:highenergycutoff}. Synchrotron backreaction becomes significant, when
the radiation reaction timescale $\tau_R=E_e/P^{\rm sync}_e$, where $P_e^{\rm
sync}$ is the total single electron synchrotron power, is comparable to, or
smaller than, the Lamour timescale $\tau_L=2\pi r_L$.
Hence, this radiation reaction becomes important for
$\gamma>\gamma_r=(3m_e^2/(e^3 \langle B^2\rangle^{1/2}))^{1/2}$ with 
\begin{align}
    \gamma_r=3\times 10^3\left(\frac{10^{-7}}{\varepsilon}\right)^{1/2}\left(\frac{M}{M_\odot}\right)^{1/2}\left(\frac{0.1}{\alpha}\right)^{5/4}.
    \label{eq:radreactioncutoff}
\end{align}
Therefore, in the regime $1\lesssim\gamma<\min(\gamma_r,\gamma_c)$, the
power-law electron and positron kinetic spectrum results in a synchrotron photon
spectral power-law $P(\nu)\propto \nu^{-s}$ with spectral index $s=0.9$
\cite{rybicki} (assuming $p=2.8$ \cite{Comisso:2018kuh}), 
making up the non-thermal tail of the
high-energy component of the emitted photon spectrum, while above this range 
the synchrotron spectral index is modified. The synchrotron spectrum from a
single electron or positron in this non-thermal distribution with $\gamma\leq
\gamma_r$ peaks at emission frequencies
\begin{align}
    \nu_{\rm peak}=12 \ \text{keV} \left(\frac{\gamma}{10^2}\right)^2\left(\frac{\varepsilon}{10^{-7}}\right)\left(\frac{M_\odot}{M}\right)\left(\frac{\alpha}{0.1} \right)^{5/2},
    \label{eq:peaksynchspectrum}
\end{align}
where the value of $\gamma$ is chosen inspired by simulations presented in~\cite{Cerutti:2015hvk} and a dedicated PIC simulation will be helpful to determine the exact spectrum, and an electron with kinetic energy of $m_e\gamma_r$ radiates mostly at $\nu_r=6.4 \ \text{MeV}$.

In summary, in the superradiance cloud, the electromagnetic fields lose energy 
predominantly through magnetic reconnection in
a strong ambient magnetic field with strength on the order of~\eqref{eq:ambientmagneticfield}. 
We can expect that this
efficiently accelerates electrons and positron to large boost factors,
$\gamma\sim \mathcal{O}(10^3)$, as given by the minimum of the values in \eqref{eq:highenergycutoff} and
\eqref{eq:radreactioncutoff}, and that these high-energy particles then radiate
synchrotron photons in the process, with spectrum ranging from
a few keV up to MeV [see
\eqref{eq:peaksynchspectrum}]. Therefore, it is likely that there will be strong non-thermal
high-energy component of the emission spectrum from the kinetically mixed
superradiance clouds.

We turning now to the low-frequency, i.e., radio, end of the 
spectrum, where the emission mechanisms are far less well-understood. 
Even in the well-studied pulsar case, this a topic of
debate. Low-frequency electromagnetic phenomena such as pulsar radio emissions
and fast-radio bursts are thought to be sourced through a shock induced
synchrotron maser emission mechanism in the pulsar wind, reconnection driven
radio emission, or near field processes \cite{Zhang:2020qgp}.
In, for instance,
Refs.~\cite{Mahlmann:2022nnz,Philippov:2019qud}, plasmoids forming from the 
discontinuous reconnection of the pulsar current sheet was demonstrated to
result in the emission of fast magnetosonic waves, plausibly escaping as radio
emission to infinity. Therefore, the efficient magnetic reconnection of the
superradiant plasma, some of which occurs through the discontinuous
reconnection channel, suggests that the cloud is a source of continuous radio
flux as part of the total power output\footnote{We note that, near the BH, 
the plasma frequency is on the order of 
a GHz, but it is expected to decrease exponentially away from the BH with
the superradiance cloud density (see Sec.~\ref{sec:quasi-steady}).}. 
Furthermore, as discussed above in
Sec.~\ref{sec:emissionperiodicity}, the plasma performs periodic longitudinal
motion along the BH spin axis with frequency $1/\mu$, resulting in collisions
of regions with enhanced dissipation density within the equatorial plane close
to the central BH. If these collisions, at the microphysical level, manifest as
colliding shock waves and trigger a synchrotron maser mechanism in the process
(see, e.g. Refs.~\cite{Metzger:2019una,Plotnikov:2019zuz}), then one would
expect periodically enhanced radio flux from these shocks. Hence, the na\"ive
expectation is that the plasma-filled superradiance cloud is a source of
continuous radio flux, together with periodic peaks in the radio power with
pulse period set by the dark photon mass $1/\mu$.

\section{Multimessenger Signals}
\label{sec:observation}

The system studied thus far motivates a novel target for multimessenger searches: a new bright,  possibly periodic, source with specific, unusual properties. In this section, we summarize the relevant dynamics and observational signatures of the kinetically-mixed dark photon superradiance cloud.
The numerical simulations performed in this work give us an estimate of the total electromagnetic power emitted, but do not directly provide the spectrum of the emitted radiation, nor conclusively establish its  periodicity. 

Nevertheless, the unique properties of the system and the analogy with the well-studied neutron star pulsars allow us to identify promising search strategies based on our system's combination of electromagnetic and GW emission. Given reasonable assumptions, outlined below, we expect current and planned telescopes and GW observatories to  reveal dark photons in the $10^{-14}-10^{-11}\ \rm{eV}$ mass range, with kinetic mixing below the current cosmological bound $\varepsilon \lesssim 3\times 10^{-7}$.

The evolution of our new pulsar begins with the birth of a new, rotating, BH. Around this BH, the superradiance instability populates a cloud of dark photons in $\mathcal{O}(100)$ superradiance times\footnote{
The results in this section are obtained using the gravitational waveform model \texttt{SuperRad} \cite{Siemonsen:2022yyf} for the superradiant vector cloud, with approximate expressions for the timescales, etc.~given to guide the reader.}, 
\begin{equation}
  t_{\rm growth} \sim  \ln(M_c/\mu) \tau_{\rm SR} \approx 10^4\, {\rm s} \left(\frac{M}{10\ M_{\odot}}\right)\left(\frac{0.7}{a_{*}}\right)\left(\frac{0.1}{\alpha}\right)^7. \label{eq:signalrise}
\end{equation}
The resulting large electromagnetic fields will, for large enough values of the kinetic mixing parameter, $\varepsilon \gtrsim 10^{-10}$,  build up a dense plasma of charged particles in the last few $e$-folds before saturation (see Sec.~\ref{sec:SFQED}). The rotation of the cloud and resulting turbulent electromagnetic processes in the plasma lead to a large flux of electromagnetic emission from the system, as described in Secs.~\ref{sec:fields} and~\ref{sec:emissionpower}. The energy output is dominated by the dissipative losses in the turbulent regions, which we infer from our numerical simulations (see Eq.~\eqref{eq:lumdissipationfits}) to be
\begin{align}
    L_{\rm{EM}} &= \varepsilon^2 F(\alpha) \frac{M_c}{G M} \simeq \varepsilon^2 \frac{\alpha^2 \Delta a_*}{G} \nonumber\\
    &\simeq 4\times 10^{41} {\rm{erg}/\rm{s}} \left(\frac{\varepsilon}{10^{-7}}\right)^2 \left(\frac{\alpha}{0.1}\right)^2 \left(\frac{\Delta a_*}{0.1}\right) , \label{eq:luminosity}
\end{align}
where $F(\alpha) = 0.13 \alpha - 0.19 \alpha^2$ is a polynomial fit to the simulations, and we used the mass of the superradiance cloud at its maximum and the small $\alpha$ limit. The luminosity can be up to five orders of magnitude brighter than the Crab pulsar's bolometric luminosity \cite{hester2008crab}, and up to ten orders of magnitude brighter than the solar luminosity. In the following, we will assume the luminosity to be given by Eq.~\eqref{eq:luminosity} also for $\alpha < 0.1$, below the smallest simulated value.

\begin{figure*}[!ht]
    \centering
    \includegraphics[width=\textwidth]{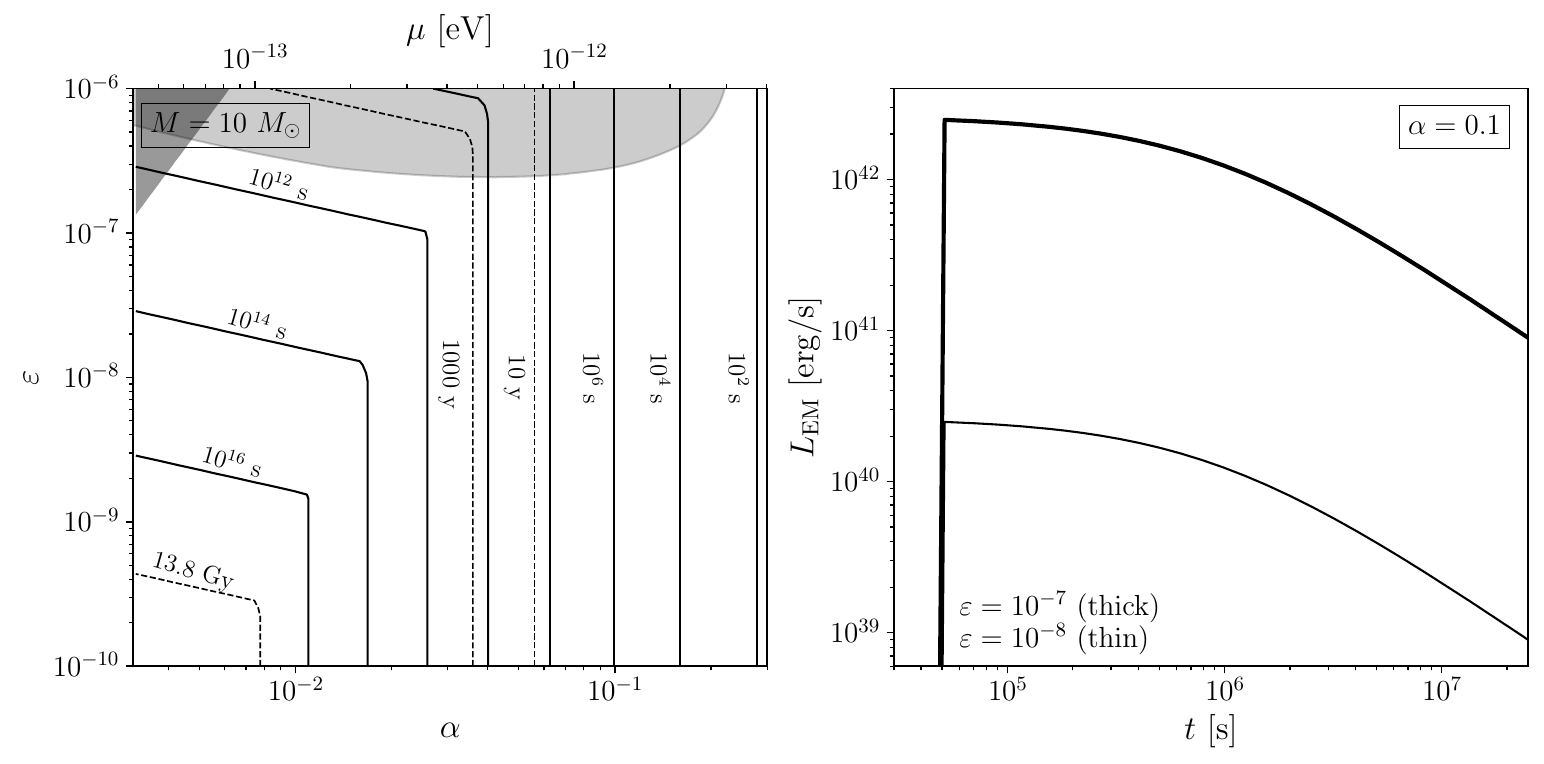}
    \caption{{\it (left)} Lifetime of the superradiance cloud as a function of the kinetic mixing parameter $\varepsilon$ and the gravitational coupling $\alpha$ for a BH with an initial mass of $10\ M_{\odot}$ and spin of $a_* = 0.9$. At large $\alpha$, the cloud decays through GW emission, and the lifetime is independent of $\varepsilon$. When $\alpha$ is too small, the power emitted in electromagnetic radiation overcomes the GW power and the cloud depletes faster for larger $\varepsilon$ (see Eq.~\ref{eq:signaldecay}). In both regimes, the lifetime is proportional to the BH mass, $\tau \propto M$, so the transition is independent of the value chosen. The initial BH spin determines the largest value of $\alpha$ that satisfies the superradiance condition, but otherwise has a mildly effect on the lifetime of the cloud. In the dark gray shaded region $\tau_{\rm{SR}} > \tau_{\rm{EM}}$, while the light shading corresponds to parameters excluded by measurements of the CMB spectrum by COBE/FIRAS \cite{Fixsen:1996nj,Caputo:2020bdy}. {\it (right)} Time evolution of the superradiance cloud's electromagnetic luminosity (see Eq.~\ref{eq:luminosity}) for two different values of $\varepsilon$, for $M = 10\ M_{\odot}$ and $a_* = 0.9$ (the luminosity is independent of $M$ and only mildly dependent on $a_*$, while the decay time will increase for heavier BHs). After the spinning BH is formed, the energy emitted in radiation quickly grows exponentially with the superradiance cloud, and later slowly decreases due to the cloud mass decay through GW emission.}
    \label{fig:lifetime}
\end{figure*}

The cloud slowly decays through emission of GWs and electromagnetic radiation on a timescale generally dominated by the gravitational dissipation of the cloud, except at small dark photon masses and large mixing, as shown in the left panel of Fig.~\ref{fig:lifetime}. The observable electromagnetic signal lasts  for $ \sim \rm{min}\left\lbrace \tau_{\rm GW}, \tau_{\rm EM}\right\rbrace$, 
\begin{eqnarray}
    \tau_{\rm GW} & \approx & \frac{G M}{17 \alpha^{11 }\Delta a_*} \approx 10^6 \, {\rm s} \left(\frac{M}{10 M_{\odot}}\right)\left(\frac{0.1}{\alpha}\right)^{11}\left(\frac{0.1}{\Delta a_*}\right), \nonumber \\
    \tau_{\rm EM} & \approx & \frac{G M \ln 2}{\varepsilon^2 F(\alpha)} \approx 10^{11}{\rm{s}} \left(\frac{M}{10 M_{\odot}}\right)\left(\frac{10^{-7}}{\varepsilon}\right)^{2}\left(\frac{10^{-2}}{{\it{F}}(0.1)}\right). \nonumber \\
    \label{eq:signaldecay}
\end{eqnarray}
The decay is power-law in time when more energy is released in GWs than electromagnetic radiation and exponential otherwise,
\begin{equation}
   \frac{ M_c(t)} {M_c(t_0)}=
    \begin{cases}
        [1+(t-t_0)/\tau_{\rm{GW}}]^{-1} & \ \  \tau_{\rm GW} \ll \tau_{\rm EM}\\
        e^{-(t-t_0)/\tau_{\rm EM} \ln 2} &  \ \ \tau_{\rm GW} \gg \tau_{\rm EM}.
    \end{cases}
\end{equation}
At small $\alpha$ and large enough $\varepsilon$, the superradiance growth time $\tau_{\rm SR}$ can become slower than $\tau_{\rm EM}$, possibly preventing the cloud from reaching its full size. A detailed study of the cloud saturation in this case is beyond the scope of this work and we always require $\tau_{\rm{SR}} < \tau_{\rm EM}$, which applies to most of the open parameter space (see Fig.~\ref{fig:lifetime}).

To summarize, the evolution of a kinetically mixed cloud is fully fixed by the  dark photon mass $\mu$, which sets the overall fundamental scale, and the dimensionless couplings $\varepsilon$ and $\alpha$.  Over the lifetime of the cloud, the electromagnetic and GW signals grow exponentially---at a rate fixed uniquely by the dark photon mass and $\alpha$---until the cloud reaches its maximum size, and then decrease on longer timescales as the cloud disappears.\footnote{The electromagnetic signal is only present for large enough cloud sizes $M_c$ and kinetic mixings $\varepsilon$ which are needed to generate the plasma, Sec.~\ref{sec:SFQED}.}  If the GW radiation dominates, there is a unique  relation between the growth and decay timescales, $\tau_{\mathrm{SR}}$ and $\tau_{\mathrm{GW}}$. For $\alpha\ll 1$ (for which the GW timescale is given by Eq. \eqref{eq:tgw}) it takes the approximate form 
\begin{equation}
\tau_{\mathrm{SR}}\simeq 0.98 \, \tau_{\mathrm{GW}}^{7/11} r_g^{4/11} \bigg[ \, a_* \bigg(\frac{\Delta a_*}{0.5}\bigg)^{7/11}\bigg] .
\label{eq:timescalerelation}
\end{equation} 
We show the characteristic electromagnetic luminosities and their time evolution due to GW emission in the right panel of Fig.~\ref{fig:lifetime}.

Clearly, the observational prospects of this system depend heavily on the spectral shape of the electromagnetic radiation; unfortunately, 
our simulations do not give us this information. However, given the similarities of our system to pulsars, which have been observed across electromagnetic bands in many systems, and have PIC simulations in agreement with aspects of the observations, we can make educated guesses as to the expected emission in different bands. 

In analogy with neutron star pulsars, one dominant emission mechanism could be synchrotron radiation. As discussed in  Sec.~\ref{sec:emissionspectrum}, the  boost factor of electrons and positrons in the cloud peaks in the range  $\gamma\sim\mathcal{O}(10-100)$, giving  typical electron radiation frequencies ranging from keV to a few MeV.
 Depending on the value of the kinetic mixing and BH mass, most of the spectrum would fall within the range of X-ray telescopes, such as Chandra \cite{chandraspec} and Swift \cite{Burrows:2005gfa} of $\mathcal{O}(0.1 - 10 \,{\rm keV})$ or Fermi-GBM ($8 \,{\rm keV}$ to $1\, {\rm MeV}$)~\cite{2009ApJ...697.1071A}.

GW emission from the system is monochromatic with frequency $f_{\rm{GW}} = \omega/\pi$, and the electromagnetic emission is also expected to have periodicity on timescales of $1/\omega$ (Sec~\ref{sec:emissionperiodicity}). Here, $\omega$ is the energy per dark photon, given by its rest mass with $\Delta \omega/\mu\approx-\alpha^2/2-\mathcal{O}(\alpha^4)$ corrections due to the gravitational potential energy of the BH~\cite{Baumann:2019eav} and $\Delta \omega_c/\mu \approx-(5/8)\alpha^2 M_c/M$ the gravitational self-energy of the cloud~\cite{Siemonsen:2022yyf}. The combination of the decrease of the BH mass as the cloud grows, and the decrease of the cloud mass as it decays, result in a monotonically increasing frequency correction.
In other words, the cloud period \textit{decreases} during the whole evolution of the cloud \cite{Siemonsen:2022yyf}, in stark contrast with conventional pulsars, for which the period increases in time.  Thus, rotating BHs can host an anomalously bright ``pulsar" which spins up over time \cite{Arvanitaki:2014wva}.

This new type of pulsar has several unique features and a peculiar evolution history. While there are a variety of signatures that can be looked for, here we highlight two distinct observational prospects: searching for the emergence of a bright electromagnetic source from a known rotating BH remnant (with or without a periodic component), or searching for continuous GWs emitted by anomalous pulsars. These two observational strategies are best-suited to regimes of large and small $\alpha$, respectively.

Given the signal uncertainties, we propose discovery oriented searches rather than exclusions. It is possible that the absence of a large number of ultraluminous X-ray sources  already places limits on dark photon parameter space;  however, given uncertainties in natal BH spin distributions, as well as in the emission spectrum of the dark photon cloud, such a constraint would not be robust. Similarly, we can speculate that such kinetically mixed superradiance clouds may account for some of the ultraluminous X-ray sources already observed~\cite{Swartz:2004xt,Pintore:2018eci}.

\subsection{Electromagnetic follow-ups of black hole mergers}\label{sec:EWfromMerger}

\begin{figure*}[!t]
    \centering
    \includegraphics[width=\textwidth]{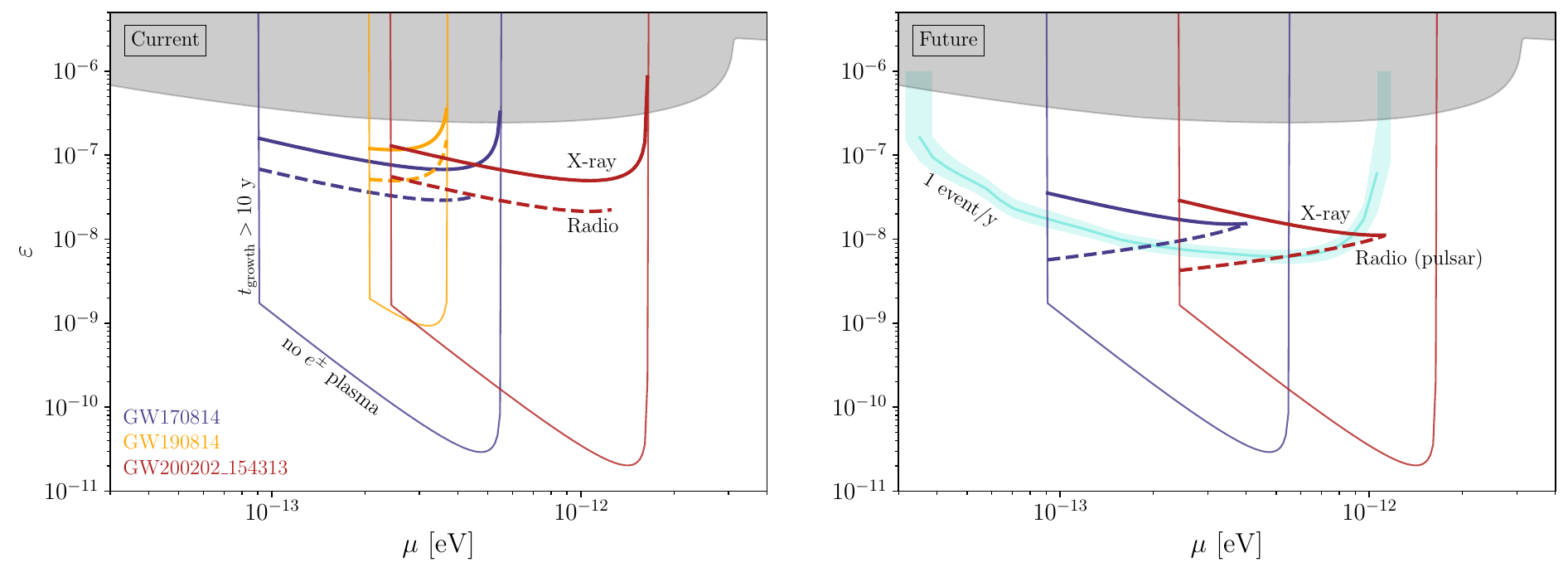}
    \caption{Range of dark photon kinetic mixing parameter $\varepsilon$ and mass $\mu$ producing a visible signal for electromagnetic follow-up observations of LVK compact binary merger events with a BH remnant. As an example, we choose the three best target events: GW170814 (blue), GW190814 (yellow), GW200202\_154313 (red) (measured parameters are given in Table \ref{tab:ligo_events}). The thin solid lines show the regions where a signal could exist, which are bounded from below by the requirement that the visible electric field is large enough to produce the plasma (see Sec.~\ref{sec:SFQED} and Eq.~\eqref{eq:eps_min_schw}), to the left by the requirement that the cloud grows within 10 years (see Eq.~\eqref{eq:signalrise}), and to the right by the superradiant condition for the fastest-growing bound state (see Sec.~\ref{sec:sr}). The reach is further limited on the right by the signal duration falling below a minimum observational time.  The gray shaded region is excluded by measurements of the CMB spectrum by COBE/FIRAS \cite{Fixsen:1996nj,Caputo:2020bdy}. {\it (left)} Current prospects for an X-ray \cite{Burrows:2005gfa} (solid) and radio transient \cite{Dobie:2021khu} (dashed) search. {\it (right)} Future prospects for an X-ray search (solid) and a radio search for a pulsating source (dashed). See the text for more details. The cyan contour corresponds to one merger event per year visible in the X-ray, with shaded band indicating the error due to the uncertainty in the BH merger rate.}
    \label{fig:ligomergers}
\end{figure*}

\begin{table*}[t]
    \centering
    \begin{tabular}{c|c|c|c|c}
    \hline \hline
        Name & Final BH mass $[M_\odot]$ & Final BH spin & Distance [Mpc] & $\Delta \Omega\ [\rm{deg}^2]$ \\
    \hline
        GW170814 \cite{LIGOScientific:2018mvr} & 53.2 & 0.72 & 600 & 87\\
    \hline
        GW190814 \cite{LIGOScientific:2020ibl} & 25.7 & 0.28 & 230 & 19\\
    \hline
        GW200202\_154313 \cite{LIGOScientific:2021djp} & 16.76 & 0.69 & 410 & 170\\
    \hline \hline
    \end{tabular}
    \caption{List of example compact binary merger events observed by LIGO-Virgo-KAGRA that are promising candidates for a dark photon superradiance search through electromagnetic follow up observations and corresponding central values of their parameters. See Fig.~\ref{fig:ligomergers} for the observational prospects. } 
    \label{tab:ligo_events}
\end{table*}

Most of the compact binary mergers detected by the LVK observatories result in a BH remnant with a mass between 10 to a 100 $M_\odot$ and high spin, due to the capture of a significant component of the binary's orbital angular momentum \cite{Buonanno:2007sv}. Electromagnetic follow-up observations of the mergers could reveal the existence of a dark photon superradiance cloud around the BH remnant. We select three illustrative events from the LSC O1-3 catalogs \cite{LIGOScientific:2018mvr, LIGOScientific:2020ibl, LIGOScientific:2021djp} and forecast the sensitivity of radio and X-ray searches in Fig.~\ref{fig:ligomergers}. The best targets and their inferred parameters are listed in Table~\ref{tab:ligo_events}; they correspond to the closest mergers with the best sky localization and one high mass and one low mass event, to cover the widest range of dark photon masses. We also include the possible neutron star–BH merger
GW190814 \cite{LIGOScientific:2020zkf}, for which an electromagnetic counterpart was not found in the radio \cite{Dobie:2019ctw, Dobie:2021khu, Gourdji:2021yrk}, optical, and near-infrared \cite{GravityCollective:2021kyg, deWet:2021qdx, Thakur:2020yvu, Ackley:2020qkz, Vieira:2020lze}, and X-ray \cite{Page:2020tnx}. These null results could exclude new dark photon parameter space, provided a more reliable prediction for the emission spectra from the superradiance system.

For our projections, we assume an $\mathcal{O}(1)$ of the superradiance cloud luminosity \eqref{eq:luminosity} is emitted into X-rays and a $\mathcal{O}(10^{-4})$ fraction into radio frequencies, in analogy with the spectrum of standard pulsars~\cite{Kaspi:2017fwg,Manchester:2004bp}. For the follow-up to detect the EM emission, the cloud needs to grow within a reasonable observational timescale; for concreteness we impose the requirement $t_{\mathrm{growth}}<10$~years (c.f. Eq. \eqref{eq:signalrise}), 
which translates into a large-$\alpha$ requirement given by $\alpha\gtrsim 0.036, 0.04$, and $0.03$ for our three selected candidates. The left panel of Fig.~\ref{fig:ligomergers} shows how current instruments could already measure a signal, taking a radio flux sensitivity of $40\ \mu\rm{Jy}$ at $944\ \rm{MHz}$ for an observation time of 10 h, achievable by ASKAP searching for transient events \cite{Dobie:2021khu}, and an X-ray detection sensitivity of $2 \times 10^{-14}\ \rm{erg}\ \rm{cm}^{-2}s^{-1}$ in $10^{4}\ \rm{s}$, attainable by Swift-XRT \cite{Burrows:2005gfa} as it scans through the sky (eROSITA has similar performance \cite{Predehl:2017}).
Better reach can be achieved with Chandra, which however requires prior angular localization.
For instance, for sources with angular localization from radio observations, Chandra could probe almost an order of magnitude smaller X-ray fluxes \cite{Jaodand:2019gcn}. 

The sky localization is expected to improve during O4 with the Advanced LIGO, Virgo, and KAGRA network, with about $10\%$ of the events localized within $5\ \rm{deg}^2$ \cite{KAGRA:2013rdx}, improving the prospects of detecting an electromagnetic counterpart. In the right panel of Fig.~\ref{fig:ligomergers}, we assume smaller positional errors that would allow a Chandra-like X-ray search sensitive to $10^{-15}\ \rm{erg}\ \rm{cm}^{-2}s^{-1}$ fluxes in $10^5\ \rm{s}$. In the same panel we also indicate with a cyan contour the region of parameter space above which more than one X-ray event per year from BH mergers could be observed  with the same X-ray sensitivity. To obtain this contour we made use of the BH merger rate as a function of primary BH mass measured by LVK \cite{LIGOScientific:2021psn}, assuming a final BH spin of 0.7, and a final mass equal to twice the primary mass. The shaded band around the cyan contour indicates the uncertainty in the merger rate.

The searches discussed so far in this subsection are aimed at a steady source that shines for as long as the required observational time, without assuming any periodicity.  If we further assume partial or total periodicity in the electromagnetic emission power, a search for a long lasting pulsating radio signal could be performed. We estimate that the prospect of such a search  with a sensitivity of $10\ \mu\rm{Jy}$ at $1\ \rm{GHz}$ with $500\ \rm{MHz}$ bandwidth for 15 minute observations daily over the lifetime of the cloud,\footnote{Private communications with Kendrick Smith.} which is comparable to the performances of FAST~\cite{Nan:2011um} and slightly better than CHIME~\cite{2021ApJS..255....5C}. 

The superradiance cloud could also have higher energy emission, up to $\gamma$-rays. Telescopes with nearly all sky coverage, such as {\it Fermi}-LAT~\cite{2009ApJ...697.1071A}, are well suited to perform follow-up observations of compact binary coalescences (which have been done and are planned during O4 \cite{Ren:2022gld}). Current flux sensitivities result in reach in kinetic mixing comparable to current constraints, but a signal could be detected in the event of an exceptionally close merger.

In the event of a positive detection of a new luminous source following a binary BH merger, there are non-trivial cross checks that can be used to confirm the superradiance origin of the signal.  First, by measuring the peak luminosity of the different sources, the luminosity's unique dependence on the parameter $\alpha$---Eq.~\eqref{eq:luminosity}---can be verified. For a dark photon with a given mass and kinetic mixing, the luminosity will only depend on the BH mass and spin, which are measured in the merger.
Secondly, the measured light curve should be consistent with an exponential growth as dictated by superradiance, and power-law decay, as expected from GW emission, with growth and decay timescales satisfying the non-trivial relation Eq.~\eqref{eq:timescalerelation}.
Thirdly, if a pulsating electromagnetic signal is observed, the measurement of the period, given by $2\pi/\mu$, can be used to extract the dark photon mass and verify the signal growth and decay times dependence on the parameter $\alpha$---see Eqs.~\eqref{eq:signalrise} and \eqref{eq:signaldecay}. Multiple sources will have the same period up to Doppler shifts of $\mathcal{O}(10^{-3})$, and gravitational potential corrections of up to a few percent. 
Finally, the superradiance system also emits continuous GWs with frequency $\omega/\pi$, which could be detected with a search targeted on the luminous source \cite{Arvanitaki:2016qwi,Isi:2018pzk,Chan:2022dkt,Siemonsen:2022yyf}. The GW has the same signal growth and decay times as the electromagnetic emission, which, together with the period coincidence, constitute unmistakable signatures of dark photon superradiance.

The discovery of such an ultraluminous source depends crucially on our ability to localize the newly formed BH. In particular, for sources with short duration (large $\alpha$), it is important that the source is located hours before the merger to
allow telescope observation coincident with the merger. Space-based, mid-band detectors can locate stellar mass BH merger events similar to GW150914 to an angular area of $\sim 0.01 \ \rm{deg}^2$ an hour before the merger~\cite{Graham:2017lmg,Seymour:2022teq}. Similar angular localization can be reached with the Laser Interferometer Space Antenna (LISA) and similar space missions for intermediate mass and supermassive BH mergers~\cite{Cutler:1997ta,Ruan:2019tje}. These missions could significantly improve the chances of finding the X-ray and radio signals from the dark photon superradiance cloud.

\subsection{Gravitational follow-ups of anomalous pulsars}\label{sec:GWfromPulsar}

If the superradiance electromagnetic emission is periodic, old galactic BHs dressed with a dark photon cloud could look like a neutron star pulsar, and be detected by ongoing surveys of pulsating sources. Such ``fake" pulsars appear to be rotating at a single frequency $f$ set by the dark photon mass, spin up over time, and emit continuous GWs with frequency $f_{\rm{GW}} = 2f$. Their GW strain is up to several orders of magnitude larger than the neutron star spin down limit and can be searched for with targeted continuous GW analyses. 

We select two types of potential candidates in the ATNF pulsar catalogue \cite{Manchester:2004bp,pulsarlist}. The first type is a set of frequency multiplets, {\it i.e.~}sources $i$ with at least one other pulsar $j$ that satisfies $|f_i -f_j|/f_i < 10^{-3}$, which takes into account the spread in frequency due to Doppler shifts of $\mathcal{O}(10^{-3})$. The second type is a source with $\dot{f}_{\rm{obs}} > 0$ that is not known to be in a binary system, to avoid spurious positive spin frequency derivatives due to the source's acceleration in a binary orbit.

Since all these pulsars are within our Galaxy, the BH formation event must have occurred long enough ago so that it was not observed. We therefore impose $\tau_{\rm{GW}} > 10^3\ \rm{yr}$, which translates into a maximum value for $\alpha\lesssim 0.03-0.05$ for the selected sources. For each system, the dark photon mass is fixed to be $\mu = 2\pi f$ (gravitational corrections to the dark photon energy are negligible in the small $\alpha$ limit). A lower bound on $\alpha$ is set by the smallest possible BH mass, which we take to be 3 $\rm{M}_{\odot}$, resulting in $\alpha \gtrsim 0.05-0.005$ across the range of frequencies considered. We take the initial BH spin to be 0.5. A better choice for the BH spin would be to sample it from the spin distribution of the Galactic BH population; however, this is not well known.

\begin{figure*}[!t]
    \centering
    \includegraphics[width=\textwidth]{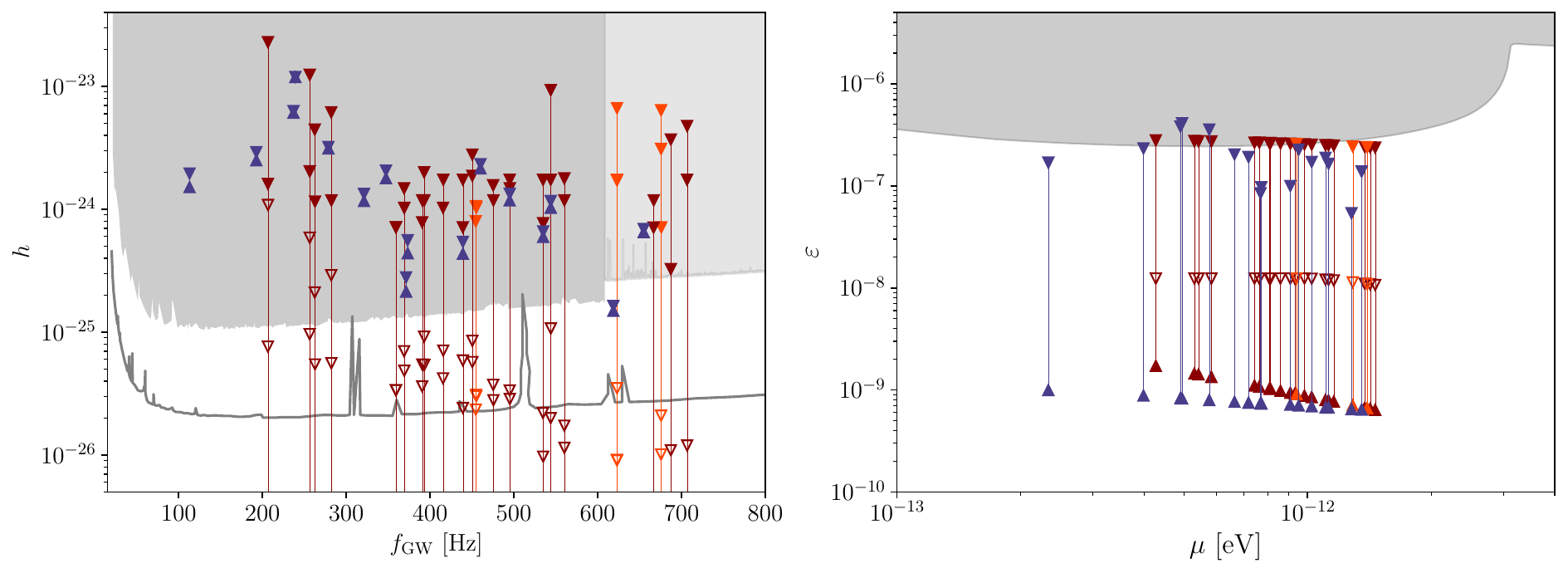}
    \caption{{\it (left)} We show the projected gravitational wave strain of observed pulsating sources whose luminosity could be powered by the kinetically mixed dark photon superradiance cloud. The potential candidates are selected from the ATNF pulsar catalogue \cite{Manchester:2004bp}, as described in Sec.~\ref{sec:GWfromPulsar}, and are frequency doublets (dark red), frequency triplets (light red), and pulsars with positive measured frequency derivative (blue). The filled (empty) triangles correspond to the largest possible strain that a source in a frequency multiplet could produce if the cloud were created $10^3$ ($10^6$) yr ago. The strain could take any value below that upper bound (thin solid lines) down to a minimum strain outside of the range shown here. The blue down-pointing and up-pointing triangles denote the range of strains allowed assuming that the spin-up is due to gravitational wave emissions from the cloud. The $95\%$ C.L. upper limits on the signal strain amplitude from Refs.~\cite{KAGRA:2021tse} and \cite{KAGRA:2021una} are shown in dark and light gray shading, respectively. The solid gray line corresponds to the expected sensitivity of a targeted search with LIGO Livingston derived in \cite{LIGOScientific:2021quq}. {\it (right)} Range of kinetic mixing parameters allowed for each pulsar, with frequency doublets (dark red), frequency triplets (light red), and pulsars with positive measured frequency derivative (blue). For the frequency multiplets, filled (empty) down-pointing triangles correspond to the largest $\varepsilon$ that allows the cloud to decay through GW emission for at least $10^3$ ($10^6$) yr. For the sources that are spinning up, down-pointing triangles give the largest $\varepsilon$ that allows the cloud to decay through GW emission up to their current age (which is fixed by $\dot{f}_{\rm{obs}}$). For all the sources, up-pointing triangles denote the smallest mixing parameter that allows for plasma pair production in the cloud (see Sec.~\ref{sec:SFQED} and Eq.~\eqref{eq:eps_min_schw}).}
    \label{fig:anomalous_pulsars}
\end{figure*}

We only consider sources with $f > 50\ \rm{Hz}$, since for a stellar mass BH, smaller frequencies would correspond to a very small $\alpha$, and correspondingly small signal. Above this threshold, the ATNF catalogue contains 229 sources with measured $f$, distance, and with positive or unknown $\dot{f}_{\rm{obs}}$. Among these, there are 3 frequency triplets and 26 frequency doublets (19  of the  doublets have a non-zero allowed range of $\alpha$ between the upper and lower bound), statistically  compatible with the number of accidental multiplets expected for a uniform frequency distribution between $50$ and $500$ Hz. For each fake pulsar in a multiplet, the emitted GW strain cannot be uniquely predicted, since the BH mass and age (that sets the remaining superradiance cloud mass, and thus affects the GW emission power) are unknown. In the left panel of Fig.~\ref{fig:anomalous_pulsars}, we show the maximum possible strain for the allowed range of $\alpha$ values, for two example BH ages of $10^3$ and $10^6$ years. As pointed out previously, we do not consider younger systems. These young systems are excluded by all-sky searches for continuous GWs~\cite{KAGRA:2021tse, KAGRA:2021una}, but strains below the current bounds are possible for older systems, and within the reach of a targeted search \cite{LIGOScientific:2021quq}. 

With the same selection criteria as above, but without requiring frequency multiplets, we find 20 sources with $\dot{f}_{\rm{obs}} > 0$, with values between $4\times 10^{-17} $ Hz/s and $5\times 10^{-14}$ Hz/s. Only 16 of these sources have a non-zero range of $\alpha$ values  between the upper and lower bound. Interestingly, four of these sources (J0024-7204Z \cite{Freire:2017mgu}, B0021-72G \cite{1995MNRAS.274..547R, Freire:2017mgu}, J1801-0857C \cite{2011ApJ...734...89L, Pan:2021zge}, and B0021-72M \cite{1995MNRAS.274..547R, Freire:2017mgu}) \textit{also} belong to a frequency doublet. The frequencies of the four doublets are approximately 219.6, 247.5, 267.4, and 272.0 Hz. These candidates could be further strengthened (or disfavored)  by performing spin derivative measurements of the other component of each of the doublets, which at present do not have $\dot{f}$ measurements,  and that are J0514-4002D \cite{Ridolfi:2022gmc}, J1824-2452J \cite{Freire:2008}, J0024-7204ad \cite{Ridolfi:2021idl}, and J0125-2327 \cite{Morello:2018goa}, respectively.

If the spin-up is due to the superradiance cloud decay through GW emission, we expect an intrinsic spin frequency derivative \cite{Siemonsen:2022yyf}
\begin{equation}
    \dot{f}_{\rm{int}} \simeq \frac{5}{8\pi} \alpha \mu^2 G P_{\rm{GW}}
    .
    \label{eq:fdot}
\end{equation}
The observed frequency derivative could differ from the intrinsic one due to additional positive contributions from acceleration along the line of sight. Thus, one should interpret the measured spin derivative $\dot{f}_{\rm{obs}}$ only as an \textit{upper limit} on the intrinsic spin derivative.  Here, however, our intention is to provide a first example of how spin derivative measurements could be used to discover anomalous pulsars, so in what follows we simply neglect accelerations along the line of sight and assume $\dot{f}_{\rm{obs}} \simeq \dot{f}_{\rm{int}}$. With this assumption, the measured value of  $\dot{f}_{\rm{obs}}$ fixes the power emitted in GWs $P_{\rm{GW}}$ for any given $\alpha$. Since the power emitted decreases as the cloud's mass decays, there is a minimum $\alpha$ that allows for a spin up rate as large as the one observed when the cloud is as young as possible, i.e., at least 1000 yr. This lower bound on $\alpha$ is stronger than the one described above from the minimum BH mass for $f \lesssim 270\ \rm{Hz}$ . Therefore, the observable strain is \textit{predicted} to be within a small range given by the small spread of $\alpha$ values allowed, as shown on the left of Fig.~\ref{fig:anomalous_pulsars} with the short segments bounded by the downward and upward pointing blue triangles.

We find that most sources should have already been seen by all-sky searches for continuous waves, but one candidate pulsar with a frequency around 600 Hz remains unconstrained,
which could be an interesting candidate for a targeted search. We note that if the assumption $\dot{f}_{\rm{obs}} \simeq \dot{f}_{\rm{int}}$ is correct, the four previously mentioned events that are frequency multiplets and have $\dot{f}_{\mathrm{obs}}>0$ would be excluded, since their $\dot{f}_{\mathrm{obs}}>0$ measurement leads to a strain prediction that is already ruled out by existing GW searches. These sources, however, could certainly be compatible with data if $\dot{f}_{\rm{int}} < \dot{f}_{\rm{obs}}$, due to the source accelerations along the line of sight that we have neglected.

For each anomalous pulsar, the kinetic mixing parameter could lie within a range of allowed values, which are shown in the right panel of Fig.~\ref{fig:anomalous_pulsars}. In the prediction of the strain described above, we assumed that the cloud decays through GW emission, and that the power emitted in electromagnetic radiation is subdominant at all times, giving an upper bound on $\varepsilon$ of around a few times $10^{-7}$ for a system that is 1000 years old (see Fig.~\ref{fig:lifetime})---a different time evolution for the cloud is, in principle, allowed and would give a different observable strain. The lower bound on $\varepsilon$ shown in Fig.~\ref{fig:anomalous_pulsars} comes from requiring the superradiance cloud to pair produce the plasma, as described in Sec.~\ref{sec:SFQED}. This bound, however, can be relaxed since the BH was formed long ago, and charged particles could be slowly accreted and build up the plasma over time (see Sec.~\ref{sec:quasi-steady}). In this case, $\varepsilon$ as small as $10^{-12}$ could produce a pulsating source that is luminous enough to be observed at galactic distances. Finally, we notice that a few of the pulsating sources used here have a measured luminosity, which could be used to fix the value of $\varepsilon$ if the emission spectra were known (see Sec.~\ref{sec:emissionspectrum}). Taking the fraction of total luminosity that goes in the radio band to be between $10^{-4}$ and $10^{-6}$ gives $\varepsilon$ between approximately $10^{-12}$ and $10^{-11}$~\cite{Manchester:2004bp,Kaspi:2017fwg}.

\subsection{Concluding remarks}

In the last two subsections, we discussed two detection strategies which take advantage of the multimessenger signals from a dark photon superradiance cloud around a spinning BH. In both cases, a combination of the electromagnetic and GW observations allow us to perform highly non-trivial cross checks to uniquely identify our system and measure the dark photon mass and kinetic mixing parameter. Our analysis can be strengthened by dedicated numerical studies which 
\begin{itemize}
    \item extend our analysis to smaller gravitational coupling $\alpha$ and larger conductivity $\sigma/\mu$,
    \item provide robust information about the electromagnetic emission spectrum,  
    \item show robust evidence for or against periodicity in the electromagnetic emission in different electromagnetic bands. 
\end{itemize}

In addition to the signals mentioned above, our system may host  a plethora of phenomena, including ultraluminous X-ray sources (ULX)~\cite{Swartz:2004xt}, as well as transient processes such as fast radio bursts (FRB)~\cite{CHIMEFRB:2019gig,Petroff:2019tty} and X-ray flares and giant flares~\cite{1980ApJ...237L...1C}. Follow-up GW observations of these events could identify their origin as a dark photon superradiance cloud. With  electromagnetic observations alone, more dedicated numerical studies are needed to determine the similarities and differences in the spectral properties and transient dynamics between our system and a neutron star pulsar or magnetar~\cite{Philippov:2014mqa}, or other astrophysical sources. If  BH superradiance is discovered with continuous GW searches at LVK and future GW observatories, follow-up electromagnetic observations can discover the kinetically mixed phenomena presented in this paper or put robust constraints on the dark photon parameter space.

Finally, note that in this work we have exclusively focused on signals from stellar BHs, but the computations and the results for the luminosity presented in Eq.~\eqref{eq:luminosity} also apply to the case of isolated supermassive BHs, which would be of relevance for follow-up signatures of mergers detected by LISA.

\section{Discussion and Conclusions}  \label{sec:conclusions}

In this paper, we performed a detailed analytical and numerical study of the
dynamics of a kinetically-mixed, dark photon superradiance cloud and the
resulting multimessenger signatures. If a dark photon with Compton
wavelength on the order of the radius of a stellar mass black hole exists,
isolated, rotating black holes turn into very bright electromagnetic sources for
kinetic mixings on the order of $\varepsilon \sim 10^{-10}-10^{-6}$.
We summarize our approach, discuss the unique dynamics and observational
signatures of the system, and outline future directions for investigation
below.

A dark photon superradiance cloud is a dark electromagnetic field rotating at a frequency fixed by the dark photon energy. Charged particles in the interstellar medium enter the rotating cloud and, in the presence of a kinetic mixing, initiate a transient phase of cascade particle production, resulting in a dense plasma. The stable rotation of the cloud, large strength of the visible electromagnetic fields, and the appearance of the charged plasma resemble pulsar magnetospheres. We use resistive electrodynamic methods that interpolate between the vacuum and the force-free limit. Our methods are adapted from those originally developed to model resistive effects in the pulsar magnetosphere, here applied to 
a fixed Kerr spacetime and the kinetically mixed case. 

Due to the system's large conductivity, plasma currents redistribute charge in
an approximately dipolar form on large scales in order to screen the coherent
visible electromagnetic field set up by the rotating superradiance cloud. We
find that, due to the electric dominance of the oscillating superradiance cloud,
the charged plasma cannot completely screen the electromagnetic field.  Our
numerical simulations of the field electrodynamics indicate that the
differential rotation between the background fields and the plasma leads to the
emergence of small-scale turbulence in the form of magnetic field reconnection
and unscreened electric fields (Sec.~\ref{sec:Highlyconductinglimit}). We
establish that the electromagnetic emission from the system is dominated by
such small-scale turbulent dissipation into the standard model plasma, with a
peak luminosity of up to 10 orders of magnitude larger than the solar
luminosity, as described in Sec.~\ref{sec:emissionpower} (see Eqs.
\eqref{eq:powerdissipationfits} and \eqref{eq:luminosity}).  In addition, we
find a Poynting flux component to the emission.  Though
subdominant, $\sim 10^{2}-10^{4}$ times smaller than the local dissipation, this 
may also be significant for the observational signatures of the system.  These are the main
results of our paper.

Given the rotation of the superradiance cloud, we expect that at least a fraction of our system's emission is periodic (with period $T \approx 2\pi/\mu$), a property that would extend the analogy with pulsars into the observational domain.  Our simulations show some evidence for such a  periodicity, as discussed in Sec.~\ref{sec:emissionperiodicity}, but cannot conclusively verify this expectation. By analogy with the results for pulsar magnetospheres and findings from PIC simulations of turbulent plasmas, we expect that the emission spectrum contains a non-thermal X-ray component for kinetic mixing parameters $\varepsilon\gtrsim 10^{-7}$, softer spectra for smaller mixing parameters, and likely a radio component (see Sec.~\ref{sec:emissionspectrum}). Our simulations cannot directly determine the spectrum of the emitted luminosity, so given the differences between our system and pulsars, both in field profile morphology and strength, more investigations are essential to conclusively establish the spectral shape.

Based on the system's large luminosity, characteristic time evolution, and expected periodicity, we propose several search strategies for dark photon superradiance clouds. The first strategy relies on the extreme brightness of our system: electromagnetic follow-up searches targeting binary BH mergers observed by the LVK Collaboration (Sec.~\ref{sec:EWfromMerger}). Based on our expectations of the spectrum, the most promising reach is achieved by X-ray and radio observations.
In addition to the spatial and temporal correlation of a merger and the resulting electromagnetic emission, one could discriminate between superradiance clouds and standard astrophysical sources by requiring a fast exponential rise, and $1/t$ power-law fall-off of the light curve, consistent with superradiant growth and subsequent decay by GW emission. Given the measured remnant BH mass and spin from the gravitational waveform, these timescales are fully determined by the dark photon mass; for the parameters necessary to see these signals over cosmological distances, the electromagnetic power is subdominant to the GW power and does not affect the time evolution. This results in a non-trivial relation between the rise and decay times (see Eq. \eqref{eq:timescalerelation}),
which if experimentally confirmed, would provide a smoking gun signature of superradiance.

Further assuming periodicity, a variety of additional signatures can be explored. For the proposed electromagnetic follow-ups, the periodicity alone could be used to measure the dark photon mass $\mu$, as the periodicity is set by the dark photon mass up to a few-percent binding energy corrections from the BH potential. This would further strengthen the evidence for the new physics hypothesis by requiring consistency between the period-based measurement of $\mu$ and the light curve rise and decay times. 

Another observational strategy targets known pulsars, either by selecting those with positive frequency derivative, or by selecting those with the same measured period up to $\mathcal{O}(10^{-3}) $ due to Doppler shifts.
Interestingly, by surveying existing pulsar catalogues we find four candidate sources for which a partner with the same frequency exist, \textit{and} that have a positive frequency derivative measurement. 
Our computation of the GW power with the measured ``pulsar'' period and spin-up rate suggest that many of these objects would emit continuous gravitational wave with strain above the LVK threshold if their origin is dark photon superradiance. Many potential sources could be excluded with blind continuous wave searches while others could be further probed with targeted continuous wave searches (section~\ref{sec:GWfromPulsar}).  

Apart from these multi-messenger signatures of correlated EM and GW emission, at small $\alpha$ and large $\varepsilon$ the dark photon cloud depletion is dominated by EM emission, leading to persistent ultra-bright sources of X-rays. Observation of several periodic sources with periods within a few percent of each other would be strong evidence for the origin of these objects being a dark photon superradiance cloud. The absence of these sources in the Universe may already imply constraints on the dark photon parameter space assuming a natal black hole mass and spin distribution; given the uncertainties in BH properties and the EM signal spectral shape, we leave this study to future work.  

We close with a discussion of future directions. In this paper, we studied the case of an isolated BH and a single new particle with a minimal interaction, the kinetically-mixed dark photon. In the presence of a more complex dark sector, such as the existence of a dark Higgs, the dynamics may be altered further. Recently, some of the authors showed that in the higgsed dark photon scenario, vortex lines can form and deplete the cloud before it reaches its maximum size~\cite{East:2022ppo,East:2022rsi}. Further study is needed to understand how the vortex production and evolution is affected by the presence of plasma (see, e.g., Refs.~\cite{Adelberger:2003qx,East:2022rsi}).

Furthermore, many BHs are not isolated: they are surrounded by an accretion disk, leading to additional dynamics. While it is a negligible perturbation to the background dark photon field, a dense accretion disk affects the visible electromagnetic field and resulting electromagnetic dynamics and emission. BH systems affected by these dynamics include accreting supermassive black holes, X-ray binaries, etc. Preliminary studies show that even a small superradiance cloud (or a small $\varepsilon\sim 10^{-12}$ for a maximal cloud) affects the disk properties, and would invalidate the ISCO-based spin measurements~\cite{Li:2004aq,McClintock:2013vwa,Reid:2014ywa}; we leave further details to a future publication~\cite{DiskPaper}. 

This work sets the stage for the exploration of kinetically-mixed dark photon superradiance. We have established numerical simulation techniques to understand the electrodynamics of the cloud and its electromagnetic emission. The parameter space of possible signatures is vast, from weak, long-lasting signals to bright, short signals, across the electromagnetic spectrum. We have detailed several possible observations, and a more wide-ranging study of search strategies is warranted. Utilizing different simulation approaches to better understand the electromagnetic emission spectrum would be invaluable to pin down observables. Finally, given the similarities between our system and a magnetar, it is conceivable that the dark photon superradiance cloud could host a wide range of astrophysical phenomena, such as X-ray flares and fast radio bursts. Due to the differences in the magnetic field structures, further investigations are clearly needed, and would be of great interest.

\begin{acknowledgments}
We are extremely grateful to Andrei Gruzinov for many invaluable conversations and suggestions.
We thank Asimina Arvanitaki, Liang Dai, Savas Dimopoulos, Anson Hook, Matthew Johnson, Robert Lasenby, Luis Lehner, Elias Most, Peter Michelson, Michael M\"uller, Zhen Pan, Sasha Philippov, Geoffrey Ryan, Kendrick Smith, and Ling Sun for many useful discussions. MB is supported by the U.S. Department of Energy, Office of Science, Office of Basic Energy Sciences Energy Frontier Research Centers program under Award Number DE-SC0022348 and through the Department of Physics and College of Arts and Science at the University of Washington. MB, DEU, and JH thank the Institute for Nuclear Theory at the University of Washington for its kind hospitality and stimulating research environment. The  INT is supported in part by the U.S. Department of Energy grant No. DE-FG02-00ER41132. JH and CM thank the Center for Computational Astrophysics and New York University for hospitality. The Center for Computational Astrophysics at the Flatiron Institute is supported by the Simons Foundation. NS and WE acknowledge financial support by the Natural Sciences and Engineering Research Council of Canada (NSERC). DEU is supported by Perimeter Institute for Theoretical Physics and by the Simons Foundation. Research at Perimeter Institute is supported in part by the Government of Canada through the Department of Innovation, Science and Economic Development Canada and by the Province of Ontario through the Ministry of Colleges and Universities. This research was undertaken thanks in part to funding from the Canada First Research Excellence Fund through the Arthur B. McDonald Canadian Astroparticle Physics Research Institute. This research was enabled in part by support provided by SciNet (www.scinethpc.ca), Compute Canada (www.computecanada.ca), and the Digital Research Alliance of Canada (www.alliancecan.ca). Simulations were performed on the Symmetry cluster at Perimeter Institute, the Niagara cluster at the University of Toronto, and the Narval cluster at the École de technologie supérieure in Montreal.
\end{acknowledgments}

\appendix

\section{Notation} \label{app:notation}
We list the variable definitions used throughout the text in Table~\ref{tab:variabledefinitions}.

\begin{table}[h!]
    \begin{ruledtabular}
    \begin{tabular}{c|c}
        Variable & Description \\
    \hline
        $\varepsilon$ & Kinetic mixing \\
        $\mu$  & Massive vector field mass\\
        $r_c$ & Cloud's Bohr radius\\
        $a_*$ & Black hole dimensionless spin\\
        $\Delta a_*$ & Loss of black hole dimensionless spin\\
        $M$ &  Black hole mass \\
        $r_g$ & Half black hole Schwarzschild radius\\ 
        $\omega$ & Superradiance cloud angular frequency\\
        $f$ & Superradiance cloud frequency\\
        $\Omega_{\rm BH}$  & Black hole horizon frequency \\
        $M_c$ &  Superradiance cloud mass\\
        $\alpha$  & Gravitational fine structure constant\\
        $t,x,y,z$ & Cartesian Kerr-Schild coordinates\\       
    \hline
        $\sigma$ &  Plasma conductivity\\
        $\gamma_e$  & Electron/positron Lorentz factor\\
        $m_e$ &  Electron/positron mass\\
        $e$ &  Positron charge\\
        $\tau_{\rm plasma}$  & Timescale to populate the $e^\pm$ plasma\\ 
        $\omega_p$ & Plasma frequency \\
    \hline
        $A'_\mu$ &  Dark vector potential (interaction basis) \\
        $E'^i,\textbf{E}'$  & Dark electric field (interaction basis)\\
        $B'^i,\textbf{B}'$  & Dark magnetic field (interaction basis)\\
        $A_\mu$ &  Visible vector potential (interaction basis)\\
        $E^i,\textbf{E}$ & Visible electric field (interaction basis)\\
        $B^i,\textbf{B}$ & Visible magnetic field (interaction basis)\\
        $I^\mu$ &  Electromagnetic 4-current\\
        $J^i$ &  Electromagnetic spatial current\\
        $\rho_q$  & Charge density\\
        $v^i_d$  & Plasma drift velocity in Eulerian frame\\
    \hline
        $f_{\rm GW}$ & Gravitational wave frequency \\
        $P_{\rm GW}$ & Gravitational wave luminosity\\
        $\tau_{\rm GW}$ & Gravitational radiation timescale\\
        $\tau_{\rm SR}$ & Superradiance instability timescale\\
        $P_{\rm EM}$ & Visible Poynting flux\\
        $L_{\rm diss}$ & Visible energy dissipation power\\
        $\tau_{\rm EM}$ & Electromagnetic radiation timescale\\
        $\nu$ & Spectral frequency of electromagnetic emissions\\
    \end{tabular}
    \end{ruledtabular}
    \caption{List of the variables used most commonly throughout the main text, as well as a brief description.} 
    \label{tab:variabledefinitions}
\end{table}

\section{Construction of massive vector cloud} \label{app:procaconstruction}

To construct the superradiantly unstable massive vector field modes on a fixed Kerr spacetime of mass $M$ and dimensionless spin $a_*$, we follow Refs.~\cite{Lunin:2017drx,Frolov:2018ezx,Krtous:2018bvk}, as well as \cite{Dolan:2018dqv,Siemonsen:2019ebd}. Neglecting all non-linear effects both on the massive vector field $A'_\mu$, as well as in the gravitational sector, we treat $A'_\mu$ as a test field on a fixed Kerr background spacetime. In this Ricci-flat geometry, the vector field satisfies the massive vector wave equation
\begin{align}
    g^{\alpha\beta}\nabla_\alpha\nabla_\beta A'^\gamma=\mu^2A'^\gamma,
    \label{eq:vectorwaveeq}
\end{align}
with $g^{\alpha\beta}=g^{\alpha\beta}_\text{Kerr}$ and mass parameter $\mu$. The Kerr family of BH spacetimes, a special case of the Kerr-NUT-(A)dS class of spacetime, admits additional symmetries beyond stationarity and axisymmetry generated by the timelike and axial Killing fields, $\boldsymbol{\xi}$ and $\boldsymbol{\eta}$, respectively. These ``hidden" symmetries were utilized in Ref.~\cite{Frolov:2018ezx} to construct an ansatz for the field $A'_\mu$, satisfying the massive vector wave equation, that separates radial and angular dependencies. The ansatz makes use of the Killing-Yano symplectic 2-form $\boldsymbol{h}$, whose tensor components satisfy $\nabla_\mu h_{\nu\gamma} =2g_{\mu [\nu}\xi_{\gamma]}$. The vector field ansatz of frequency $\omega$ and azimuthal index $m$ reads \cite{Frolov:2018ezx}
\begin{align}
    A'^\mu=B^{\mu\nu}\nabla_\nu Z, & & Z=R(r)S(\theta)e^{-i(\omega T-m\varphi)}
    \label{eq:vectorfieldansatz}
\end{align}
in Boyer-Lindquist (BL) coordinates $(T, r, \theta, \varphi)$. The polarization tensor $B^{\mu\nu}$ is implicitly defined by $B^{\alpha\beta}(g_{\beta\gamma}+ih_{\beta\gamma})=\delta^\alpha_\gamma$ utilizing the Killing-Yano 2-form. Due to the presence of $\boldsymbol{\xi}$ and $\boldsymbol{\eta}$ in the spacetime, the temporal and azimuthal dependencies are trivially satisfied. The radial and polar dependencies are determined by solving a nonlinear ordinary differential eigenvalue problem with eigenfunctions $R(r)$ and $S(\theta)$ and complex eigenvalues $\omega$ and $\nu$ of the form, $\mathcal{D}^r_{\omega,\nu}R(r)=0$ and $\mathcal{D}^\theta_{\omega,\nu}S(\theta)=0$, where $D^{r,\theta}_{\omega,\nu}$ are second order ordinary differential operators depending on $r$ and $\theta$ only. Bound state solutions are obtained by imposing ingoing radiation boundary conditions at the horizon, and asymptotically flat boundary conditions at spatial infinity. For details on how the polar equation is solved, see Ref.~\cite{Dolan:2018dqv}. The radial solution is expanded around the outer horizon, $r\geq r_+$, with $r_\pm = M\pm \sqrt{M^2-a^2}$, in a Frobenius series of the form \cite{Dolan:2018dqv}
\begin{align}
    R_\text{near}(r)=\hat{r}^{i\kappa}(1+\hat{a}_1\hat{r}+\hat{a}_2 \hat{r}^2+\dots),
    \label{eq:nearHradialsol}
\end{align}
with $\kappa=2mr_+(\omega-m\Omega_{\rm BH})/(r_+-r_-)$, coefficients $\hat{a}_i$, and $\hat{r}=(r-r_+)(r_+-r_-)^{-1}$. The coefficients $\hat{a}_i$ can be solved for by plugging $R_\text{near}(r)$ into the radial equation $\mathcal{D}^r_{\omega,\nu}R(r)=0$. The near-horizon solution \eqref{eq:nearHradialsol} is then used to numerically integrate the radial second order ordinary differential equation outwards from $r_s=r_++\epsilon$ towards large $r\gg 10/(\alpha\mu)$, with $\epsilon=10^{-4}M$ typically, in the spirit of the shooting method. Integration cannot start at $r=r_+$, as the BL coordinates are singular on the event horizon. For further details, see Refs.~\cite{Dolan:2018dqv,Siemonsen:2019ebd}. The superradiantly unstable vector cloud can then be reconstructed with Eq.~\eqref{eq:vectorfieldansatz}.

Our numerical setup, outlined in App.~\ref{app:numericalsetup}, utilizes the Kerr spacetime in Cartesian Kerr-Schild (KS) coordinates $(t,x,y,z)$. Therefore, the relevant spacelike hypersurface is the surface of constant KS coordinate time $t$, not BL time $T$; these are two \textit{different} slices of the Kerr spacetime. Therefore, we transform the above constructed vector field $A'^\mu$ from BL coordinates to KS coordinates. The two gauges are related by
\begin{align}
\begin{aligned}
t= & \ T+\frac{M^2\log\frac{r-r_+}{r-r_-}}{\sqrt{M^2-a^2}}+M\log \Delta, \\
x= & \ \sin\theta(r\cos\bar{\phi}-a\sin\bar{\phi}),\\
y= & \ \sin\theta(a\cos\bar{\phi}+r\sin\bar{\phi}),\\
z= & \ r \cos\theta,
\label{eq:bltoks}
\end{aligned}
\end{align}
where $\Delta=a^2-2Mr+r^2$. The inverse of these relations is
\begin{align}
\begin{aligned}
T = & \ t-\frac{M^2\log\frac{r-r_+}{r-r_-}}{\sqrt{M^2-a^2}}-M\log \Delta,\\
r= & \ 2^{-1/2}\Big[-a^2+x^2+y^2+z^2 \\
& \ +\sqrt{4a^2z^2+(a^2-x^2-y^2-z^2)^2}\Big]^{1/2},\\
\phi= & \ \arctan\left[\frac{rx+ay}{-ax+ry}\right]-\frac{a\log\frac{r-r_+}{r-r_-}}{2\sqrt{M^2-a^2}}, \ \ \ x>0\\
\theta = & \ \arccos\frac{z}{r},
\label{eq:kstobl}
\end{aligned}
\end{align}
where $\bar{\phi}=\phi+(2\sqrt{M^2-a^2})^{-1}a\log[(r-r_+)(r-r_-)^{-1}]$, valid outside the event horizon $r>r_+$. Using these coordinate transformations, we transform the massive vector field to KS coordinates for $r>r_++\epsilon$. Subtleties arise for $r<r_++\epsilon$. However, it is necessary for the successful evolution of the system of equations in the interaction basis that the source term is defined (at the very least) everywhere outside the event horizon $r\geq r_+$, and while \textit{all} modes on the horizon are marginally trapped, the finite difference length scale (potentially) allows for values at points inside the horizon to numerically affect points just outside the horizon. Furthermore, the exponential blue-shift captured in the transformation rules \eqref{eq:bltoks} and \eqref{eq:kstobl} of the type $\cos\log (r-r_+)$ for $r> r_+$ leads to an exponential amplification of any truncation error in the numerical solution of the radial and angular equations $\mathcal{D}^r_{\omega,\nu}R(r)=0$ and $\mathcal{D}^\theta_{\omega,\nu}S(\theta)=0$. Hence, any power-like converging numerical truncation error is exponentially enhanced for $r\rightarrow r_+$, and dominates the solution for some $r_d$ with $r_d>r>r_+$. To address this subtlety, we employ a $C^4$-transition function $f_t(r)$ that matches the Frobenius solution $R_\text{near}(r)$, valid for $r\gtrsim r_+$ and the numerical solution $R_\text{num.}(r)$, valid for $r>r_++\epsilon$, in the overlap region $r_++\epsilon<r<r_++10^2\epsilon$: $R_\text{matched}(r)=f_t R_\text{near}(r)+(1-f_t)R_\text{num.}(r)$. In addition to these manipulations at small radii, we also need to address the exponential fall-off as $r\rightarrow\infty$. Due to finite floating point precision, the shooting method will inevitably switch from the exponentially decaying solution into the exponentially diverging solution at some large $r=r_\text{max}$. Therefore, in order to provide sensible estimates also for $r>r_\text{max}$ (which is necessary since we are working with a compactified setup that includes spacelike infinity, as outlined below in App.~\ref{app:numericalsetup}), we fit an exponential of the form $a e^{-br}$ for $b>0$ to the solution $R_{\rm num.}(r)$ in the range $r\in(0.9 r_\text{max},r_\text{max})$. With this, we obtain a $R_{\rm matched}(r)$ that is valid for $r\in(r_+,\infty)$.

Finally, the 3+1 superradiant vector variables are projected with respect to the $t$=const. spacelike hypersurface using the hypersurface normal $n^\mu$ and the projector $\gamma^\alpha{}_\beta=\delta^\alpha_\beta+n^\alpha n_\beta$ (further details can be found in App.~\ref{app:numericalsetup}). In this framework, the vector field decomposes into $\chi_\phi\equiv-n_\mu A'^\mu$ and $\chi^i\equiv\gamma^i{}_\mu A'^\mu$, which are reconstructed from the matched and extrapolated radial solution, in conjunction with the polar solution, the transformation rules \eqref{eq:kstobl} and \eqref{eq:bltoks}, and ansatz \eqref{eq:vectorfieldansatz}, everywhere in the $t=$ const. spacelike hypersurface with $r>r_+$. Lastly, we find better convergence properties of the $D_i B^i=0$ constraint close to the BH event horizon if a buffer region between the event horizon and the excision surface in the BH interior is used. To that end, we utilize second order extrapolation of all 3+1 variables along lines of constant $\theta,\varphi,t$ from $r>r_+$ to $r_-<r<r_+$ to ensure well-defined gradients at $r_+$, we utilize zeroth order extrapolation from a distance of $\delta=10^{-4}M$ away from the spin-axis to set the cloud values \textit{on} the axis in KS coordinates, since the BL coordinates also exhibit a coordinate singularities at the poles. The time-dependence of the cloud in KS coordinates is then simply $\chi_\phi, \chi_i\sim e^{i\omega t}$.
\begin{table}[t]
    \begin{ruledtabular}
    \begin{tabular}{ccc}   
    \centering
         $\alpha$ & $\omega M$ & $a_*$ \\ \hline
         0.1 & $0.099485$ & $0.382787$ \\ 
         0.2 & $0.195543$ & $0.678411$ \\ 
         0.3 & $0.283390$ & $0.857953$ \\ 
         0.4 & $0.357498$ & $0.946250$ 
    \end{tabular}
    \end{ruledtabular}
    \caption{The properties of the four clouds used in the main text. We consider only the $m=1, \hat{n}=0$ and $S=-1$ superradiant vector boson clouds in the language of Ref.~\cite{Siemonsen:2019ebd} [corresponding to $(j,n,l,m)=(1,1,0,1)$ used in Sec.~\ref{sec:sr}] around spinning BHs of mass $M$ and dimensionless spin $a_*$. The saturation condition, $\omega=\Omega_{\rm BH}$, fixes the spin $a_*$ for each $\alpha$.}
    \label{tab:clouds}
\end{table}
In this work, we consider only those superradiance clouds that arose from the fast growing modes, and subsequently saturated the superradiance condition, i.e., satisfying $\omega=m\Omega_{\rm BH}$. That is, we focus on $m=1, \hat{n}=0$, and $S=-1$ clouds, in the language of \cite{Siemonsen:2019ebd} (corresponding to $(j,n,l,m)=(1,1,0,1)$, introduced in Sec.~\ref{sec:sr}). Properties of the clouds considered in this work are given in Table~\ref{tab:clouds}. These states are expected to be the endstate of the instability to a good approximation \cite{East:2017mrj,East:2017ovw}, and hence, exhibit vanishing growth rates.

\begin{figure}[t]
    \centering
    \includegraphics[width=0.48\textwidth]{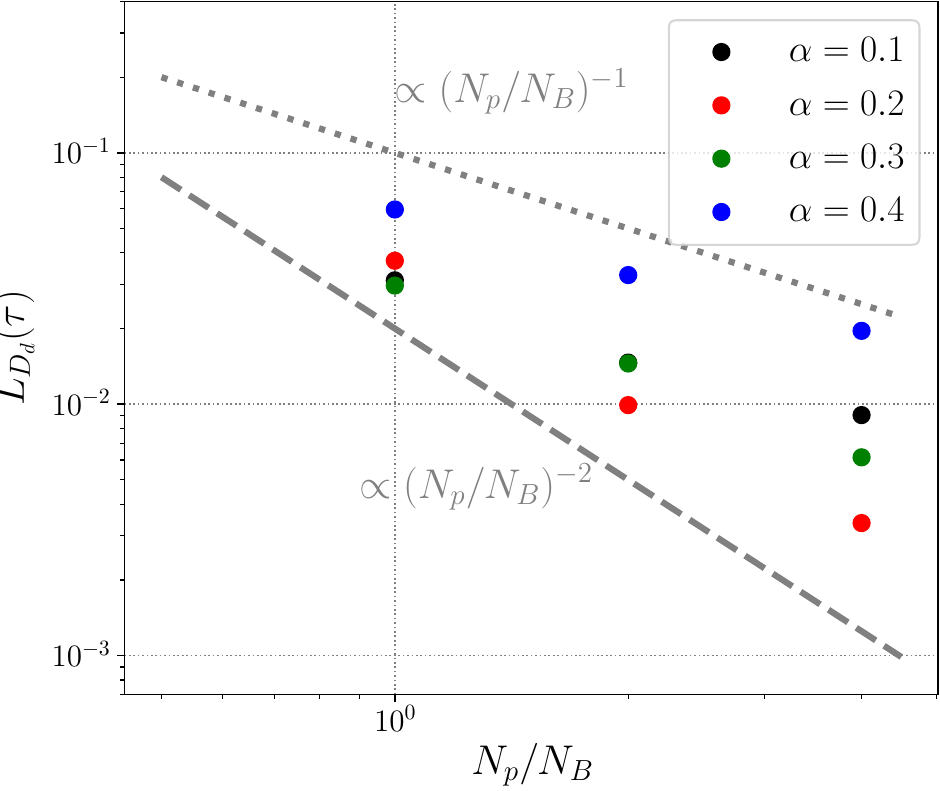}
    \caption{The norm $L_{D_d}(\tau)$ of the residual of the massive vector wave equation, defined in \eqref{eq:convnorm}, as a function of the number of grid points $N_p$ considered with respect to the base resolution $N_B$. The default resolution used for each of the configurations given in table~\ref{tab:clouds} and presented throughout the main text is $N_p/N_B=2$.}
    \label{fig:Ltauconvergence}
\end{figure}

We test our numerical implementation of the above described reconstruction of the cloud in KS coordinates by considering the numerical truncation error $\tau$ of the massive vector wave equation \eqref{eq:vectorwaveeq} in KS coordinates: $\tau=\sum_\alpha|(\square_\text{Kerr,KS}-\mu^2)(\chi_\phi n_\alpha+\chi_\alpha)|$. Inside the event horizon, the truncation error $\tau$ is divergent with decreasing grid spacing $h$, which is likely due to inaccuracies in the extrapolation procedure described above. On the event horizon, the truncation error shows marginal point-wise convergence $\tau|_{r=r_+}\sim\mathcal{O}(h^{1/2})$ or better. To quantify the convergence behavior outside the event horizon, in the coordinate domain $D$, we introduce the norm $L_D(f)$ of function $f$ as
\begin{align}
    L_D(f)\equiv\int_D d^3x\sqrt{\gamma}|f|,
    \label{eq:convnorm}
\end{align}
where $d^3x \sqrt{\gamma}=d^3x\sqrt{\det\gamma_{ij}}$ is the volume form of the spacelike hypersurface. Here, and in the following, the domain $D=D_d$ is the coordinate shell defined by the radii $10 r_c>r>r_+$ (recall, $r_c=1/(\mu\alpha)$ is the cloud's Bohr radius). The norm of the truncation error converges approximately as $L_{D_d}(\tau)\sim\mathcal{O}(h^{2})$ for $\alpha\in\{0.2,0.3\}$ and as $L_{D_d}(\tau)\sim\mathcal{O}(h^{1})$ for $\alpha\in\{0.1,0.4\}$, see \figurename{ \ref{fig:Ltauconvergence}}. We use second order accurate methods to compute the residual $\tau$. This convergence behavior can be explained by considering the shooting method underlying the reconstructed solution. As discussed above, any remaining truncation error close to the horizon is exponentially amplified, while at large radii, the shooting method inevitable switches from the exponentially decaying solution to the exponentially increasing solution due to finite floating point accuracy (see also a discussion of this in \cite{Siemonsen:2019ebd}). While our shooting method implementation makes use of higher-than double-precision floating point arithmetic, the convergence is ultimately limited, especially for small $\alpha$. Further, the extrapolation of the cloud to the spin-axis in KS coordinates, and the exponential extrapolation at large radii, can further reduce the convergence rate, especially at large $\alpha$. Therefore, even slow convergence is a good sign of correct implementation of the superradiant source terms in our compactified KS evolution implemented discussed next.

\section{Numerical evolution setup} \label{app:numericalsetup}

Our calculations are carried out on a fixed rotating BH background of mass $M$ and dimensionless spin $a_*$. We adopt the Kerr metric $g_{\alpha \beta}$ in Cartesian Kerr-Schild coordinates $(t,x,y,z)$, with line element
\begin{align}
\begin{aligned}
    ds^2= & \ -dt^2+dx^2+dy^2+dz^2 \\
    & \ +\frac{2Mr^3}{r^4+a^2M^2z^2}\Big[ dt+\frac{z}{r}dz\\
     & \ +\frac{r(xdx+ydy)}{r^2+a^2M^2}-\frac{aM(xdy-ydx)}{r^2+a^2M^2}\Big]^2,
\end{aligned}
\end{align}
where $r$ satisfies $(x^2+y^2)/(r^2+a^2)+z^2/r^2=1$. 
However, we evolve the fields using a $3+1$ space--time decomposition, making
using of several such geometric quantities. This includes the unit normal to slices of constant time, which can be decomposed
as $n^\mu=(1,-\beta^i)/N$, where $N$ and $\beta^i$ are the lapse and shift vector, respectively. 
There is also the spatial metric/projection operator $\gamma_{\mu \nu }=n_\mu n_\nu+g_{\mu \nu}$,
the extrinsic curvature tensor
$K_{ij}=-(D_t\gamma_{ij})/(2N)$, where $D_t=\partial_t-\mathcal{L}_\beta$
with $\mathcal{L}_\beta$ the Lie derivative along the shift. In
a Minkowski spacetime in Cartesian coordinates these 3+1 variables would be $N=1$,
$\beta^i=0$, $\gamma_{ij}=\delta_{ij}$, and $K_{ij}=0$. For convenience, we also define
also the covariant derivative $D_i$ defined with respect to $\gamma_{ij}$, as well as the trace
of the extrinsic curvature $K=K_{ij}\gamma^{ij}$.

The kinetically mixed Maxwell's equations, presented in covariant from in \eqref{eq:FieldeqInteraction1}, in terms of the visible electric and
magnetic fields are given by
\begin{align}
\begin{aligned}
D_t E^i=& \ N K E^i+\varepsilon^{ijk}D_j (N B_k)- N J^i+ N \varepsilon\mu^2\gamma^i{}_\mu A'^\mu, \\
D_t B^i=& \ N K B^i-\varepsilon^{ijk}D_j(N E_k), \\
D_i E^i=& \ \rho_q+\varepsilon\mu^2 n_\mu A'^\mu,\\
D_i B^i=& \ 0,
\end{aligned}
\label{eq:evolutionequations}
\end{align}
where $\varepsilon^{ijk}=\varepsilon^{ijk\alpha}n_\alpha$ is the 3-dimensional
Levi-Civita tensor. 
These evolution equations
\eqref{eq:evolutionequations} are discretized using fourth-order accurate
spatial finite-difference stencils, in conjunction with fourth-order
Runge-Kutta integration in time~\cite{East:2015pea,East:2018ayf}. All
simulations are performed on a 3D Cartesian grid that includes spatial infinity through the
use of compactified coordinates (details can be found in
\cite{Pretorius:2004jg}). Sixth-order, Kreiss-Oliger-type numerical
dissipation is applied to the evolution variables for numerical stability. This
further dissipates shorter wavelength features 
at large distances beyond the compactification scale,
minimizing reflection off spatial infinity. We use between ten and seven
mesh-refinement levels centered on the BH, with refinement ratio $2:1$, for $\alpha=0.1$ through
$\alpha=0.4$, respectively. We choose the finest level to have length roughly
twice the diameter of the BH in each linear dimension. 
The system is rescaled so that $20\mu^{-1}\alpha^{-1}=20r_c$ is roughly equal to
the compactification scale (recall $r_c$ is the cloud's Bohr radius). 
This allows us to resolve both the scale set by the BH,
as well as that set by the superradiance cloud, for sufficiently long times as to
ensure relaxation into a quasi-equilibrium state. For all cases,  
we use a grid spacing of $\Delta x\approx 0.03M$ on
the finest mesh refinement level. Due to this scaling, radiation extraction can
be done up to a distance of $r\leq 10r_c$; beyond this distance, 
high-frequency radiation modes are no longer sufficiently resolved in the wave extraction
zone due to the compactification of the domain. The Maxwell equations
with an Ohm's law become stiff in the high conductivity limit (discussed further
in App.~\ref{app:resistivitytest}). Hence, at conductivies of $\sigma/\mu>2$,
we adjust the time-step $\Delta t$ to account for this behavior. For
$\alpha=0.3$, we decrease it gradually with increasing conductivity from $\Delta t/\Delta x=0.5$ 
down to $\Delta t/\Delta x=0.075$ to achieve a
robust numerical evolution. For $\alpha\in\{0.1,0.2,0.4\}$ and $\sigma/\mu=20$,
we scale $\Delta t$, such that $\sigma \Delta t$ remains as small, or smaller
than, the value of $\sigma \Delta t$ for $\alpha=0.3$, everywhere in the relevant
computational domain. The construction of the superradiance cloud is described
in detail in App.~\ref{app:procaconstruction}. As we are neglecting
backreaction of the presence of the plasma and massless electromagnetic fields,
the superradiance cloud is not evolved numerically, rather it is a pre-prescribed 
function of time. 

The set of equations \eqref{eq:evolutionequations} is comprised of the two
constraints, the Gauss law for electric and magnetic fields, and the Faraday
equation and Ampere's law as evolution equations. 
Numerically, we damp possible
violations of the constraint equations by means of two constraint-damping fields
$\Phi$ and $\Psi$ \cite{Dedner2002,Palenzuela:2008sf}. To that end, we perform the replacements $D_tB^i\rightarrow
D_tB^i-N D^i\Phi$ and $D_tE^i\rightarrow D_tE^i-N D^i\Psi$ at the
level of the evolution equations in \eqref{eq:evolutionequations}. Furthermore,
we promote the constraint equations to evolution equations for these auxiliary
fields, following Refs.~\cite{Dedner2002,Palenzuela:2008sf}:
\begin{align}
\begin{aligned}
    D_t\Psi= & \ -N (D_iE^i-\varepsilon\mu^2n_\mu A'^\mu-\rho_q)-N\kappa\Psi,\\
    D_t\Phi= & \ -N D_iB^i-N\kappa\Phi.
    \label{eq:auxiliaryevolutionequations}
\end{aligned}
\end{align}
This ensures that any numerical violation of the constraints
$D_iE^i-\varepsilon\mu^2n_\mu A'^\mu-\rho_q=0$ and $D_iB^i=0$ are damped
exponentially over timescales $1/\kappa$. For all (resistive) force-free
simulations, the constraint $D_iE^i-\varepsilon\mu^2n_\mu A'^\mu-\rho_q=0$ is
trivially satisfies since the charge density is \textit{defined} to be
$\rho_q=D_i E^i-\varepsilon\mu^2n_\mu A'^\mu$. Hence, unless we explicitly
assume vacuum (and in particular, set $\sigma=0$), $\Psi$ is
identically zero. In all cases, the initial conditions for these auxiliary fields is
$\Phi=\Psi=0$. In addition, we perform ideal force-free simulations by means of
two ad-hoc field modifications applied at each grid point after an evolution
step \cite{Palenzuela:2008sf,Palenzuela:2010xn} (see Ref.~\cite{Alic:2012df} for a discussion):
\begin{align}
    E^i\rightarrow & \ E^j\left(\delta^i_j-\frac{B_j B^i}{B^2}\right),\\ 
    E^i\rightarrow & \ E^i\left\{1-\hat{\theta}(\lambda)+\hat{\theta}(\lambda)\frac{B}{E} \right\},
    \label{eq:forcefreeprescription}
\end{align}
where $\lambda=E^2-B^2$ and $\hat{\theta}$ is the Heaviside function. This
prescription enforces the two force-free conditions, $E_iB^i=0$ and $B^2>E^2$,
by explicitly rescaling the electric field at each grid point. The rescaling is
a form of ad-hoc numerical dissipation that is not physically motivated and
reproduces physical dissipation behavior only in special cases. Therefore, as
pointed out in the main text, the dissipation estimates provided by these
force-free evolution schemes should be interpreted with caution. 

The evolution of the system proceeds as follows.  We evolve the system towards
its equilibrium state in several steps.  Initially, we set the fields to
visible fields to zero $E^i=B^i=0$ and evolve until time $t=t_s$ assuming
vacuum $I^\alpha= 0$. With $t_s\approx 5/\mu$, this allows the system to
equilibrate at roughly $E^i=\varepsilon E'^i$ and $B^i=\varepsilon
B'^i$, which is purely the superradiance cloud's contribution to the
visible fields. During this time, we utilize both the electric and magnetic
field's Gauss constraint cleaning potentials $\Phi$ and $\Psi$. These ensure
that constraint violations in the magnetic field are kept small, as well as
efficiently remove constraint violations of the $E^i=0$ initial data on
timescales $1/\kappa\ll t_s$. 
At $t=t_s$, the resistive
current \eqref{eq:resistivecurrent} is discontinuously turned on, and the system
is evolved until the total Poynting flux at the largest radii where we extract it is relaxed to
a quasi-constant value. For $\alpha=0.4$, we found that this required the system to be
evolved for $\sim 8 t_{\rm LC}$ light crossing times of the entire cloud, defined as $t_{\rm
LC}=10r_c$, whereas for $\alpha=0.1$, we
evolved the system for $\sim 3 t_{\rm LC}$. During the evolution of the system, we
monitor the behavior of the Gauss constraint $D_i B^i=0$ throughout the entire
domain. This provides a measure for the rate of convergence of the numerical
solution, and the self-consistency of the numerical implementation.

\begin{figure*}
    \centering
    \includegraphics[width=0.46\textwidth]{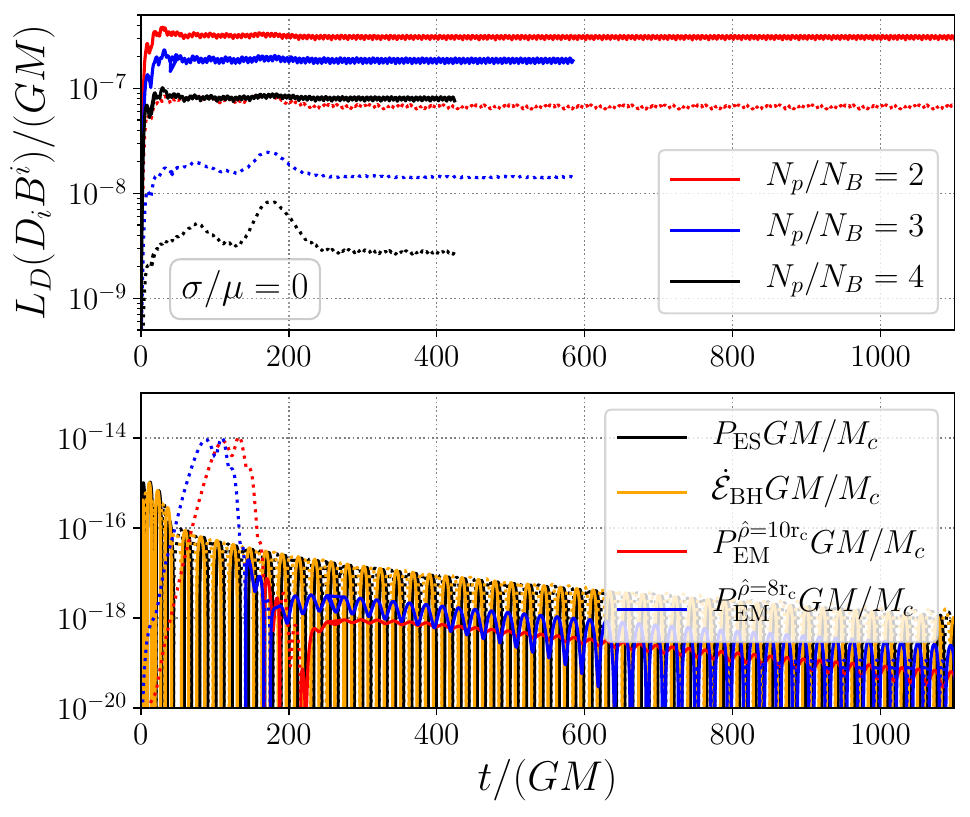}
    \hfill
    \includegraphics[width=0.4725\textwidth]{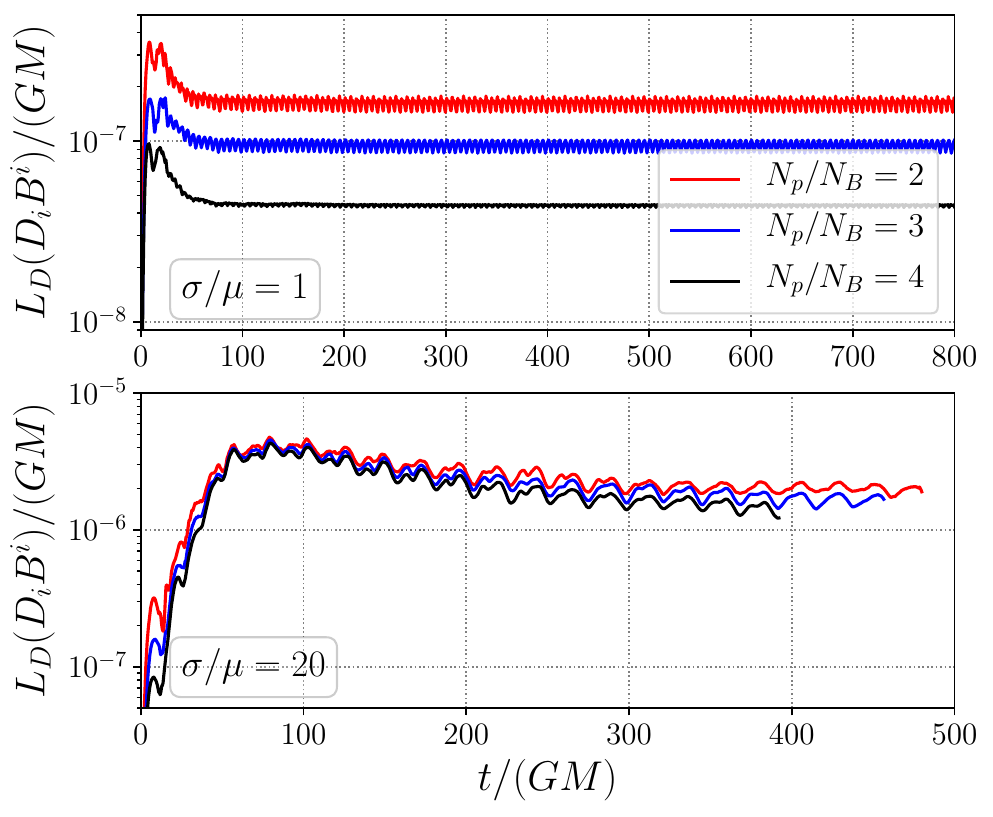}
    \caption{\textit{(right and top left)} We present the convergence of the norm \eqref{eq:convnorm} of the
    Gauss constraint $D_i B^i$ within domains $D_s$ (solid lines) and
    $D_d$ (dashed lines) for the $\alpha=0.3$ system at three different
    conductivities $\sigma$. The former is the entire domain outside the outer
    horizon with $r>1/\mu$ (i.e., neglecting the near-horizon behavior),
    whereas the latter is the entire domain outside the BH $r>r_+$. $N_p/N_B$ is defined
    as in \figurename{ \ref{fig:Ltauconvergence}}, and we set $\varepsilon=10^{-6}$. 
    \textit{(bottom left)} We plot the Poynting fluxes across the ergosurface, $P_{\rm ES}$, 
    the event horizon, $\dot{\mathcal{E}}_{\rm BH}$, as well as the flux $P_{\rm EM}$ coordinate radii $r=8r_c$ and
    $r=10 r_c$ in the $\alpha=0.3$ cloud with $I^\mu=0$ and $\varepsilon=10^{-6}$.}
    \label{fig:divBconvergence}
\end{figure*}

To test the numerical implementation of the kinetically mixed force-free
Maxwell equations, together with the reconstruction
of the massive vector field modes, we begin by considering the $\alpha=0.3$
vacuum case $\sigma=0$. To that end, we consider \eqref{eq:evolutionequations}
with $I^\mu=0$, and perform a series of simulations with increasing resolution
starting from vacuum initial data: $B_i=E_i=0$. To demonstrate the correct
implementation of the equations, we monitor the evolution of the constraints
$D_iB^i=0$. In \figurename{ \ref{fig:divBconvergence}}, we show the convergence
of this Gauss constraint with grid spacing $h$ utilizing the norm defined in
\eqref{eq:convnorm}. As can be seen there, roughly $L_{D_s}(D_iB^i)\sim
\mathcal{O}(h^{3.5})$, whereas $L_{D_d}(D_iB^i)\sim
\mathcal{O}(h^{2})$. Therefore, in the bulk of the vector cloud, the
convergence is roughly fourth order, as expected. Close to the event horizon,
convergence is slower. This may be attributed to the presence of the excision
surface close to the event horizon, as well as the lower convergence order of
the massive vector cloud residual $\tau$ just outside the horizon. The constraint
$D_iE^i-\varepsilon\mu^2 n_\mu A'^\mu=0$ (not shown here) is violated by the initial
data, but quickly becomes dominated by converging truncation error after a few periods
of the cloud. This could be improved upon, by choosing constraint satisfying
initial data. However, the goal of this work is to investigate the system
with non-vanishing charge density, and at late times, the latter constraint is
satisfied to floating point accuracy if $I^\mu\neq 0$. Moving to cases with
non-vanishing conductivity, we show in \figurename{ \ref{fig:divBconvergence}}
the convergence properties of $\alpha=0.3$ systems for $\sigma/\mu=1$ and
$\sigma/\mu=20$. The former exhibits the same convergence behavior as the
vacuum case, whereas the latter follows roughly $L_{D_d}(D_i B^i)\sim
\mathcal{O}(h^{0.6})$. This can be explained by considering the scales
of the features that need to be resolved. As we showed in the main text, the
scale of the features is roughly given by $1/\sigma$, which translates into a
length scale of $\ell\approx 0.17 M$ for $\alpha=0.3$. On the finest level, the
grid spacing, $\Delta x=0.03GM$, is sufficient to resolve these features, while
on coarser levels, numerical dissipation likely dampens these scales
efficiently. This damping is numerical, rather than physical, in nature, and does not
obey the Maxwell equations, and therefore, leads to a larger violation of the
Gauss constraint and worse convergence properties.

Apart from the convergence of the constraints, the vacuum quasi-stationary
state, after several $t_{\rm LC}$, should exhibit no energy flux across the
event horizon, as it is, by construction, synchronized with the BH angular
velocity. In practice, there are various sources of numerical error that can
spoil this property. The synchronization condition $\omega=\Omega_{\rm BH}$ can
be achieved only up to finite precision, when solving for the superradiance
cloud. Finite resolution both in the evolution scheme, as well as in the cloud
construction scheme, may also leave room for the solution to develop a small,
but finite, energy flux across the horizon and towards spatial infinity. To
quantify this, and to obtain a rough estimate for the time at which the system
is truly settled, $t_\text{settle}$, we monitor the energy fluxes 
across the horizon, the BH ergosurface, and coordinate spheres of
radius $\hat{\rho}=8 r_c$, and $\hat{\rho}=10r_c$ in \figurename{ \ref{fig:divBconvergence}}. In the
continuum limit, with $\omega=\Omega_{\rm BH}$ exactly, we expect
all these fluxes to be zero.
Therefore, the flux evolution presented in \figurename{
\ref{fig:divBconvergence}} can be used to establish $t_{\rm settle}$, i.e.,
when the system has reduced the superradiance cloud's emission powers to the
degree necessary.

We briefly comment on issues related to performing simulations of the superradiance cloud system
in the small-$\alpha$ limit. In the Newtonian limit, the massive vector wave
equation on a Kerr background is obtained by expanding in small $\alpha$ to
leading order. All spin-effects are subleading in this expansion, and the
leading contribution is solely given by the far-zone Newtonian potential of the
BH $\sim G M/r$. In this limit, the vector wave equation reduces to a radial
Schrödinger-type equation with solution \eqref{eq:srprofiles}. Within a
numerical time-domain evolution setup, the singular behavior of the Newtonian
potential at the location of the BH poses challenges. However, 
there are subtleties associated with replacing the far-zone weak-field metric by Minkowski
both in the interaction and mass eigenbases
(\eqref{eq:interlagrangian} and \eqref{eq:masslagrangian}, respectively). Within
the mass eigenbasis, the force-free condition (or resistive generalization thereof)
$F^{\alpha\beta}I_\beta=0$, depends on the electric and magnetic field components
of \eqref{eq:srprofiles}. The non-relativistic field \eqref{eq:srprofiles} and
its electric and magnetic field components are multivalued, i.e.,
discontinuous, at the origin, leading to a breakdown in the validity of
numerical schemes around the origin. Additionally, the usual force-free current
would require modification, as it requires the field \eqref{eq:srprofiles} to
satisfy the corresponding Maxwell equations on a weak-field background
spacetime. As noted above, a weak-field metric is numerically challenging to
implement, such that the choice of a flat background introduces inconsistencies
when using the (resistive) force-free current. These could be remedied,
however, by including terms involving 
higher order derivatives of the massive vector field~\eqref{eq:srprofiles} 
in the equations, but this would add further
complications at the origin. By contrast, the evolution equations in the
interaction basis depend only on $A'_\mu$, and not on its spatial derivatives;
hence, the interaction basis evolution approach allows one to evolve the system
self-consistently on a flat background. On the other hand, this choice is
accompanied by subtleties associated with the photon-dark photon interaction
term in \eqref{eq:interlagrangian}. The massless state can mix into the massive
state as it radiates towards infinity. In a weak-field metric, this mixing
prevents the massive component of the visible field $A_\mu$ from radiating to
infinity, as it is bound to the central gravitational potential. In the flat
spacetime limit, leakage of the massive state into radiation emitted to
infinity is not prevented (an illustration of this behavior is presented in
\figurename{ \ref{fig:flattests}}). Therefore, in this context, any radiated
Poynting flux is to be understood as an upper bound for the total emitted
power. All these subtleties are absent in the fully-relativistic calculations
we use as our main results, where
the relativistic clouds constructed in App.~\ref{app:procaconstruction} is
considered on a Kerr BH background spacetime as described above.

\section{Resistive force-free currents} \label{app:resistivitytest}

In this appendix, we discuss different resistive generalizations
of force-free electrodynamics used in the literature to
identify the approach most applicable in the kinetically-mixed scenario at
hand. In the main text, we demonstrated that the system is characterized by
turbulence and magnetic reconnection with efficient energy dissipation
into the plasma. In principle, there are two feasible approaches for capturing
these effects: resistive magnetohydrodynamics and kinetic 
PIC methods. PIC methods, which capture the macro- and micro-physics, are
ideally suited to tackle the magnetosphere of the kinetically mixed
superradiance cloud. However, as we are interested in the overall
electromagnetic power-output and large scale features of the system in three
dimensions, PIC simulations are prohibitively computationally expensive, especially on a
curved background BH spacetime. Full resistive magnetohydrodynamics, on the
other hand, is notoriously difficult to apply to regimes in which the plasma
mass density is far below the energy density of the electromagnetic field,
which is the case for the superradiant system considered here. 
Hence, we choose to use a 
resistive approach where the plasma dynamics is not directly tracked, and
rely only on the electromagnetic field's evolution. A few approaches have been
developed in the literature, particularly to model the resistive regions of
pulsar magnetospheres
\cite{Komissarov:2005xc,Gruzinov:2007se,Li:2011zh,Parfrey:2016caq,Kalapotharakos2012}
(see also Refs.~\cite{Palenzuela:2008sf,Palenzuela:2012my}). All are based on an
electromagnetic current $J^i$ that aims to capture the physics of a
highly conducting plasma in strong electromagnetic fields, while
being specified solely in terms of the electromagnetic fields. Generally, this
current can be decomposed into a piece describing the drift velocity of the charges,
$v_d^i$, and a contribution orthogonal to the drift velocity
\begin{align}
    J^a= \rho_q v_d^a+J^a_\perp. 
\end{align}
In all (resistive) force-free approaches, the charge density is defined using the (kinetically-mixed) electric Gauss law: $\rho_q=D_i E^i-\varepsilon\mu^2 n_\mu A'^\mu$. In the following, we briefly review the currents considered in the literature, apply these to kinetically-mixed superradiance clouds in the non-relativistic limit, and compare our findings with the vacuum and force-free limits in order to evaluate their applicability.

\begin{figure*}
    \centering
    \includegraphics[width=0.99\textwidth]{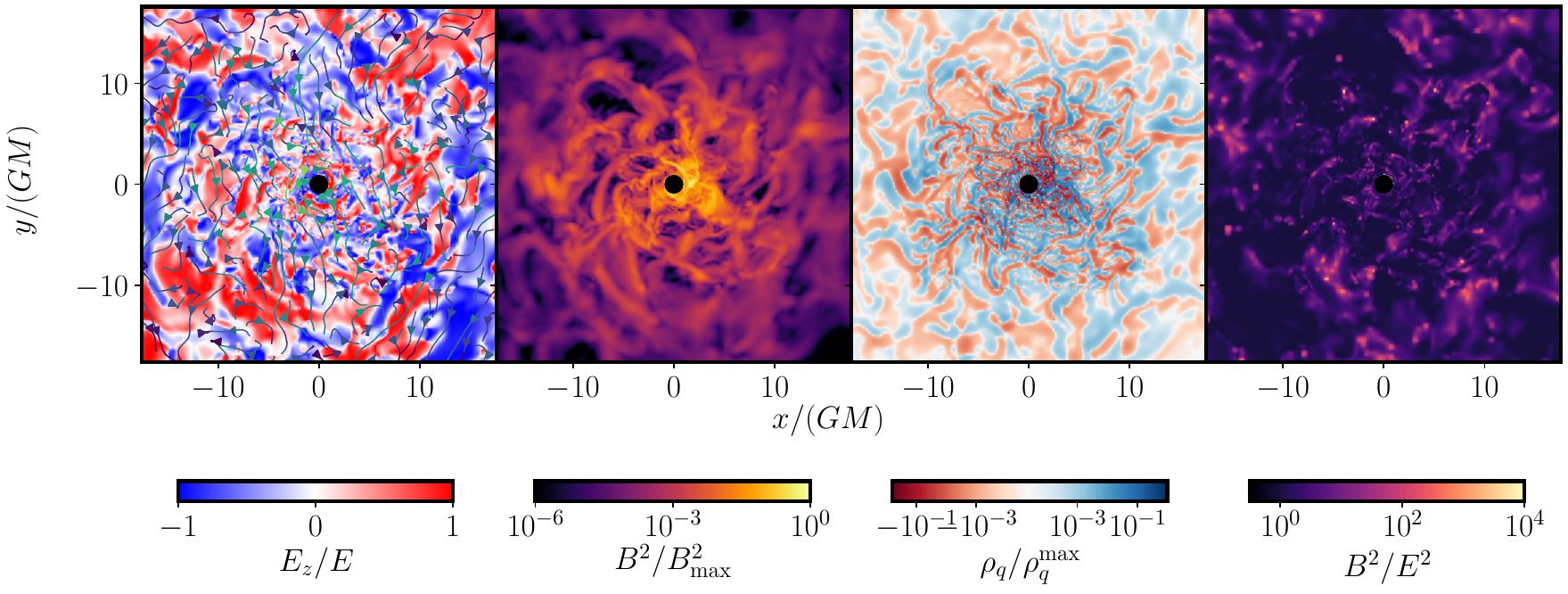}
    \caption{We show the visible electric and magnetic fields, charge density, and the ratio of electromagnetic field magnitudes obtained from a force-free simulation using the current \eqref{eq:forcefreecurrent}, with prescription \eqref{eq:forcefreeprescription} for a superradiance cloud system with $\alpha=0.3$ and a BH spin of $a_*=0.86$. In the case of the visible electric field, the field line color code is the same as in \figurename{ \ref{fig:condEFieldlines}}.}
    \label{fig:forcefreequantities}
\end{figure*}

We begin with a discussion of the commonly invoked force-free
paradigm, which assumes vanishing Lorentz force $F^{\mu \nu}I_{\nu}$, as well as $F_{\mu\nu}F^{\mu\nu}>0$ and
$F_{\mu\nu}(*F)^{\mu\nu}=0$. 
In terms of electric and magnetic fields, the last two conditions are, respectively, equivalent to
magnetic dominance $B^2>E^2$ and $E_iB^i=0$. 
With these assumptions, the corresponding
force-free current is given by (the relativistic contributions in the form of
the extrinsic curvature $K$ and covariant derivative $D_i$ are defined in App.~\ref{app:numericalsetup})
\begin{align}
\begin{aligned}
v_d^a \ & =\frac{\varepsilon^{ajk}E_j B_k}{B^2},\\
J^a_{\perp,\text{FF}} \ & =\frac{B^a}{B^2}\Big[2K B_i E^i-2K_{ij}E^i B^j +B_i \epsilon^{ijk}D_j B_k\\
& \ \ -E_i\epsilon^{ijk}D_j E_k+\varepsilon\mu^2B_i\gamma^i_\mu A'^\mu\Big].
\label{eq:forcefreecurrent}
\end{aligned}
\end{align}
The current is perpendicular to the electric field, and there is no dissipation of electromagnetic
energy. 
Therefore, the force-free
limit of ideal magnetohydrodynamics excludes resistive processes or transfer
of electromagnetic energy to the plasma (e.g.,
particle acceleration, magnetic reconnection, plasma heating, etc.), which
however, are active throughout the superradiance cloud due to the
\textit{electric} dominance of the fields $A'_\mu$ in vacuum. Notice, the
plasma drift velocity $v_d^a$ is entirely determined by the dynamics of the
electromagnetic fields. While the force-free approximation is, in principle, ideal,
it can break down at current sheets and places where magnetic dominance is lost.
Numerically, this is handled with numerical dissipation which is
particularly large in turbulence driven regimes (due to the cascade to short,
unresolved wavelengths), as well as by (as noted in App.~\ref{app:numericalsetup}) 
enforcing the force-free conditions by rescaling the visible electric field, as
shown in \eqref{eq:forcefreeprescription}. We perform a set of force-free
simulations of the $\alpha=0.3$ and $a_*=0.86$ cloud-plasma system. As the
turbulent features reach scales much below the grid scale of our simulations,
numerical dissipation and the prescription \eqref{eq:forcefreeprescription}
efficiently remove energy that was sourced by the superradiance cloud. Therefore, it
serves as an artificial source of dissipation, that nonetheless agrees well with
the $\sigma\rightarrow\infty$ extrapolations shown in \figurename{
\ref{fig:FluxJouletotal}}. Regardless, results from these force-free
simulations should be interpreted with caution and in light of the
un-physical dissipation mechanism. In \figurename{
\ref{fig:forcefreequantities}}, we show the force-free solution the system
attains at late times (with strong numerical dissipation in the bulk of the
cloud). In all cases, the magnetic Gauss constraint is non-convergent, while
the time-averaged outgoing Poynting flux and energy injection from the
superradiance cloud are roughly consistent across resolutions to within an
$\mathcal{O}(1)$-factor (the Poynting flux estimates is shown in \figurename{
\ref{fig:FluxJouletotal}}). For all the numerical resolutions we considered, features emerged
on the grid scale, suggesting the endstate of the pair production cascade is a bulk turbulent state.
As can be seen in \figurename{ \ref{fig:forcefreequantities}}, no large scale electric field and
charge separation persists, while $E_z/E\sim\mathcal{O}(1)$. 
The numerical implementation, by construction, removes any violation of $B^2>E^2$ at
each grid point after each timestep, such that the ratio $B^2/E^2$, shown in
last panel in \figurename{ \ref{fig:forcefreequantities}}, is strictly larger than
unity. Similarly, the violations of $E_iB^i=0$ are at the level of floating
point error.

In the context of resistive magnetohydrodynamics, a macroscopic resistivity is introduced by means of a suitably chosen Ohm's law with conductivity $\sigma$. In order to recover the force-free approximation in the $\sigma\rightarrow\infty$ limit and to maintain a form for the electrodynamics that does not require one to also keep
track of the fluid dynamics, all resistive force-free approaches assume the drift velocity $v^i_d$ of the charges is altered as \cite{Gruzinov:2007se} (see also Ref.~\cite{Komissarov:2005xc} for a similar approach)
\begin{align}
\begin{aligned}
v^i_d=& \frac{\varepsilon^{ijk}E_j B_k}{B^2+E_0^2},\qquad E_0^2=B_0^2+E^2-B^2,\\
B_0^2=& \frac{1}{2}\left[B^2-E^2+\sqrt{(B^2-E^2)^2+4(E_iB^i)^2}\right] .
\end{aligned}
\end{align}
Thus, even for fields with $E^2>B^2$, the drift velocity $v^i_d$ is bounded by
the speed of light due to the additional electric field contribution $E_0^2$ in
the denominator compared with the force-free prescription
\eqref{eq:forcefreecurrent}. Non-vanishing $E_i B^i$ can only reduce the
resulting drift velocity further. This ensures that around current sheets
within a strongly magnetized plasma, the characteristic speeds remain physical.
Three distinct methods to construct $J^i_\perp$ have been considered in the
literature. In Ref.~\cite{Gruzinov:2008um}, the Ohm's law was applied in the
frame of vanishing charge density, referred to as \textit{(A)} in the
following. In Ref.~\cite{Li:2011zh} (see also Ref~\cite{Lyutikov:2003cz}), the
Ohm's law was applied in the minimal velocity (with respect to the ``lab" or
simulation frame) fluid frame, labeled as
\textit{(B)} in what follows. Lastly, the approach of
Ref.~\cite{Parfrey:2016caq} introduces resistive effects with a prescription
driving $E_i B^i$ towards $J^i B_i/\sigma$ over some arbitrary timescale
$1/\kappa$, called approach \textit{(C)} from here on. Beyond the drift
velocity, the three approaches \textit{(A)}, \textit{(B)}, and \textit{(C)}
differ. 

Comparing the three currents, method \textit{(A)} is manifestly covariant, but lacks a well-defined vacuum limit, while both \textit{(B)} and \textit{(C)} exhibit $J^i\rightarrow 0$ as $\sigma\rightarrow 0$. Since the superradiant system is well-understood a priori only in the vacuum limit, we focus on \textit{(B)} and \textit{(C)} in this discussion. Explicitly, the orthogonal component of the current \textit{(B)} constructed in Ref.~\cite{Lyutikov:2003cz, Li:2011zh} reads
\begin{align}
    J^a_{\perp,(B)}=\sigma E_0 \sqrt{\frac{B^2+E_0^2}{B_0^2+E_0^2}}\frac{E_0 E^a+B_0 B^a}{B^2+E_0^2}.
    \label{eq:currentii}
\end{align}
The prescription, \textit{(C)} modifies the force-free contribution as \cite{Parfrey:2016caq}
\begin{align}
    J^a_{\perp,(C)}=\frac{\sigma}{(\sigma+\kappa)}\left(J^a_{\perp,\text{FF}}+\kappa E^iB_i\frac{B^a}{B^2}\right).
    \label{eq:currentiii}
\end{align}
Here, the driving timescale $1/\kappa$ can be understood by contracting the above current by $B^a$, using the Maxwell equations to arrive at
\begin{align}
D_t(E_iB^i)=-\kappa N\left(E^i-\frac{1}{\sigma}J^i \right)B_i,
\label{eq:drivingequation}
\end{align}
such that $E_i B^i$ is driven towards $J_i B^i/\sigma$. Note, we assume the resulting system to be hyperbolic (see e.g., Refs.~\cite{Komissarov:2002my,Pfeiffer:2013wza} for a discussion).

Physically, these currents describe the interaction of an effective plasma with the visible electromagnetic fields, assuming $E^2,B^2\gg \rho_p,P_p$, where $\rho_p$ and $P_p$ are the plasma's mass density and pressure. The $\sigma\rightarrow 0$ limit corresponds to the vacuum limit. To understand this, consider the charge conservation $\nabla_\mu I^\mu=0$. In the Eulerian frame, together with \eqref{eq:currentdeceuler}, this leads to
\begin{align}
    D_t\rho_q=N\rho_q K-D_i(N J^i)=N\rho_q K-D_i(N\rho_q v_d^i)=0.
\end{align}
Hence, if the initial conditions satisfy $\rho_q=0$, then the system does not acquire a non-trivial charge distribution dynamically, i.e., in a medium with small conductivity charges cannot separate. This implies that the resistive currents above reduce to $I^\mu=0$, assuming the initial data is neutral. Therefore, the $\sigma=0$ regime is the vacuum limit of the system. Moving away from this limit to non-zero, but small conductivities, $\sigma\ll\mu$, the effective fluid coupling to the visible electromagnetic fields is an efficient insulator. The current $I^\mu$ is timelike, and $J^i$ is advection dominated. Due to the residual conductivity, the charges in the insulating fluid can move along the electric field with mobility $\sim \sigma/\mu$, i.e., the charge mobility in the fluid frame is conductivity suppressed. However, since the system is advection dominated, no large charge gradients can build up, unless the fluid is compressible. In our case, the fluid velocity in the Eulerian frame is $D_iv_d^i\neq 0$, resulting in potential charge pile-up in regions of large $v_d$-gradients and compressibility. At intermediate resistivity, $\sigma\sim \mu$, the insulating fluid transitions to a moderately conducting plasma. Here, the current $I^\mu$ is both locally spacelike and timelike in different places, and the system has advection and conduction dominated regions. For $\sigma\gg\mu$, the plasma turns into a highly-conducting plasma with only residual resistivity. Here, the advection of the fluid is a negligible contribution to the overall charge distribution. Large scale charge separation is enabled by large conduction currents along the electromagnetic fields. In this regime, the conductivity sets the diffusion length scale $\ell=1/\sigma$ that governs residual resistive features such as current sheets and tearing modes. Finally, assuming that in the $\sigma\rightarrow\infty$ limit, the system becomes largely magnetically dominated and $E_i B^i\rightarrow 0$ while $\sigma E_0$ remains finite, then all three currents reduce to the familiar and physically well-defined force-free limit. This is discussed further in the context of the superradiance system in App.~\ref{app:lowconductivityregime}.

\begin{figure}
    \centering
    \includegraphics[width=0.47\textwidth]{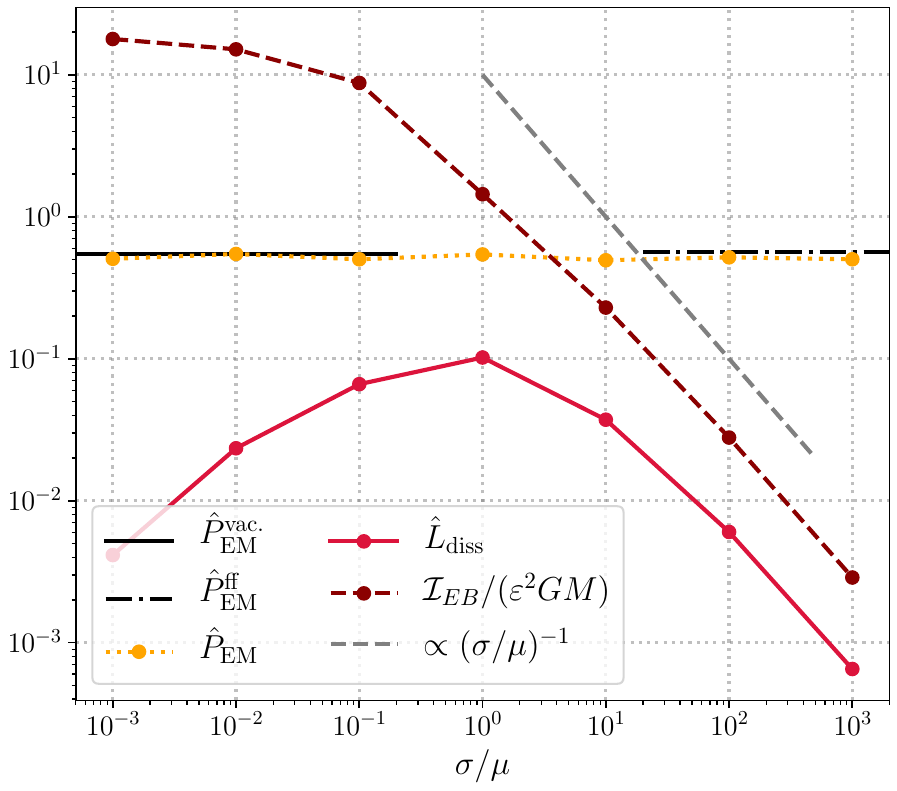}
    \caption{We show the rescaled energy emission rates, $\hat{P}=P(G/\varepsilon^2)(M/M_c)$, of the Poynting flux $P_\text{EM}$ (extracted at $r=10 r_c$), the Joule heating $L_\text{diss}$, and $\mathcal{I}_{EB}=\int d^3\sqrt{\gamma}|E_iB^i|$, as functions of conductivity $\sigma/\mu$ for model \textit{(C)} with a superradiance cloud of $\alpha=0.3$ on a Minkowski background with magnetic guide field $B_z=3\times 10^{4}B'_{\rm max}$. The corresponding Poynting fluxes in the vacuum $P_{\rm EM}^{\rm vac.}$ and force-free limits $P_{\rm EM}^{\rm ff}$ are indicated for reference. }
    \label{fig:flattests}
\end{figure}

In all cases, the conductivity $\sigma$ is to be understood as a proxy for a
class of dissipative processes and is chosen to be constant in space and time 
(primarily due to the lack of a physically motivated prescription for the
spatial dependence of conductivity in this setup), as as typically done, for
instance in \cite{Li:2011zh,Mahlmann:2020yxn} (see \cite{Parfrey:2016caq} for a
notable exception). The advantage of current \textit{(C)} is its numerical
properties in the high-conductivity limit. There, due to the prefactor 
$\sigma/(\sigma+\gamma)$ multiplying the orthogonal component, the magnitude of the
source of the Maxwell equations remains small, ensuring that the evolution
equations do not become stiff. This ultimately allows us to evolve the system even
at relatively large conductivities with moderate resolution within an explicit
forward integration scheme. However, a drawback of approach \textit{(C)} is
that $J^i_{\perp, (C)}$ diverges wherever $B^2=0$. In a magnetically dominated
pulsar magnetosphere, this does not lead to problematic behavior, while in the
case of an electrically dominated superradiance cloud, this causes issues at
moderate and high conductivities, since within the equatorial plane, the
magnetic field of the superradiance cloud smoothly transitions through zero. We
tested explicitly, that this magnetic null line causes the current
\eqref{eq:currentiii} to diverge in the intermediate and high conductivity
regime, leading to non-convergent features orbiting in the equatorial plane
(particularly on mesh-refinement boundaries). As the resistive methods outlined
above are designed to remove non-converging behavior in, for instance, current
sheets, and we require convergence of our numerical implementation in order to
validate our findings, approach \textit{(C)} is not well-suited to tackle the
kinetically-mixed superradiance cloud without modification. Therefore, we
resort to approach \textit{(B)} and current \eqref{eq:currentii} to model
resistive processes and the electromagnetic field geometries throughout the
superradiance cloud. This evolution method has a stiffness problem at large
conductivities, as outlined in App.~\ref{app:numericalsetup}, which
ultimately limits our ability to explore the $\sigma/\mu> 20$ parameter space.

We briefly illustrate the shortcomings of performing simulations on Minkowski
spacetime, and the extent to which current \textit{(C)} can be used in the
context of a magnetic guide field removing magnetic null lines. To that end, we
consider a $\alpha=0.3$ superradiance cloud of the form \eqref{eq:srprofiles}
on a fixed Minkowski background. The constant magnetic guide field is
initialized at the beginning of the simulations as $B^i_0=(0,0,B_z)^i$ (where
$\hat{z}$ is the spin-direction of the cloud), with magnitude $B_z=3\times 10^{4} B'_{\rm max}\gg
\varepsilon |\textbf{E}'| M$. We test that the following results are
independent of the choice of $B_z$, as long as the guide field magnitude is
larger than a threshold, $B_z>B_t$. Below this threshold, the electric field
$\varepsilon \mathcal{E}'$ starts dominating around the origin of the cloud.
With this construction, a series of simulations is performed varying the
conductivity from $\sigma/\mu=10^{-3}$ to $10^3$ within the context of the
resistive methods \textit{(C)} introduced above. In addition, we also study the
vacuum limit $J^a=0$, as well as the force-free limit on this flat
background. 

In Fig.~\ref{fig:flattests}, we show the behavior of the total power output of
the system as function of bulk conductivity in the model \textit{(C)} in
\eqref{eq:currentiii}. Let us compare these quantities to those obtained on
Kerr spacetime \textit{without} a guide field and using model \textit{(B)} 
[given by Eq.~\eqref{eq:currentii}] shown in \figurename{ \ref{fig:FluxJouletotal}} and
\figurename{ \ref{fig:limitandcurrent}}. The flat spacetime guide field setup
recovers the correct bulk dissipation component $L_{\rm diss}=L^{\rm bulk}_{\rm
diss}$, both in amplitude and in conductivity dependence, while the turbulent
component $L_{\rm diss}^{\rm turb}$ is absent. The latter is due to the
magnetic guide field removing any magnetically diffusive regions that might
form due to turbulence. The behavior of $\mathcal{I}_{EB}$ in \figurename{
\ref{fig:flattests}} is entirely analogous to the corresponding quantity in
\figurename{ \ref{fig:limitandcurrent}}. Lastly, the outgoing Poynting flux in
\figurename{ \ref{fig:flattests}} is constant across decades of conductivities,
and agrees well with both the vacuum and the force-free limits. This
illustrates the leakage discussed above due to the lack of gravitational
confining potential in flat spacetime, filtering out the massive propagating
states. This demonstrates that, within the interaction basis, the flat space
solution cannot be used to estimate physical observables associated with the
outgoing Poynting flux. Hence, we cannot use model \textit{(C)} without a guide
field, as discussed above. However, using a guide field also does not give
the correct answer, as this 
artificially removes the turbulent dynamics characterizing the
high-conductivity limit of the system.

\section{Charge distribution and small conductivity regime} \label{app:lowconductivityregime}

\begin{figure*}[t]
    \centering
    \includegraphics[width=0.98\textwidth]{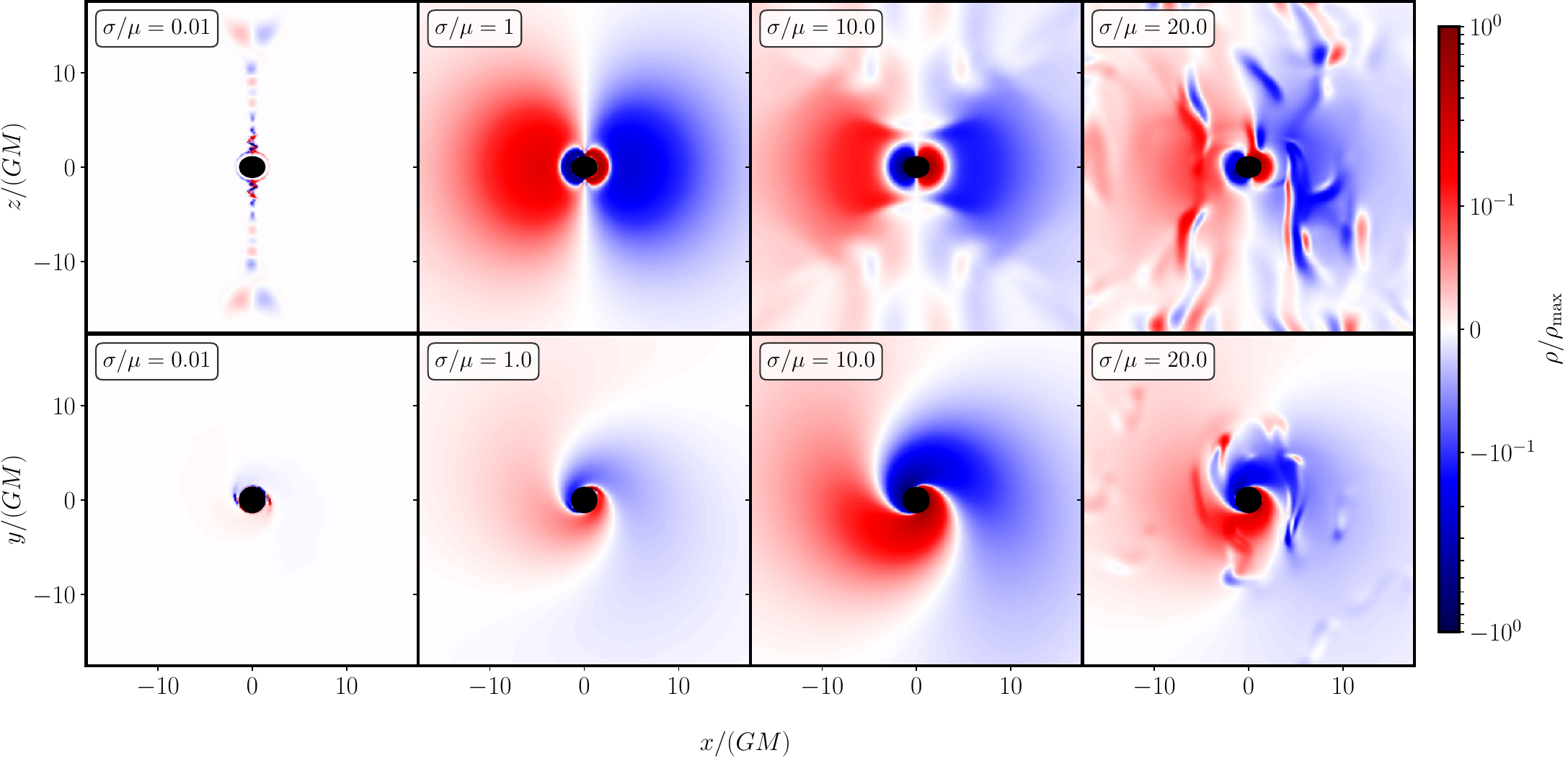}
    \caption{The charge distribution $\rho_q$ for various small to large conductivities in the equatorial plane (bottom row) and in a plane spanned by the BH spin and an arbitrary superradiance cloud phase (top row). The slices at varying conductivities correspond to the same superradiance cloud phase. We focus on a $\alpha=0.3$ and $a_*=0.86$ BH-cloud system.}
    \label{fig:chargedistribution}
\end{figure*}

In \figurename{ \ref{fig:chargedistribution}}, we illustrate the spatial charge
distribution $\rho_q$ of the solution at low and intermediate conductivities. At low
conductivity, the largest charge separation occurs along the spin-axis of the
BH. This may be interpreted as follows: Charge separation is suppressed at high resistivity, $1/\sigma \gg 1$.
However, any residual conductivity can separate charges on scales $\sim\sigma/\mu$ (assuming vanishing charge diffusion). Any separated charge distribution 
advects with the drift velocity of the fluid. In the presence of sufficiently large fluid
velocity gradients (with finite fluid compressibility) large charge
densities may build up. In the superradiance cloud context, 
regions of high velocity gradients coincide with regions
where the charge density is largest for $\sigma/\mu=10^{-2}$. Once
the charge density is accumulated, and the fluid velocity varies on scales larger
than the charge distribution scale, the latter is frozen into the flow of the
former and is carried away from the BH along the spin axis. At moderate and large 
conductivity, $\sigma/\mu\gtrsim 1$, the resistivity is sufficiently small as to enable large scale charge
separation. While for $\sigma/\mu=1$, the charge density follows roughly the superradiant electric field morphology (compare with \eqref{eq:density}), for $\sigma/\mu=20$, small scale features begin to appear, likely driven by the turbulent dynamics inside the plasma.

\begin{figure}[t]
    \centering
    \includegraphics[width=0.48\textwidth]{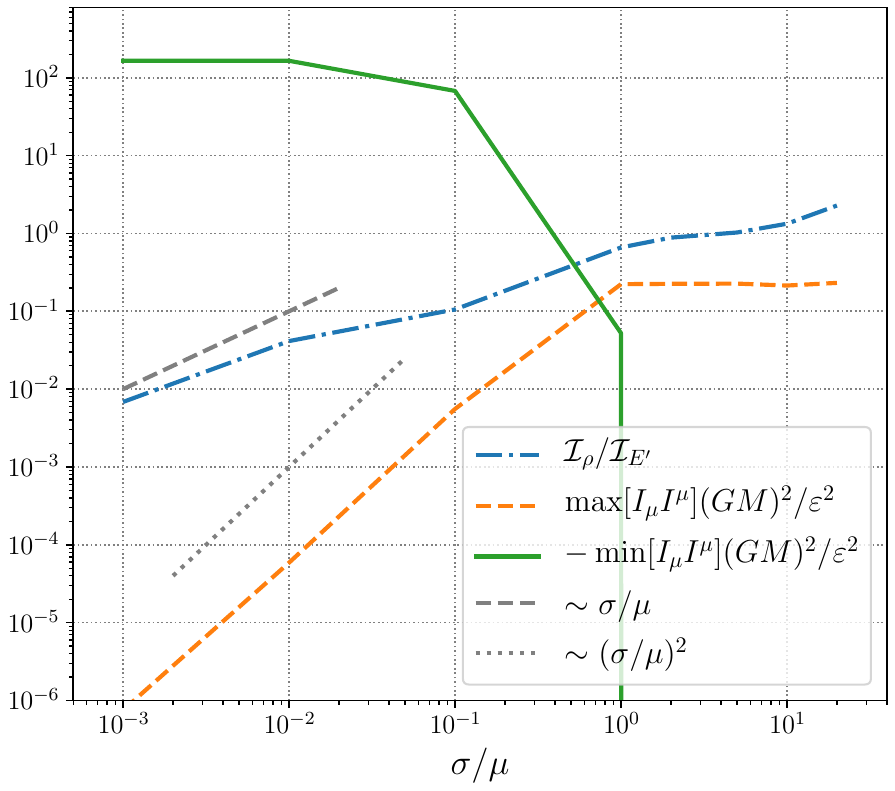}
    \caption{We show the charge separation $\mathcal{I}_{\rho}$ [defined in \eqref{eq:chargeseparation}] as a function of conductivity in units of $\mathcal{I}_{E'}$ [defined in \eqref{eq:superradiantchargeseparation}], as well as the global maximum $\max [I_\mu I^\mu]$ of the electromagnetic 4-current \eqref{eq:resistivecurrent} and global minimum $-\min [I_\mu I^\mu]$ (recall, we are using the $-+++$ signature). We focus on a $\alpha=0.3$ and $a_*=0.86$ BH-cloud system and consider conductivities $\sigma/\mu\in\{0.01,0.1,1,2,5,10,20\}$.}
    \label{fig:currentdiag}
\end{figure}

Let us demonstrate explicitly that at large resistivity, the plasma cannot charge
separate across scales larger than the charge separation scale $\sim\sigma$ and
that the superradiant electric field is screened efficiently at large conductivity. To that end, we define the quantity
\begin{align}
    \mathcal{I}_\rho=\int_Dd^3x\sqrt{\gamma}|\rho_q|,
    \label{eq:chargeseparation}
\end{align}
measuring the charge separation in a coordinate volume $D$ extending out to $\hat{\rho}=10r_c$ within the slice $\Sigma_t$ with
volume form $\sqrt{\gamma}d^3x$. This is compared with the charge separation required
screen the superradiant electric field $E'_i$ entirely [see also \eqref{eq:density}]:
\begin{align}
    \mathcal{I}_{E'}=\int_Dd^3x\sqrt{\gamma}\varepsilon|D_i E'^i|.    
    \label{eq:superradiantchargeseparation}
\end{align}
We show the behavior of $\mathcal{I}_\rho$ as
a function of conductivity, compared with $\mathcal{I}_{E'}$, 
in the right panel of \figurename{ \ref{fig:currentdiag}} (recall, $\mathcal{I}_{E'}$ is conductivity independent). 
For $\sigma/\mu\ll 1$, we find that
$\mathcal{I}_\rho\rightarrow 0$, indicating that the system tends to the vacuum
solution set by the superradiance cloud. At intermediate
conductivity, the charge separation scales roughly as
$\mathcal{I}_\rho\sim\sigma/\mu$. At large conductivity, $\mathcal{I}_\rho\sim
\mathcal{I}_{E'}$, supporting the conclusion that, for $\sigma/\mu\gg 1$, the
solution exhibits large scale charge separation that screens the field $E'_i$ 
efficiently, even in the turbulent regime. Furthermore, we also show the behavior of 
the norm $I_\mu I^\mu$ of the electromagnetic 4-current \eqref{eq:resistivecurrent}
in \figurename{ \ref{fig:currentdiag}}. As outlined in 
App.~\ref{app:resistivitytest}, for
$\sigma\ll\mu$, the current is fluid advection dominated, where a residual
charge distribution is flowing with the fluid on timelike trajectories.
Conversely, for large conductivities the conduction part of the current starts
to dominate, the current becomes spacelike, and the solution begins to 
asymptote towards a conductivity independent value of $\max[ I_\mu I^\mu]$.

\begin{figure*}[t]
    \centering
    \includegraphics[width=1\textwidth]{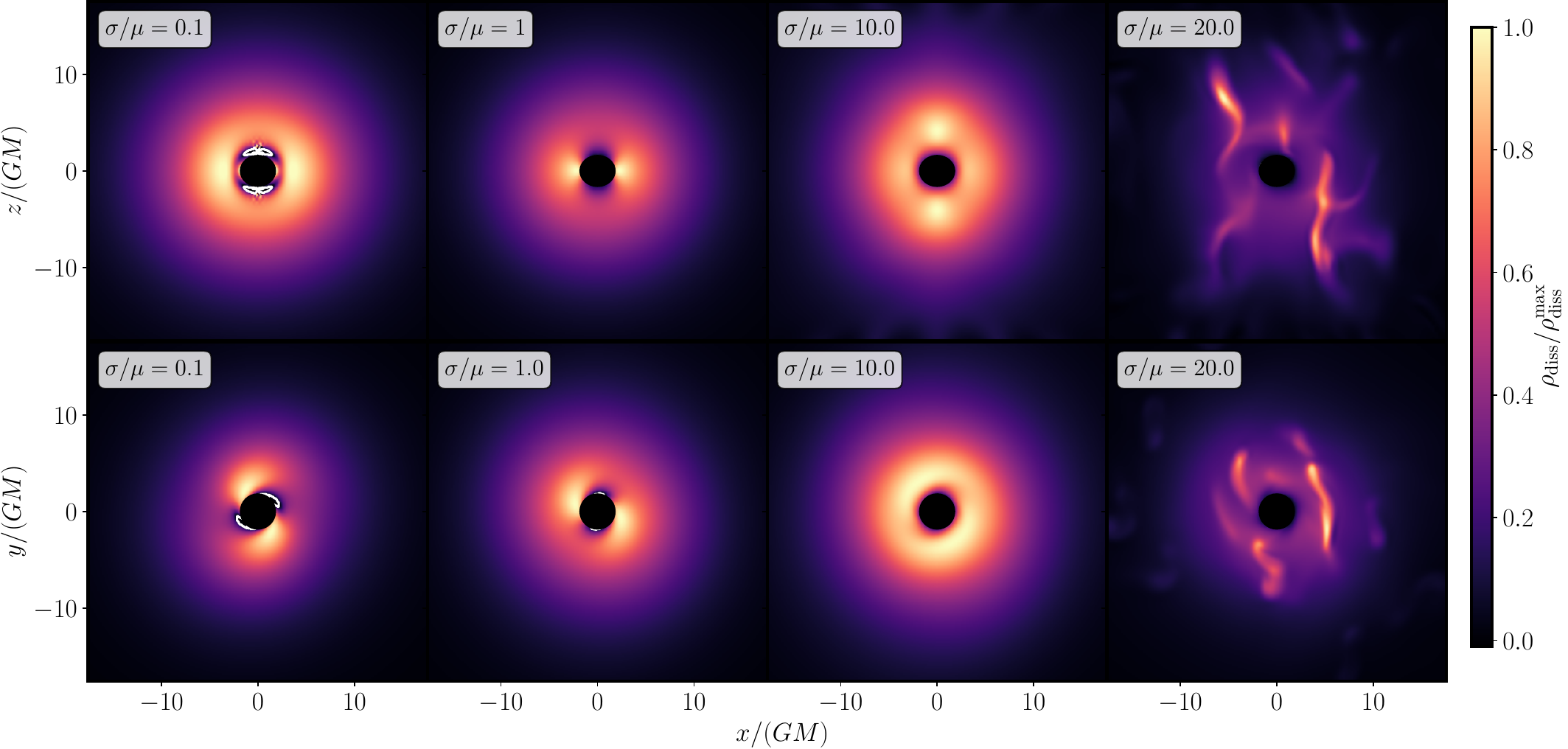}
    \includegraphics[width=1\textwidth]{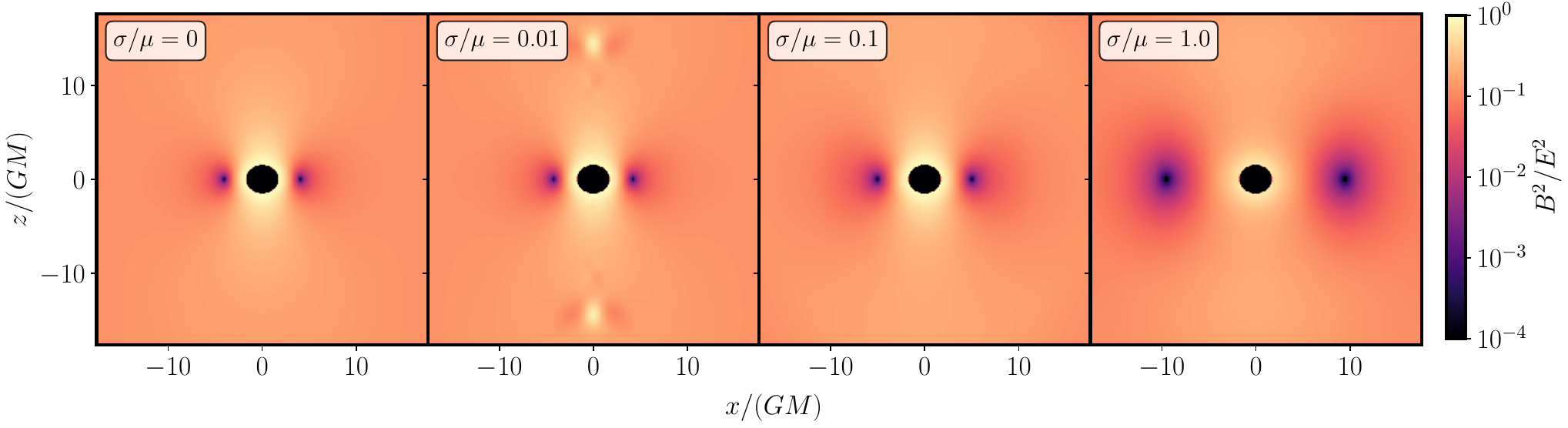}
    \caption{\textit{(top and middle row)} We show the energy dissipation density $\rho_{\rm diss}$, defined in \eqref{eq:dissipationdensity}, as a function of conductivity in a slice spanned by the BH spin and an arbitrary superradiance cloud phase \textit{(top row)}, as well as in the equatorial plane of the BH \textit{(middle row)}. The color scale is normalized by the global maximum dissipation density at each conductivity. All slices correspond to the same superradiance cloud phase. White contour lines indicate where the density goes through zero. We focus on a $\alpha=0.3$ and $a_*=0.86$ BH-cloud system. \textit{(bottom row)} We show the ratio of visible electromagnetic fields, $B^2/E^2$, in a plane spanned by the BH spin and an arbitrary superradiance cloud phase for various small and moderate conductivies. The $\sigma/\mu>1$ regime is shown in \figurename{ \ref{fig:condBsqoverEsq}}. We compare the plasma cases to the vacuum case, i.e., $\sigma/\mu=0$.}
    \label{fig:dissipationspatialdistribution}
\end{figure*}

In \figurename{ \ref{fig:dissipationspatialdistribution}}, we show the spatial
distribution of $\rho_{\rm diss}$ at small and large $\sigma/\mu$. For
$\sigma/\mu <10$, the dissipation density roughly follows the shape of the
superradiance cloud. This is consistent with \eqref{eq:dissipationdensity},
since the visible electric field is dominated by the superradiance electric
field component $E'^i$ for $\sigma/\mu\ll 1$. Hence, any small dissipation
density traces out the superradiance cloud's electric field $E'^i$; in the main
text, this component is referred to as $L_{\rm diss}^{\rm bulk}$. This is the
relativistic result corresponding to the approximation
\eqref{eq:nonrelcurrentdensity}. In \figurename{
\ref{fig:dissipationspatialdistribution}}, for $\sigma/\mu = 10$, the
dissipation density deviates from the superradiance cloud's electric field;
more precisely, the dissipation density is set by the superradiance cloud for
$r<r_*$, and set by the plasma dynamics for $r>r_*$, with $r_*$ defined in
\eqref{eq:rstar}. Therefore, the appearance of $r_*$ marks the breakdown of
approximation \eqref{eq:nonrelcurrentdensity}, and the onset of the
reconnection driven regime, leading to the turbulent dissipation component
$L_{\rm diss}^{\rm turb}$. The latter dominates over the dissipation density
component provided by the contribution of the dark photon electric field to the
visible electric fields in practically all regions outside the BH, for
$\sigma/\mu = 20$, as can be seen in \figurename{
\ref{fig:dissipationspatialdistribution}}. The dissipation density develops
features on the scale $1/\mu$, set by the boson mass, and $1/\sigma$, set by
the conductivity. This is consistent with the discussion in
Sec.~\ref{sec:Highlyconductinglimit}, where local magnetic field line twisting,
on scales of $1/\mu$ and scales of the entire cloud, are relaxed by
reconnection events, dissipating energy through a locally enhanced $\rho_{\rm
diss}$. For completeness, we show the ratio of visible electromagnetic fields
in \figurename{ \ref{fig:dissipationspatialdistribution}} at low to moderate
conductivities. This completes the low-conductivity regime of the behavior
shown in \figurename{ \ref{fig:Bsqlargesigma}}. As evident from \figurename{
\ref{fig:dissipationspatialdistribution}}, the field structure is affected at
the $\mathcal{O}(1)$-level only at intermediate conductivities,
$\sigma\sim\mu$. For $\sigma/\mu<1$, the two magnetic null lines inside the
equatorial plane are unchanged. In the case of $\sigma/\mu=10^{-2}$, the charge
distribution accumulating along the spin-axis of the BH, leaves mild imprints
on $B^2/E^2$. 

\section{Dark photon basis}\label{appendix:darkphoton}

A kinetic mixing between the SM photon and a dark massive photon enters as $\mathcal{L}=\mathcal{L}_\text{SM}+\mathcal{L}_\text{Proca}+\varepsilon F'_{\mu\nu}F^{\mu\nu}/2$ at low energies. Under the field redefinition ${A}_\mu\rightarrow {A}_\mu+\varepsilon {A}'_\mu=:\mathcal{A}_\mu$ this turns into the Lagrangian in the mass eigenbasis
\begin{align}
\begin{aligned}
    \mathcal{L}_\mathrm{mass}=-&\frac{1}{4}\mathcal{F}_{\mu\nu}\mathcal{F}^{\mu\nu}-\frac{1}{4}\mathcal{F}'_{\mu\nu}\mathcal{F}'^{\mu\nu}\\
    & -\frac{\mu^2}{2}\mathcal{A}'_\mu \mathcal{A}'^\mu+I_\mu (\mathcal{A}^\mu+\varepsilon \mathcal{A}'^\mu),
    \label{eq:masslagrangian}
\end{aligned}
\end{align}
or, using ${A}'_\mu\rightarrow A'_\mu+\varepsilon A_\mu=:{A}'_\mu$ into the interaction basis
\begin{align}
\begin{aligned}
    \mathcal{L}_\mathrm{inter}=-&\frac{1}{4}{F}_{\mu\nu}{F}^{\mu\nu}-\frac{1}{4}{F}'_{\mu\nu}{F}'^{\mu\nu}\\
    &-\frac{\mu^2}{2}{A}'_\mu {A}'^\mu-\varepsilon\mu^2 {A}'_\mu {A}^\mu+I_\mu {A}^\mu.
    \label{eq:interlagrangian_app}
\end{aligned}
\end{align}
In both bases, the current $I_\mu$ is the current of the SM charged particles. These lead to the field equations, in the mass eigenbasis,

\begin{align}
\begin{aligned}
    \nabla_\alpha \mathcal{F}^{\alpha\beta}=& \ -I^\beta,\\
    \nabla_\alpha \mathcal{F}'^{\alpha\beta}=& \ \mu^2 \mathcal{A}'^\beta-\varepsilon I^\beta,
    \label{eq:FieldeqMass}
\end{aligned}
\end{align}
and interaction basis

\begin{align}
\begin{aligned}
    \nabla_\alpha {F}^{\alpha\beta}=& \ -I^\beta+\varepsilon\mu^2{A}'^\beta,\\
    \nabla_\alpha {F}'^{\alpha\beta}=& \ \mu^2{A}'^\beta+\varepsilon\mu^2 {A}^\beta,
    \label{eq:FieldeqInteraction}
\end{aligned}
\end{align}
when working to leading order in the kinetic mixing $\varepsilon$. The mixing of the SM and the dark fields at the level of the electromagnetic current $I^\alpha$ is manifest in \eqref{eq:masslagrangian}. Hence, both the SM fields and the dark fields can accelerate charged particles. This is reflected in the energy-momentum conservation of \eqref{eq:masslagrangian}:
\begin{align}
\begin{aligned}
    \nabla_\alpha \mathcal{T}^{\alpha\beta}=& \ - \mathcal{F}^{\beta\gamma}I_\gamma,\\
    \nabla_\alpha  \mathcal{T}'^{\alpha\beta}=& \ -\varepsilon  \mathcal{F}'^{\beta\gamma}I_\gamma.
    \label{eq:EnergyeqMass}
\end{aligned}
\end{align}
In the mass eigenbasis, no energy is transferred from the dark to the SM field, while both transfer energy to and from charge particles. In the interaction basis, we have
\begin{align}
\begin{aligned}
    \nabla_\alpha {T}^{\alpha\beta}=& \ -{F}^{\beta\gamma}(I_\gamma-\varepsilon \mu^2 {A}'_\gamma),\\
    \nabla_\alpha {T}'^{\alpha\beta}=& \ \varepsilon\mu^2 {F}'^{\beta\gamma}{A}_\gamma,
    \label{eq:energymomentumconservation_app}
\end{aligned}
\end{align}
where the energy transfer between fields is manifest.

Furthermore, at leading order in $\varepsilon$ this implies the Lorenz condition on ${A}'_\mu$, as well as the current conservation 
\begin{align}
    \nabla^\mu {A}'_\mu=0, & & \nabla_\mu I^\mu=0.
\end{align}

In the main text, in particular after Sec.~\ref{sec:cloudsummary}, we work out all the dynamics in the interaction basis, which is most convenient for the analysis since, inside the dense plasma, ${A}_\mu$ has equations of motion which are potentially sensitive to the scale $\sigma$ and $\omega_p$, both of which are much larger than the dark photon mass $\mu$. This manifests in the simulation as short distance turbulent dynamics of the field ${A}_\mu$, while the background ${A}'_\mu$ has only dynamics on length scales of order $1/\mu$. Therefore,  inside this dense plasma, the interaction basis of the plasma modes ${A}_\mu$ and the dark photon ${A}'_\mu$ is also the {\it mass} basis. Clearly, the interaction basis is more convenient for our simulations. In the following, we emphasize some of the important physical intuition that is hidden in Eqs.~\eqref{eq:FieldeqMass},~\eqref{eq:FieldeqInteraction},~\eqref{eq:EnergyeqMass} and~\eqref{eq:energymomentumconservation_app} and clarifying some points of confusion.

A first confusion comes from searching for static solutions by inspection. From Eqs.~\eqref{eq:FieldeqMass} and~\eqref{eq:EnergyeqMass}, it seems apparent that there is a solution of $I_{\mu} = 0$ while Eqs.~\eqref{eq:FieldeqInteraction} and~\eqref{eq:energymomentumconservation_app} would naively suggest that there is a solution of $I^\beta = \varepsilon\mu^2{A}'^\beta$. Both of these two solutions we have in fact discussed in the main text. The solution $I_{\mu} = 0$ is the vacuum solution, which corresponds to a dark photon cloud with zero charged plasma. Such a solution is, however, unstable due to pair production instabilities described in Sec.~\ref{sec:SFQED}. The solution of $I^\beta = \varepsilon\mu^2{A}'^\beta$ in the interaction basis corresponds to the naive physical picture of a rotating electric dipole, which is not viable due to the fact that the cloud is electrically dominated (with size that is much larger than the light cylinder radius).

A second confusion pertains to the common lore that in a dense plasma, the effect of the dark photon is suppressed by the ratio of the dark photon mass and the plasma mass of the photon, usually in the form of $(\mu/\omega_p)^2$. However, this suppression assumes that the dark photon field weakly perturbs a dense fluid of SM particles, which is not true in our case. Rather, in the superradiance cloud, the dark photon cloud energy density scales as $|\vect{E}'|^2$, the visible electric field energy density scales as $\varepsilon^2 |\vect{E}'|^2$, while the charged particle energy density we obtain in the simulation is $\mathcal{O} (m_e\mu\varepsilon |\vect{E}'|)$. Given that $\varepsilon |\vect{E}'| \approx m_e^{3/2} \mu^{1/2}$ when pair production starts, the pair produced plasma carries energy density that is at most $\mathcal{O}((\mu/m_e)^{1/2})$ of the energy density of the electromagnetic field. In this case, the dark photon field is no longer a small perturbation and the SM plasma, as a result, is very far from an equilibrium state at zero field. As a result, the intuition developed in Refs.~\cite{Dubovsky:2015cca,Chaudhuri:2014dla} fails.

\section{Flux discussion} \label{app:fluxdiscussion}

\begin{figure*}[t]
    \centering
    \includegraphics[width=0.32\textwidth]{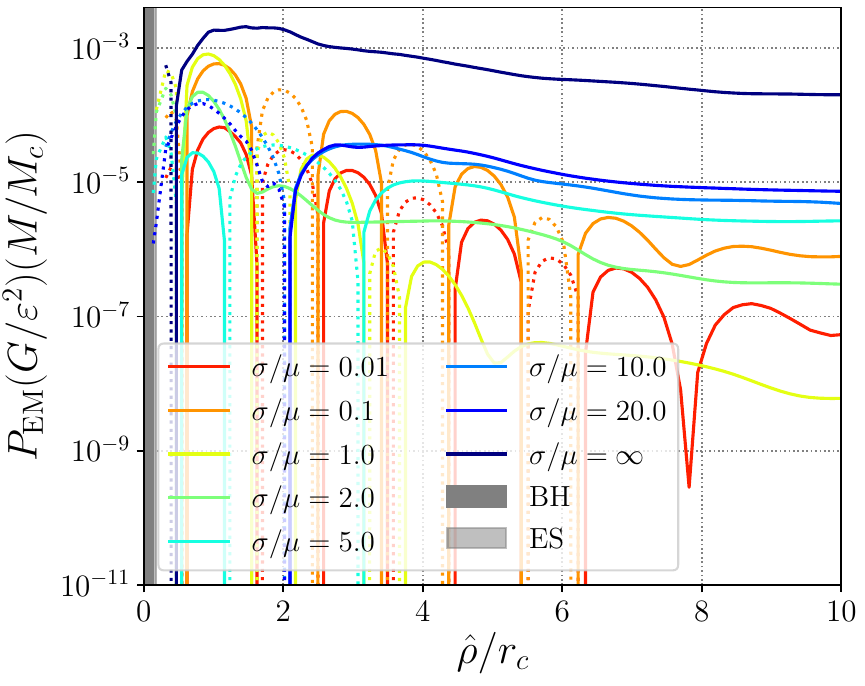}
    \hfill
    \includegraphics[width=0.32\textwidth]{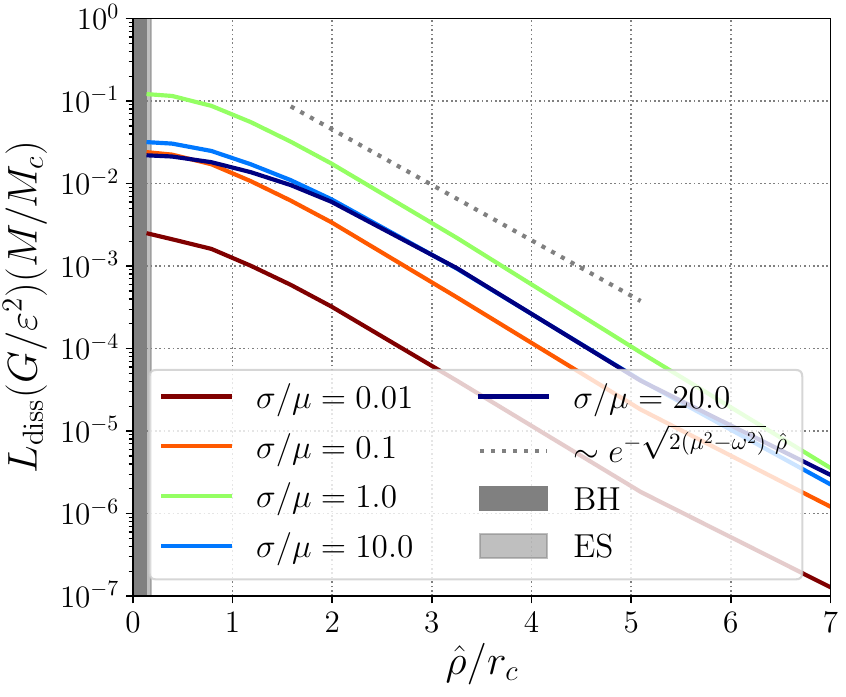}
    \hfill
    \includegraphics[width=0.32\textwidth]{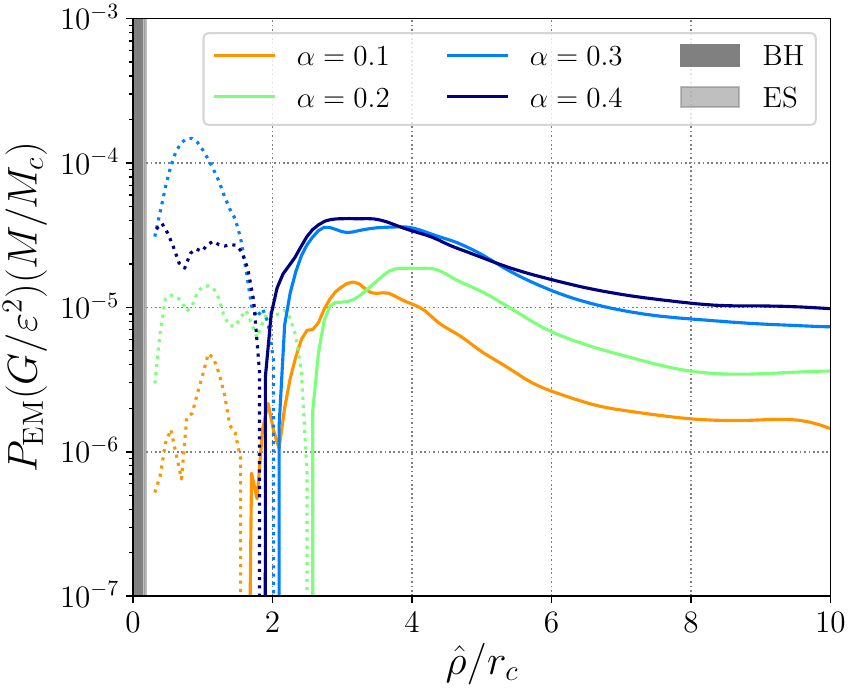}
    \caption{\textit{(left and middle)} The total (time-averaged) electromagnetic power emitted or injected into the plasma as a function of coordinate radius $\hat{\rho}$ in terms of the cloud's Bohr radius $r_c$ for various plasma conductivities $\sigma/\mu$. The flux obtained in the force-free context is labelled as $\sigma/\mu=\infty$. Here, we focus on the $\alpha=0.3$ case with a BH spin of $a_*=0.86$, and indicate the BH horizon and ergosurface by shaded regions labelled BH and ES, respectively. \textit{(left)} We show the total time-averaged visible Poynting flux $P_{\rm EM}$, defined in \eqref{eq:poyntingflux}, through spheres at radii $\hat{\rho}$ starting from the horizon and extending to large distances [here solid (dashed) lines indicate locally radially outwards (inwards) going fluxes]. \textit{(middle)} We show the total dissipation power $L_{\rm diss}$, defined in \eqref{eq:dissipativelosses}, integrated from $\hat{\rho}\rightarrow\infty$ to $\hat{\rho}$. Recall that our simulations assume spatially constant plasma conductivity $\sigma$. \textit{(right)} Here, we show the total time-averaged Poynting flux $P_{\rm EM}$ for all values of $\alpha$, keeping $\sigma/\mu=20$ fixed.}
    \label{fig:FluxJouleradiusAll}
\end{figure*}

In Sec.~\ref{sec:emissionpower}, we extrapolated the electromagnetic power through the Poynting flux $P_{\rm EM}$ and the energy dissipation $L_{\rm diss}$ from our numerical data at moderate conductivities $\sigma/\mu\leq 20$ to very large conductivities $\sigma\rightarrow\infty$. In \figurename{ \ref{fig:FluxJouleradius}}, we showed, however, only those scenarios with conductivities resulting in qualitatively different behavior of the Poynting flux. Therefore, for completeness, we show the electromagnetic emission power for all values of the conductivity considered in \figurename{ \ref{fig:FluxJouleradiusAll}}, as function of the radial coordinate distance $\hat{\rho}$ from the BH. The exponential decay and oscillatory behavior of $P_{\rm EM}$ in the low-conductivity regime reflects the  exponential decay and oscillatory behavior of the electromagnetic waves in a medium with low conductivity. These electromagnetic waves are eigenstates of the Helmholtz equation in spherical coordinates, with eigenvalues of $\pm i \sqrt{\sigma \mu + i \mu^2}$. The eigenfunctions are spherical Hankel functions of the first kind, which have a similar spatial dependence (this oscillatory behavior can be observed also in e.g. $B^2$ in the equatorial plane of the BH-cloud system for $\sigma/\mu<1$, which is not shown here). Such exponential decay and oscillatory behavior is evident in \figurename{ \ref{fig:FluxJouleradiusAll}} close to the BH at low and intermediate conductivities. In this regime, low frequency oscillations on large scales dominate, which gives rise to the same oscillatory features in $P_{\rm EM}$. Finally, in the right panel of \figurename{ \ref{fig:FluxJouleradiusAll}}, we show the total time-averaged Poynting flux for $\sigma/\mu=20$ as a function of coordinate distance from the BH for each of the values of $\alpha$ considered in this work.

\bibliographystyle{apsrev4-1}
\bibliography{bib.bib}

\end{document}